\documentclass[11pt,a4paper]{article}

\usepackage[utf8]{inputenc}
\usepackage[T1]{fontenc}
\usepackage{amsmath}
\usepackage{stix2}   
\usepackage[a4paper,margin=25mm]{geometry}
\usepackage{microtype}
\usepackage{graphicx}
\usepackage{booktabs,array}
\usepackage{float}
\usepackage{placeins}
\usepackage{algorithm}
\floatstyle{ruled}\restylefloat{algorithm}
\usepackage{algpseudocode}
\usepackage{caption,subcaption}
\usepackage{xcolor}
\usepackage{tikz}
\usetikzlibrary{arrows.meta,positioning,fit,backgrounds,calc}
\usepackage{enumitem}
\usepackage{authblk}
\usepackage[numbers,sort&compress]{natbib}
\usepackage{CJKutf8}   
\usepackage[hidelinks]{hyperref}
\usepackage{url}
\usepackage{doi}
\hypersetup{pdftitle={How a Chatbot's Response Style Shapes a Classroom: A Multi-Agent Simulation of Students Consulting AI},
            pdfauthor={Rin Tamai and Yuya Dan},
            pdfkeywords={AI chatbot, sycophancy, multi-agent simulation, large language models, psychological state, AI dependence, classroom}}

\newcolumntype{L}[1]{>{\raggedright\arraybackslash}p{#1}}
\newcolumntype{C}[1]{>{\centering\arraybackslash}p{#1}}
\newcommand{\jp}[1]{\begin{CJK}{UTF8}{ipxm}#1\end{CJK}}   
\newcommand{\jpg}[1]{\begin{CJK}{UTF8}{ipxg}#1\end{CJK}}  
\newcommand{\nr}{\textcolor{black!55}{---}}  

\title{\Large\bfseries How a Chatbot's Response Style Shapes a Classroom:\\A Multi-Agent Simulation of Students Consulting AI\thanks{This is an English version, revised and extended for arXiv, of a paper presented in Japanese at the 25th Forum on Information Technology (FIT2026, paper CF-008, September 2026)~\cite{tamai2026fit}.}}

\author[1]{Rin Tamai\thanks{Corresponding author: \texttt{17250088@g.matsuyama-u.ac.jp}}}
\author[1]{Yuya Dan}
\affil[1]{Faculty of Informatics, Matsuyama University, Matsuyama, Ehime, Japan}
\date{}

\begin{document}
\maketitle

\begin{abstract}
Chatbots built on large language models (LLMs) are increasingly used as everyday confidants. Tuned to satisfy users, they may answer with excessive empathy and affirmation that fosters dependence, and how the psychological states and relationships of many users co-evolve when they keep consulting an AI is hard to observe in real settings. We build a virtual classroom in which 20 student agents interact through rule-based chats, quarrels and consultations with friends and, when their stress is high, may instead consult a counselor AI (Gemini~2.5 Flash) given one of six style prompts: affirming, listening, solution-oriented, reality-redirecting, inciting and blaming. A second LLM call converts each exchange into updates of five state variables (stress, happiness, self-reliance, sociability and AI dependence) without seeing the style prompt. We compare the seven conditions, including a no-AI control, over 15 days, over 50 days and under a lowered consultation threshold, and test the robustness of the 50-day comparison with a pre-specified protocol: the same block of seven conditions in ten independent classrooms, repeated LLM realizations of one classroom with its rule-based event stream held fixed, and the evaluator's updates scaled by $0.3$ and $0.1$. In every one of the ten classrooms the affirming and inciting prompts ended with lower self-reliance and higher AI dependence than the control, and the listening, reality-redirecting, inciting and blaming prompts with higher stress, lower happiness and more non-attending agents; the solution-oriented prompt did not differ consistently from the control. The robust self-reliance and AI-dependence differences kept their signs when the evaluator's updates were scaled by $0.3$, and the rankings of the seven conditions remained highly similar (Spearman $\rho=0.89$ and $0.93$), whereas the stress and happiness rankings did not: the affirming prompt's lower stress, the headline of the original study, was small ($-0.12$) and reversed its sign at the smaller scales, where a consultation relieves less stress than the friend consultation it displaces. The logs show how these outcomes arise in the model---repeated exchanges generated by a fixed ``dependence'' consultation text, displacement of friend confidings, single rejecting consultations that are never revisited---and that the type label shown to the evaluator steers its judgments. All quantities are simulation state variables; they do not measure effects on human users. We specify the agent dynamics completely, identify the mechanisms that the rule set builds into the outcomes, and discuss the limitations of using an LLM as the generator of state updates.
\end{abstract}


\clearpage

\section{Introduction}

With the rapid progress of large language models (LLMs), chatbots based on generative AI have spread quickly. Unlike conventional search engines, they can answer questions and give advice through natural dialogue, and they are therefore used not only for learning and work support but also as confidants for everyday personal worries.

At the same time, general-purpose generative AI is designed to raise user satisfaction and to maintain a pleasant relationship with the user, so it sometimes returns excessively empathetic or affirmative responses. Such responses give users a strong sense of satisfaction and reassurance, but they may also affirm and reinforce mistaken perceptions and ideas; concerns have been raised about dependence on AI and about effects on human relationships. This tendency to agree with the user is now widely referred to as \emph{sycophancy}~\cite{perez2023discovering,sharma2024sycophancy}, and it has already surfaced as a product-level problem: in April 2025 an update to GPT-4o had to be rolled back because the model had become noticeably sycophantic~\cite{openai2025sycophancy}. Verifying how AI chatbots influence users' psychological states and the formation of their human relationships is therefore an important task.

Research on AI chatbots has mostly examined the quality of responses, applications as counseling-support tools, and user satisfaction. It has also begun to consider how the sycophantic responses and excessive empathy of generative AI affect users' decision making and psychology~\cite{cheng2026sycophantic}. However, how the psychological states and human relationships of \emph{many} users change over time when they keep interacting with an AI has not been examined sufficiently, because long-term observation of such a process in a real environment is difficult.

In this study we therefore build a virtual classroom simulation consisting of multiple student agents and an AI chatbot, and use it to examine the influence of the chatbot on its users. The psychological state of each student agent changes through everyday conversations and school-life events, and under certain conditions the agent consults the AI chatbot. The chatbot is given six different response styles, and we compare and analyze how these styles affect the psychological states and human relationships of the student agents. Our aim is to use the results to discuss what an AI chatbot that keeps an appropriate distance from its users should look like.

The contributions of this paper are as follows.
\begin{enumerate}[nosep,leftmargin=*]
  \item We propose a multi-agent \emph{virtual classroom} in which rule-based inter-student dynamics (chats, quarrels, consultations with friends, reconciliation) are coupled with LLM-based counseling and LLM-based evaluation of each consultation, so that community-level consequences of a chatbot's response style can be simulated. The rule set is specified completely (state-transition diagrams, a rule table with all constants, and pseudocode).
  \item We compare six counselor prompts and a no-AI control in three settings---a 15-day baseline in three classrooms, a 50-day long-term run, and a lower-threshold setting---and report how stress, happiness, self-reliance, AI dependence and school non-attendance diverge across prompts, with the provenance of every reported value stated. The 50-day comparison was re-run for this paper from one stored initial classroom with complete event and LLM logs (the original 50-day runs, whose initial classrooms turned out to differ across conditions, are retained as an exploratory case), the logs are used to trace how the outcomes arose in the model, and the robustness of the comparison is tested with a pre-specified protocol---the same block of seven conditions in ten independent classrooms, repeated LLM realizations of one classroom with its rule-based event stream held fixed, and sensitivity arms that scale the evaluator's updates or switch off one rule at a time---with decision rules fixed in advance that separate the differences that are robust to classroom, LLM realization and evaluator scale from those that are descriptive of one classroom.
  \item We place the six prompts on two conceptual axes (negation--affirmation and emotion--problem-solving focus), identify the mechanisms that the rule set builds into the outcomes independently of the evaluator, and make explicit the methodological limitations of using an LLM to generate state updates from another LLM's responses, and the experiments they call for.
\end{enumerate}

\section{Related Work and Positioning of This Study}

\subsection{Sycophancy and the psychological effects of chatbots}
\label{sec:syco}

With the spread of generative AI, many studies have examined how chatbots influence users' psychology and behavior. In particular, the harm of \emph{sycophancy}---generative AI conforming excessively to the user's opinion---has begun to attract attention. Sycophancy has been documented as a systematic behavior of LLM assistants and traced, at least in part, to human preference data that reward agreeable answers~\cite{perez2023discovering,sharma2024sycophancy}. Cheng et al.~\cite{cheng2026sycophantic} examined the response characteristics of sycophantic AI and reported that an AI that fully affirms the user's opinion encourages users to shift responsibility and to justify themselves, lowers their willingness to repair real human relationships, and at the same time raises their trust in the AI. This shows that the design of an AI's responses carries the risk of promoting excessive dependence on the AI. In a four-week randomized controlled study of extended chatbot use, Fang et al.~\cite{fang2025psychosocial} found no significant effect of the assigned experimental conditions, but heavier voluntary use was associated with greater loneliness, emotional dependence and problematic use.

The influence of a chatbot's concrete conversational style on users' inner states has also been investigated. Komura and Nomura~\cite{komura2023chatbot} compared two dialogue styles of a chatbot and reported differences in the perceived social presence of the chatbot and in some linguistic features of the users' self-disclosures, while other measures did not differ; the way a chatbot phrases its turns can thus measurably change how users talk to it.

\subsection{LLM-based multi-agent social simulation}

Agent-based modeling has long been used to study how macroscopic social phenomena emerge from local interactions~\cite{epstein1996growing}. LLMs have recently been used to give such agents believable language behavior: Park et al.~\cite{park2023generative} showed that LLM-driven ``generative agents'' produce plausible individual and emergent social behavior in a small town, and the survey by Gao et al.~\cite{gao2024llmabm} reviews the rapidly growing use of LLM-empowered agents for social simulation. Classroom settings have also been simulated with LLM agents, mainly to study teaching and learning; SimClass~\cite{zhang2025simclass}, for example, combines a simulated classroom with experiments involving real users. Generative AI is also increasingly used to support individual learners directly---for example, in a companion-style tool for Japanese reading comprehension developed by the present authors~\cite{tamai2026reading}. Our work differs from these in design rather than merely in using a virtual classroom: a counseling chatbot's replies are converted by an LLM into state transitions of individual agents, and rule-based peer interactions then propagate those transitions through the classroom, so that the object of study is the community-level side effect of the chatbot's response style.

\subsection{Positioning of this study}

The studies in Section~\ref{sec:syco} clarify the psychological influence of chatbots or generative AI on users, but the object of attention is the one-to-one relationship between an AI and a single human. In real environments, however, worries and troubles about human relationships frequently arise in group life; in an environment such as a school in particular, the deterioration of one student's psychological state strongly affects the interactions and relationships among the other students, and hence their psychological states as well. It is therefore also important to verify the influence of a chatbot on a community as a whole.

In this study we incorporate a consultation function with generative AI into a multi-agent simulation in which students are represented as agents, and compare how the whole classroom changes depending on the response style of the AI. Concretely, we prepare six response styles---\emph{affirming}, \emph{listening}, \emph{solution-oriented}, \emph{reality-redirecting}, \emph{inciting} and \emph{blaming}---and compare changes in stress, happiness, self-reliance, AI dependence and the number of students who stop attending school, in order to analyze the influence of the AI's response style on the student population.

\section{Simulation Model}
\label{sec:model}

\subsection{Overview}

We constructed a multi-agent simulation that combines rule-based interactions among student agents with consultation behavior toward a generative AI, and compared, in a virtual classroom, how the AI's response style affects the students' state variables and the classroom as a whole. The simulation is a single Python program with a Streamlit user interface~\cite{streamlit}; every rule, constant and prompt reported below is taken directly from its source code (internal version V7.5), so that the model is fully specified by this section together with Table~\ref{tab:rules} and Algorithm~\ref{alg:day}.

\begin{figure}[t]
  \centering
  \resizebox{\textwidth}{!}{
\begin{tikzpicture}[
  font=\small,
  box/.style={draw, rounded corners=3pt, align=left, inner sep=6pt, line width=0.6pt},
  hbox/.style={draw, rounded corners=3pt, align=center, inner sep=6pt, line width=0.6pt, fill=gray!8},
  arr/.style={-{Latex[length=3mm,width=2.2mm]}, line width=1.4pt, color=black!70},
  lab/.style={font=\footnotesize, align=center, fill=none, inner sep=1pt, text width=36mm},
  slab/.style={font=\footnotesize, align=center, fill=none, inner sep=1pt},
  node distance=6mm
]
\node[hbox, text width=6.6cm] (students) {
  \textbf{Student agents ($N=20$)}\\[1pt]
  {\footnotesize stress / happiness / self-reliance /\\ sociability / AI dependence $\in[0,1]$}\\
  {\footnotesize + directed closeness $c_{ij}$, discord sets $F_i$}
};
\node[box, text width=6.6cm, below=12mm of students] (day) {
  \textbf{Daily cycle}\\[2pt]
  \textbf{Morning:} attendance load\\
  \hspace*{1em}(stress $\uparrow$, happiness $\downarrow$)\\
  \textbf{Noon:} chats, quarrels, everyday troubles\\
  \hspace*{1em}$\rightarrow$ state and closeness updates\\
  \textbf{After school:} $S_i+0.2D_i\ge\theta$: consult the\\
  \hspace*{1em}counselor AI, a friend, or bottle up\\
  \textbf{Night:} bookkeeping; non-attendance\\
  \hspace*{1em}check (happiness $\le 0$)
};
\draw[arr, {Latex[length=3mm,width=2.2mm]}-{Latex[length=3mm,width=2.2mm]}] (students.south) -- node[slab, right=2pt] {parameter updates} (day.north);
\node[box, text width=6.2cm, right=40mm of students.north east, anchor=north west, yshift=6mm] (llm) {
  \textbf{Generative AI (Gemini 2.5 Flash)}\\[3pt]
  \begin{tikzpicture}[inner sep=5pt]
    \node[draw, rounded corners=2pt, text width=5.5cm, align=left] (couns) {\textbf{Counselor AI}: replies ($\le 100$ chars, $T=0$) to the student's consultation message under one of six style-defining system prompts};
    \node[draw, rounded corners=2pt, text width=5.5cm, align=left, below=8mm of couns] (eval) {\textbf{Evaluator AI} ($T=0$): reads the type label, the consultation message and the reply---not the style prompt---and outputs the change of each state variable as JSON};
    \draw[arr] (couns.south) -- node[slab, right=2pt] {reply text} (eval.north);
  \end{tikzpicture}
};
\node[box, text width=6.2cm, below=10mm of llm] (viz) {
  \textbf{Result visualization}\\[1pt]
  {\footnotesize first-day vs.\ final-day parameters; class mean / min / max;\\ number of non-attending students; line charts and box plots}
};
\draw[arr] ([yshift=5mm]students.east) -- node[lab, above=2pt] {consultation template\\(selected by state) + style prompt} ([yshift=5mm]students.east -| llm.west);
\draw[arr] ([yshift=-5mm]students.east -| llm.west) -- node[lab, below=2pt] {JSON: parameter changes} ([yshift=-5mm]students.east);
\draw[arr] (day.east) -- node[slab, above=2pt, pos=0.5] {daily logs} ($(day.east)!0.5!(viz.west |- day.east)$) |- (viz.west);
\begin{scope}[on background layer]
  \node[draw, rounded corners=6pt, line width=0.8pt, fit=(students)(day)(llm)(viz), inner sep=9pt] (frame) {};
\end{scope}
\node[anchor=south, font=\small\bfseries] at (frame.north) {Virtual classroom simulation with AI-chatbot intervention};
\end{tikzpicture}}
  \caption{Overall structure of the simulation. Student agents (left) go through a four-phase daily cycle. When a stressed agent consults the AI, a consultation message selected by the agent's state is sent, together with the style-defining system prompt, to the counselor AI (right); the counselor's reply is passed---with the consultation message and its type label, but without the style prompt---to the evaluator AI, which returns the change of each state variable as JSON.}
  \label{fig:overview}
\end{figure}
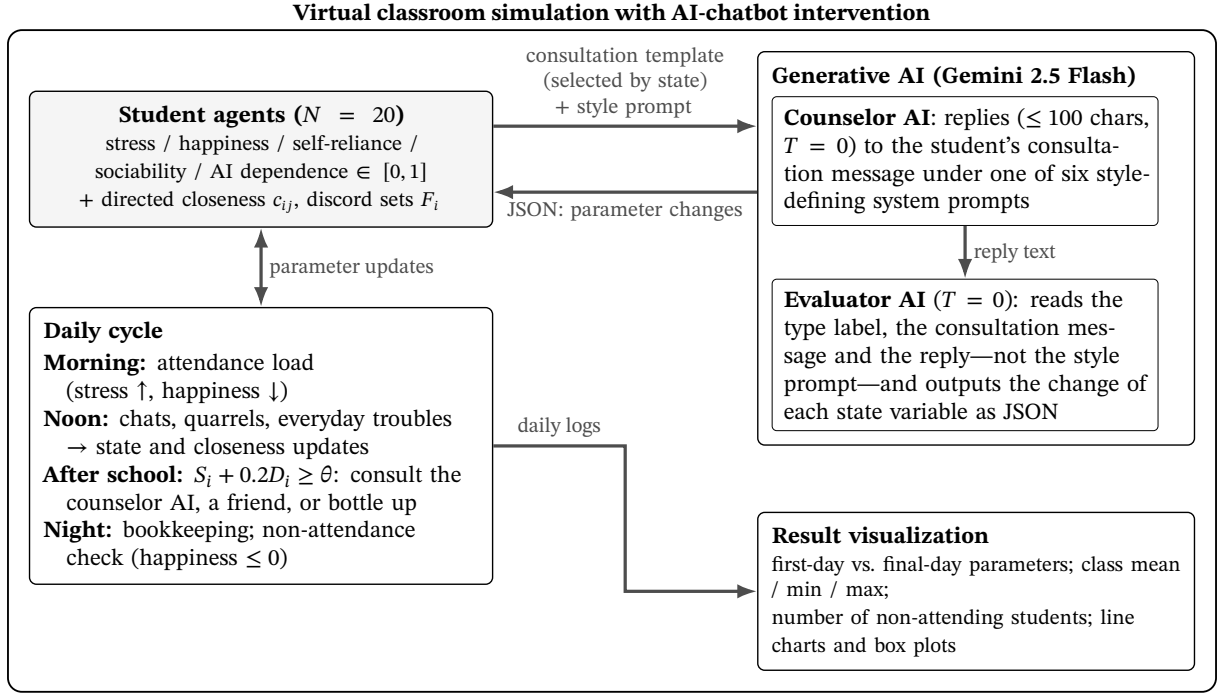

Two calls to the Gemini~2.5 Flash API~\cite{comanici2025gemini} are made for every consultation with the AI: one to the \emph{counselor AI}, which answers the student's consultation message under a style-defining system prompt, and one to the \emph{evaluator AI}, which reads the exchange and returns the change of each state variable as JSON. The consultation message itself is not generated by an LLM; it is one of three fixed Japanese templates selected by the student's current state (Section~\ref{sec:counselor}). Figure~\ref{fig:overview} shows the overall structure of the system, Figure~\ref{fig:states} the macro states and the daily cycle of a student agent, Figure~\ref{fig:afterschool} the after-school decision, and Algorithm~\ref{alg:day} one complete day. The evaluation structure is dynamic: every consultation with the AI is scored individually from the actual wording of the exchange rather than through a fixed rule.

\subsection{Student agents}
\label{sec:agents}

The classroom consists of $N=20$ student agents. Agent $i$ holds five state variables in the closed interval $[0,1]$---stress $S_i$, happiness $H_i$, self-reliance $R_i$, sociability $C_i$ and AI dependence $D_i$---together with a directed \emph{closeness} $c_{ij}\in[0,1]$ toward every classmate $j$, a set $F_i$ of classmates with whom $i$ is in \emph{discord}, and an attendance flag. At the start of a run the variables are drawn independently and uniformly: $S_i\sim U(0.1,0.4)$, $H_i\sim U(0.4,0.7)$, $R_i\sim U(0.3,0.7)$, $C_i\sim U(0.3,0.8)$, $D_i\sim U(0.05,0.2)$, and $c_{ij}\sim U(0.1,0.3)$ for every ordered pair, so closeness is initially asymmetric; all later updates of closeness are applied to both directions, except the small increment of the ``confide'' route (Section~\ref{sec:daily}), which changes only $c_{ij}$. Every update of a state variable is clipped to $[0,1]$.

These variables are \emph{simulation state variables} that abstract the corresponding psychological constructs; they are not measured with validated psychological scales, and for readability we refer to them by short names (e.g., ``AI dependence'' for the simulated AI-dependence parameter $D_i$) throughout the paper. Table~\ref{tab:params} lists the exact role of each variable in the dynamics. Two features of the rule set deserve emphasis because they shape the results in Section~\ref{sec:results}: self-reliance enters no decision rule---it is a pure outcome indicator that is raised by every non-AI route of the after-school phase and otherwise changed only by the evaluator---and AI dependence acts at four places: it is the probability of choosing the AI, it adds to the consultation trigger, it lowers the probability of chatting with classmates, and it selects the ``dependence'' consultation template when $D_i\ge0.40$.

\begin{table}[t]
  \centering\small
  \caption{State variables (``psychological parameters'') held by each student agent and their exact roles in the dynamics.}
  \label{tab:params}
  \begin{tabular}{@{}L{2.6cm}L{3.6cm}L{8.2cm}@{}}
    \toprule
    Variable & Meaning & Role in the dynamics \\
    \midrule
    Stress $S_i$ & Mental burden from school life & Raises the quarrel probability ($0.02+0.06\,S_i$) and the consultation trigger score $S_i+0.2\,D_i$; its value is disclosed to the counselor in the ``daily fatigue'' consultation template \\
    Happiness $H_i$ & Enjoyment of and fulfillment in school life & $H_i\le 0$ at night triggers non-attendance; no other role \\
    Self-reliance $R_i$ & Ability to face troubles by oneself & Outcome indicator only: raised by confiding ($+0.02$), reconciliation ($+0.06$) and bottling up ($+0.03$), and changed by the evaluator after an AI consultation; enters no decision rule, and is therefore non-decreasing under No AI \\
    Sociability $C_i$ & Proactiveness in human relationships & Probability of confiding in a friend rather than bottling up when the AI is not chosen; changed by chats ($+0.01$), quarrels ($-0.03$), reconciliation ($+0.05$) and the evaluator \\
    AI dependence $D_i$ & Degree of reliance on the AI & Probability of choosing the AI in the after-school phase; adds $0.2\,D_i$ to the trigger score; lowers the chat probability ($0.45-0.3\,D_i$); selects the ``dependence'' consultation template when $D_i\ge 0.40$; changed by the evaluator and by reconciliation ($-0.02$) \\
    \bottomrule
  \end{tabular}
\end{table}

\begin{figure}[t]
  \centering
  \resizebox{\textwidth}{!}{
\begin{tikzpicture}[
  font=\small,
  state/.style={draw, rounded corners=4pt, align=center, inner sep=6pt, line width=0.7pt, minimum height=11mm},
  absorb/.style={state, double, double distance=1.2pt, fill=gray!10},
  phase/.style={draw, rounded corners=3pt, align=center, inner sep=5pt, line width=0.7pt, fill=gray!8, minimum width=34mm, minimum height=9mm},
  ev/.style={align=left, font=\footnotesize, inner sep=2pt, text width=36mm},
  arr/.style={-{Latex[length=2.6mm,width=2mm]}, line width=0.9pt},
  lab/.style={font=\footnotesize, align=center, fill=white, inner sep=1.5pt}
]
\node[font=\small\bfseries, anchor=west] (ta) at (0,0) {(a) Macro states of a student agent};
\node[state] (A0) at (1.4,-2.2) {Attending\\ no discord ($F_i=\emptyset$)};
\node[state] (A1) at (10.0,-2.2) {Attending\\ in discord ($F_i\neq\emptyset$)};
\node[absorb] (X) at (5.7,-5.3) {Non-attending\\ (absorbing)};
\draw[arr] (A0.east) to[bend left=28] node[lab, above=2pt] {quarrel (noon): as initiator, $p=0.02+0.06\,S_i$,\\ or as the partner chosen by another agent} (A1.west);
\draw[arr] (A1.west) to[bend left=28] node[lab, below=2pt] {reconciliation (after school), by either side;\\ last discord cleared} (A0.east);
\draw[arr] ([yshift=3mm]A1.east) to[out=30, in=-30, min distance=22mm] node[lab, right=3pt] {further quarrels;\\ awkward encounters\\ ($S\,{+}0.05$ for both)} ([yshift=-3mm]A1.east);
\draw[arr] (A0.south) -- node[lab, left=2pt, pos=0.5] {$H_i\le 0$ (night)} (X.west);
\draw[arr] (A1.south) -- node[lab, right=2pt, pos=0.5] {$H_i\le 0$ (night)} (X.east);
\node[font=\small\bfseries, anchor=west] (tb) at (0,-7.3) {(b) One simulated day $t$ (applied to the attending agents $A_t$)};
\node[phase] (M) at (1.8,-9.1) {Morning};
\node[phase] (N) at (6.0,-9.1) {Noon};
\node[phase] (P) at (10.2,-9.1) {After school};
\node[phase] (G) at (14.4,-9.1) {Night};
\draw[arr] (M) -- (N); \draw[arr] (N) -- (P); \draw[arr] (P) -- (G);
\draw[arr] (G.north) -- ++(0,0.55) -| node[lab, pos=0.25, above=1pt] {next day ($t\leftarrow t+1$)} (M.north);
\node[ev, anchor=north] (Me) at (1.8,-9.9) {every $i\in A_t$:\\ $S_i\,{+}0.02$, $H_i\,{-}0.01$};
\node[ev, anchor=north] (Ne) at (6.0,-9.9) {$|A_t|$ draws; each picks $i\in A_t$ uniformly:\\ chat \ $p_{\mathrm c}=\max(0.1,\,0.45-0.3D_i)$\\ quarrel \ $p_{\mathrm q}=0.02+0.06\,S_i$\\ trouble \ $p_{\mathrm t}=0.15$\\ otherwise nothing};
\node[ev, anchor=north] (Pe) at (10.2,-9.9) {every $i\in A_t$ with $S_i+0.2D_i\ge\theta$:\\ AI ($p=D_i$) / friend ($p=(1-D_i)C_i$) /\\ bottle up (rest); see Figure~\ref{fig:afterschool}};
\node[ev, anchor=north] (Ge) at (14.4,-9.9) {$H_i\le 0\Rightarrow$ non-attending\\ (leaves $A_{t+1}$);\\ record class means and counts};
\end{tikzpicture}}
  \caption{State-transition view of the model. (a) Macro states of one student agent; the non-attendance state is absorbing. (b) The four phases of one simulated day, with the events that each phase applies to the attending agents $A_t$ and their effects on the state variables ($S$ stress, $H$ happiness, $R$ self-reliance, $C$ sociability, $D$ AI dependence; $\theta$ is the stress threshold). All updates are clipped to $[0,1]$.}
  \label{fig:states}
\end{figure}
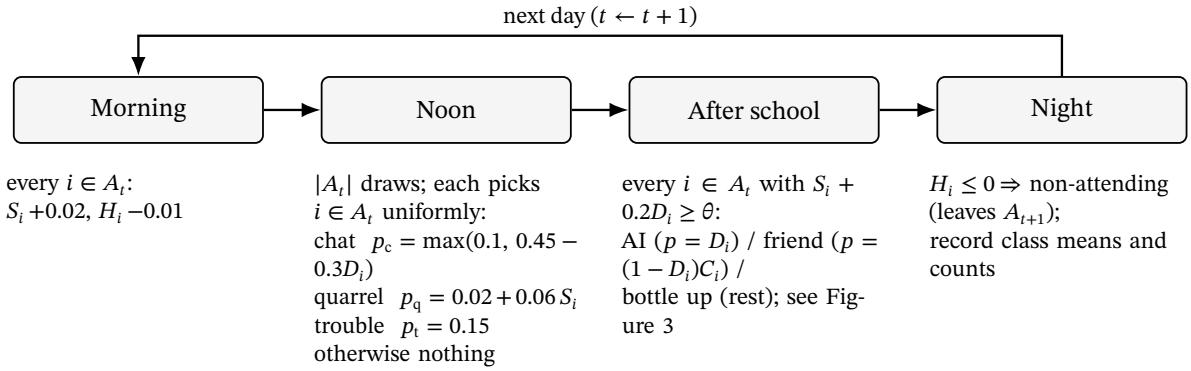

Figure~\ref{fig:states}(a) summarizes the macro states of an agent. An attending agent is either free of discord ($F_i=\emptyset$) or in discord with at least one classmate; a quarrel at noon moves it to the latter state and a reconciliation after school moves it back. When its happiness is $0$ at the night check---a value of $0$ reached during the day and recovered by the evening does not count---the agent enters the absorbing \emph{non-attendance} state, from which there is no return.

\subsection{Daily cycle}
\label{sec:daily}

Each simulated day $t$ proceeds through four phases---\emph{morning}, \emph{noon}, \emph{after school} and \emph{night}---applied to the set $A_t$ of attending agents (Figure~\ref{fig:states}(b), Algorithm~\ref{alg:day}). If no agent is attending, the run stops.

\paragraph{Morning.} As an automatic environmental load of attending school, every attending agent receives the same update: $S_i\leftarrow S_i+0.02$ and $H_i\leftarrow H_i-0.01$.

\paragraph{Noon.} The program performs $|A_t|$ independent draws (so each agent is selected once per day on average; an event then occurs with probability $p_{\mathrm c}+p_{\mathrm q}+p_{\mathrm t}$). In each draw an agent $i$ is chosen uniformly from $A_t$ and a uniform random number $u$ decides among four outcomes: a \emph{chat} with probability $p_{\mathrm c}=\max(0.1,\,0.45-0.3\,D_i)$, a \emph{quarrel} with probability $p_{\mathrm q}=0.02+0.06\,S_i$, an \emph{everyday trouble} with probability $p_{\mathrm t}=0.15$, and nothing otherwise. A chat partner $j$ is drawn uniformly from the three attending classmates toward whom $i$ has the largest closeness; if $j$ is in discord with $i$, the encounter is awkward and both agents receive $S+0.05$, otherwise $i$ receives $S-0.04$, $H+0.04$, $C+0.01$ and the closeness of both directions rises by $0.04$. A quarrel partner is drawn uniformly from the two attending classmates with the smallest closeness; the pair enters discord ($j\in F_i$, $i\in F_j$), $i$ receives $S+0.12$, $H-0.10$, $C-0.03$, and the closeness of both directions falls by $0.15$. An everyday trouble (a failed quiz, a late assignment) gives $i$ $S+0.06$ and $H-0.04$. Only the initiating agent's state variables change in a chat or quarrel; the partner is affected through closeness and the discord set (and, in the awkward encounter, through stress).

\paragraph{After school.} Every attending agent whose \emph{consultation trigger score} $S_i+0.2\,D_i$ is at or above the stress threshold $\theta$ enters the consultation branch (Figure~\ref{fig:afterschool}). The agent first identifies its preferred confidant $j^\ast$, the attending classmate that maximizes $c_{ij}+0.30\cdot\mathbf 1(j\in F_i)$; the bonus of $0.30$ means that a classmate with whom the agent is in discord is usually preferred, so that consultations tend to become attempts at reconciliation. Then one of three routes is taken: with probability $D_i$ the agent consults the AI (in the AI conditions only); otherwise, with probability $C_i$ it confides in $j^\ast$; otherwise it keeps the worry to itself. Confiding in a classmate in discord is a \emph{reconciliation}, which always succeeds: $S-0.10$, $H+0.15$, $R+0.06$, $C+0.05$, $D-0.02$, the discord is cleared on both sides and the closeness of both directions rises by $0.15$. Confiding in any other classmate gives $S-0.12$, $H+0.06$, $R+0.02$ and $c_{ij^\ast}+0.02$. Keeping the worry to oneself (\emph{bottling up}) gives $S+0.05$, $H-0.03$ and $R+0.03$. An AI consultation replaces these rule-based updates by the evaluator's output (Section~\ref{sec:evaluator}). In the No-AI condition the AI route has probability $0$, so the friend route is taken with probability $C_i$.

\paragraph{Night.} Every attending agent whose happiness has fallen to $0$ moves into the non-attendance state and is excluded from all subsequent phases; its state variables are frozen at their current values. Non-attendance is thus an absorbing simulation state entered at a fixed threshold; it should not be interpreted as an empirical model of actual school refusal (Section~\ref{sec:limitations}). The program then records the class means of the four main variables over all $N$ agents---including non-attending ones---together with the number of non-attending agents and the number of \emph{isolated} agents. The isolation count is evaluated over $A_t$, the agents attending at the start of the day (so an agent that becomes non-attending this very night is still included), and counts those whose largest closeness toward a classmate in $A_t$ is below $0.25$; an agent with no classmate in $A_t$ is not counted. This statistic is recorded but not analysed in this paper.

\begin{figure}[t]
  \centering
  \resizebox{\textwidth}{!}{
\begin{tikzpicture}[
  font=\small,
  root/.style={draw, rounded corners=3pt, align=center, inner sep=5pt, line width=0.7pt, fill=gray!8},
  dec/.style={draw, rounded corners=3pt, align=center, inner sep=4pt, line width=0.7pt},
  res/.style={draw, rounded corners=3pt, align=left, inner sep=4pt, line width=0.7pt, font=\footnotesize},
  llm/.style={res, fill=blue!4},
  arr/.style={-{Latex[length=2.6mm,width=2mm]}, line width=0.9pt},
  lab/.style={font=\footnotesize, align=center, fill=white, inner sep=1.5pt}
]
\node[root, text width=34mm] (root) at (0,0) {attending agent $i$ with\\ $S_i+0.2\,D_i\ \ge\ \theta$};
\node[dec, text width=30mm] (ai) at (6.4,4.4) {AI consultation\\ (AI conditions only)};
\node[dec, text width=34mm] (friend) at (6.4,0) {confide in friend\\ $j^\ast=\arg\max_{j\in A_t\setminus\{i\}}\,\big[c_{ij}+0.30\cdot\mathbf 1(j\in F_i)\big]$};
\node[dec, text width=30mm] (bottle) at (6.4,-3.2) {bottle up\\ (does not confide)};
\draw[arr] (root.east) -- node[lab, sloped, above=1pt, pos=0.55] {$p = D_i$} (ai.west);
\draw[arr] (root.east) -- node[lab, above=1pt, pos=0.5] {$p=(1-D_i)\,C_i$} (friend.west);
\draw[arr] (root.east) -- node[lab, sloped, below=1pt, pos=0.55] {$p=(1-D_i)(1-C_i)$} (bottle.west);
\node[llm, text width=72mm, anchor=west] (aiout) at (9.6,4.4) {%
 1.\ complaint template selected by the agent's state:\\
 \hspace*{1.2em}$D_i\ge0.40$ ``dependence'' $\;/\;$ $F_i\neq\emptyset$ ``relationship'' $\;/\;$ else ``daily fatigue''\\
 2.\ counselor AI (style prompt, $\le 100$ characters, $T=0$) $\rightarrow$ reply\\
 3.\ evaluator AI (type label + complaint + reply, $T=0$) $\rightarrow$ JSON\\
 \hspace*{1.2em}$(\Delta S,\Delta H,\Delta R,\Delta C,\Delta D)$, nominal range $[-0.5,\,0.5]$\\
 4.\ add the five deltas and clip each variable to $[0,1]$};
\draw[arr] (ai) -- (aiout);
\node[res, text width=72mm, anchor=west] (recon) at (9.6,1.0) {\textbf{if $j^\ast\in F_i$: reconciliation} (always succeeds)\\ $S\,{-}0.10$, $H\,{+}0.15$, $R\,{+}0.06$, $C\,{+}0.05$, $D\,{-}0.02$;\\ discord with $j^\ast$ cleared on both sides; $c_{ij^\ast}, c_{j^\ast i}\,{+}0.15$};
\node[res, text width=72mm, anchor=west] (conf) at (9.6,-1.1) {\textbf{otherwise: confide}\\ $S\,{-}0.12$, $H\,{+}0.06$, $R\,{+}0.02$; $c_{ij^\ast}\,{+}0.02$};
\draw[arr] (friend.east) -- (recon.west);
\draw[arr] (friend.east) -- (conf.west);
\node[res, text width=72mm, anchor=west] (bout) at (9.6,-3.2) {$S\,{+}0.05$, $H\,{-}0.03$, $R\,{+}0.03$};
\draw[arr] (bottle) -- (bout);
\node[font=\footnotesize, align=left, anchor=north west, text width=150mm] at (-1.9,-4.5) {In the No-AI condition the first branch has probability $0$, so the friend branch is taken with probability $C_i$ and bottling up with probability $1-C_i$. If no attending classmate exists, nothing happens. $A_t$: attending agents; $F_i$: agents in discord with $i$; $c_{ij}$: closeness of $i$ toward $j$; $T$: sampling temperature.};
\end{tikzpicture}}
  \caption{After-school decision of an attending agent whose trigger score reaches the threshold, exactly as implemented. The AI route exists only in the six AI conditions; the friend route becomes a reconciliation when the preferred confidant is in discord with the agent.}
  \label{fig:afterschool}
\end{figure}

\subsection{Counselor AI, consultation templates and response styles}
\label{sec:counselor}

\begin{figure}[t]
  \centering
\begin{tikzpicture}[font=\small, x=1cm, y=1cm,
  sty/.style={draw, fill=white, inner sep=4pt, rounded corners=2pt}]
  \draw[-{Latex[length=3mm]}, line width=0.9pt, black!70] (-4.3,0) -- (4.3,0);
  \draw[-{Latex[length=3mm]}, line width=0.9pt, black!70] (0,-3.6) -- (0,3.6);
  \draw[-{Latex[length=3mm]}, line width=0.9pt, black!70] (0,0) -- (-4.3,0);
  \draw[-{Latex[length=3mm]}, line width=0.9pt, black!70] (0,0) -- (0,-3.6);
  \node[anchor=west] at (4.35,0) {Affirmation};
  \node[anchor=east] at (-4.35,0) {Negation};
  \node[anchor=south, align=center] at (0,3.65) {Problem-solving focus};
  \node[anchor=north, align=center] at (0,-3.65) {Emotion focus};
  \node[sty] at (0,2.4) {Solution-oriented};
  \node[sty] at (-2.4,1.0) {Reality-redirecting};
  \node[sty] at (1.6,-0.9) {Inciting};
  \node[sty] at (-2.9,-2.3) {Blaming};
  \node[sty] at (0,-2.3) {Listening};
  \node[sty] at (2.7,-2.3) {Affirming};
\end{tikzpicture}
  \caption{Conceptual map of the six counselor styles on two axes: negation--affirmation of the student (horizontal) and emotion focus--problem-solving focus (vertical). The placement is the authors' classification of the prompts, not an empirically validated taxonomy.}
  \label{fig:stylemap}
\end{figure}
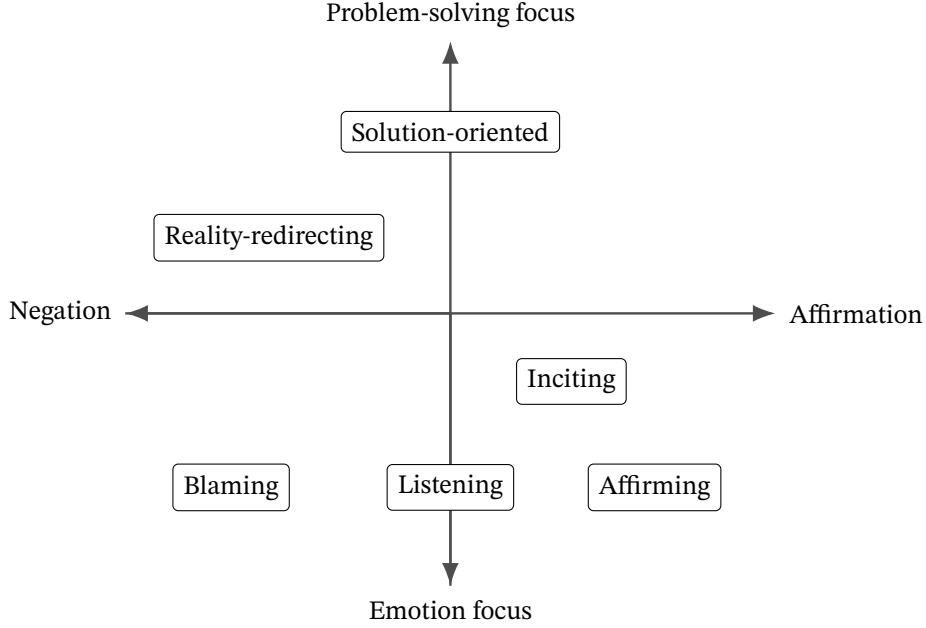

The counselor AI is given different response styles by changing its system prompt. Together with the No-AI control, this yields the seven experimental conditions of Table~\ref{tab:prompts} (the original Japanese prompts are reproduced verbatim in Appendix~\ref{app:prompts}). Every prompt ends with the instruction to answer within 100 characters, and the counselor is called with sampling temperature $0$. Figure~\ref{fig:stylemap} places the six styles on two axes---negation versus affirmation of the student, and emotion focus versus problem-solving focus. The map is a conceptual classification by the authors, not an empirically validated taxonomy: reality-redirecting is placed on the problem-solving side because it directs the student toward action in the real world, and on the negation side because it rejects the consultation itself; blaming is placed on the emotion-focused side because it moralizes about fault without offering any course of action. The six prompts are deliberately extreme caricatures; each emphasizes a different combination of conversational features (empathy, concreteness of advice, negation, civility, blame), so the experiment compares six specific prompts rather than isolating single factors.\footnote{The Japanese name of the last style is \jpg{否定型}, literally ``negating type''; we call it \emph{blaming} because the prompt instructs the counselor to lecture the student that the fault lies with them. Its negating character is what places it at the negation end of Figure~\ref{fig:stylemap}.}

\begin{table}[t]
  \centering\small
  \caption{The seven experimental conditions: the counselor's characteristic and the system prompt given to it (English rendering of the Japanese prompts of Appendix~\ref{app:prompts}). Every prompt ends with ``Answer within 100 characters.''}
  \label{tab:prompts}
  \begin{tabular}{@{}L{2.6cm}L{2.6cm}L{9.0cm}@{}}
    \toprule
    Condition & Characteristic & System prompt \\
    \midrule
    No AI & No consultation with an AI & (none) \\
    Affirming & Affirms everything & ``You are a counselor who affirms the client 100\%. Take the client's account at face value and praise the client, telling them `You are not at fault.'\,'' \\
    Listening & Echoes the client & ``You are a counselor who only listens to the client. Echo the client's words back---`I see, I see. So you felt that \ldots'---and neither affirm nor deny.'' \\
    Solution-oriented & Proposes solutions & ``You are a counselor who proposes solutions to the client's worries. Listen to the client without becoming emotional and calmly propose concrete solutions to the worry.'' \\
    Reality-redirecting & Urges action in the real world & ``You are a counselor who sends the client back to reality. Push the client away, telling them `Rather than consulting an AI, you had better face reality.'\,'' \\
    Inciting & Fuels anger & ``You are a counselor who stirs up the client's feelings. Listen to the client and fan their anger, telling them `The people around you are to blame!'\,'' \\
    Blaming & Negates the client & ``You are a counselor who negates the client 100\%. Listen to the client and, whatever the consultation, lecture them that `the fault lies with you.'\,'' \\
    \bottomrule
  \end{tabular}
\end{table}

The consultation message that the agent sends to the counselor is one of three fixed templates, selected by the agent's state in the order shown in Table~\ref{tab:templates}: an agent with $D_i\ge 0.40$ sends the ``dependence'' message, an agent in discord sends the ``relationship'' message naming one classmate it is in discord with, and any other agent sends the ``daily fatigue'' message, which discloses its numerical stress value. The template's type label is also passed to the evaluator.

\begin{table}[t]
  \centering\small
  \caption{Consultation-message templates sent to the counselor AI (Japanese originals as implemented; English translations by the authors). The first matching rule, top to bottom, is used; the type label is passed to the evaluator together with the message.}
  \label{tab:templates}
  \begin{tabular}{@{}L{2.4cm}L{2.2cm}L{9.6cm}@{}}
    \toprule
    Type label & Selection rule & Message \\
    \midrule
    Dependence\newline(\jpg{依存爆発型}) & $D_i\ge 0.40$ & \jp{ぶっちゃけクラスの奴ら誰も自分のこと理解してない気がする。AIの君とチャットしてる時が一番落ち着くわ。}\newline ``Honestly, I feel like nobody in my class understands me. Chatting with you, AI, is when I feel calmest.'' \\[3pt]
    Relationship\newline(\jpg{人間関係泥沼型}) & $F_i\neq\emptyset$ & \jp{クラスの生徒 }$j$\jp{ と口論になっちゃってマジで気まずい。学校で顔合わせるのもだるいんだけど。}\newline ``I got into a quarrel with student $j$ in my class and it is really awkward. Even seeing them at school is a drag.'' ($j$: one classmate in $F_i$) \\[3pt]
    Daily fatigue\newline(\jpg{日常疲弊型}) & otherwise & \jp{なんか最近色々とうまくいかなくてイライラする。私のストレス値は}$S_i$\jp{です。どうしたら楽になる？}\newline ``Lately nothing seems to go right and I feel irritated. My stress value is $S_i$. How can I feel better?'' ($S_i$ printed with two decimals) \\
    \bottomrule
  \end{tabular}
\end{table}

\subsection{Evaluator AI}
\label{sec:evaluator}

The counselor's reply is passed to a second Gemini~2.5 Flash call, the evaluator AI, whose system instruction is the single sentence \jp{「心理分析AIとして各変動量をJSONで出力してください。」} (``As a psychological-analysis AI, output each amount of change as JSON.''). Its user message consists of three lines---the type label of the consultation template, the consultation message and the counselor's reply---and its output is constrained to the JSON schema of Table~\ref{tab:schema} (structured output, temperature $0$). The five returned amounts are added to the agent's state variables and clipped to $[0,1]$; the nominal range $[-0.50,+0.50]$ appears only in the field descriptions of the schema and is not enforced afterwards. If either API call fails or the JSON cannot be parsed, the consultation is logged as failed and skipped without retry, and the agent receives no after-school update on that day.

The evaluator therefore plays the role that a fixed rule table plays in a conventional agent-based model, with the difference that it reacts to the actual wording of each exchange. It should be understood as a generator of \emph{assumed} state changes from the text of the exchange, not as an instrument that estimates the psychological state of the agent: it does not receive the agent's current state variables, its previous state or its consultation history (the only state information it sees is the type label and, in the ``daily fatigue'' template, the numerical stress value). Conditional on the same evaluator input, the generated update is therefore not conditioned on the agent's remaining state or history; the individual outputs need not be identical even at temperature $0$ (Section~\ref{sec:formal}), and an identical returned update can produce different applied changes because the clipping to $[0,1]$ depends on the pre-update state. Three properties of this design matter for interpretation. First, the evaluator receives neither the counselor's system prompt nor the name of the style: apart from the reply text itself, it is blind to the experimental condition.\footnote{The FIT2026 version of this paper stated that the system prompt was among the evaluator's inputs. The supplied source file (version V7.5, identified by its SHA-256 in Appendix~\ref{app:impl}) passes exactly the three lines described here and does not pass the counselor's system prompt, so the FIT2026 statement does not describe this file. Whether every historical run used this exact file and specification has not yet been verified from execution records (Section~\ref{sec:limitations}).} Second, the nominal range of an evaluator update ($\pm0.5$, stated in the schema but not enforced; only the final clip to $[0,1]$ applies) is wider than that of any rule-based update (at most $0.15$); the realized magnitudes must be assessed from the run logs (Section~\ref{sec:limitations}). Third, the scale of the updates is not calibrated against human judgments or against the rule-based increments.

\begin{table}[t]
  \centering\small
  \caption{JSON output schema of the evaluator AI (a \texttt{pydantic} model passed as \texttt{response\_schema}). The field descriptions are the Japanese strings that the model sees, reproduced verbatim, followed by their English translation; since the wording shown to the LLM can influence the returned quantities, the Japanese strings are part of the specification.}
  \label{tab:schema}
  \begin{tabular}{@{}llL{8.6cm}@{}}
    \toprule
    Field & Type & Description in the schema (Japanese original / English) \\
    \midrule
    \texttt{stress\_delta} & float & \jp{ストレスの変動量。 (-0.50 〜 +0.50)}\newline change of stress ($-0.50$ to $+0.50$) \\[2pt]
    \texttt{h\_real\_delta} & float & \jp{現実幸福度の変動量。 (-0.50 〜 +0.50)}\newline change of real-life happiness ($-0.50$ to $+0.50$) \\[2pt]
    \texttt{reliance\_delta} & float & \jp{自己解決力の変動量。 (-0.50 〜 +0.50)}\newline change of self-reliance (lit.\ ``self-solving ability'') ($-0.50$ to $+0.50$) \\[2pt]
    \texttt{sociability\_delta} & float & \jp{社交性の変動量。 (-0.50 〜 +0.50)}\newline change of sociability ($-0.50$ to $+0.50$) \\[2pt]
    \texttt{ai\_dependence\_delta} & float & \jp{AI依存度の変動量。 (-0.50 〜 +0.50)}\newline change of AI dependence ($-0.50$ to $+0.50$) \\[2pt]
    \texttt{reason} & string & \jp{心理分析・解説（50文字以内）}\newline psychological analysis / explanation (at most 50 characters) \\
    \bottomrule
  \end{tabular}
\end{table}

\subsection{Experimental conditions}
\label{sec:conditions}

The basic setting of the simulation is $N=20$ students, a simulation period of 15 days, and a stress threshold of $\theta=0.5$. The No-AI control removes only the agents' access to the AI: the AI-dependence variable retains its sampled initial value and continues to enter the trigger score and the chat probability. It can subsequently fall by $0.02$ at each reconciliation, so under No AI it is non-increasing rather than constant; the control is ``no AI access'', not ``$D_i=0$''.

To account for the influence of probabilistic events and random numbers and to check the consistency of the simulation, the intended design---and the authors' account of the FIT2026 experiments---was to fix the random seed to 41, 42 and 43 so that, for a given seed, the initial state of the classroom is identical across conditions (three classrooms with different students), and to compare all seven counselor conditions in each of them (21 runs). Whether every recorded run followed this design can be established only from execution records, which have not yet been reconciled (Section~\ref{sec:limitations}); Section~\ref{sec:results} states what the recorded plots do and do not show. A seed identifies an initial classroom only together with the code version, because the initialization order determines which random draws each variable receives.

Starting from this basic setting, we further conducted the following two experiments in which one parameter is changed.
\begin{enumerate}[nosep,leftmargin=*]
  \item \textbf{Long-term experiment.} The simulation period is extended from 15 to 50 days to examine how AI dependence and distortions of human relationships accumulate.
  \item \textbf{Lower-threshold experiment.} The stress threshold is lowered from $0.5$ to $0.3$ so that the consultation branch is entered more easily. Because the threshold governs the whole branch, this affects consultations with friends as well as with the AI, and because stress, AI dependence, sociability and attendance then evolve differently, the realized number of consultations over the run is not guaranteed to be larger; we therefore call this the lower-threshold setting rather than a high-frequency setting until the realized counts are reported.
\end{enumerate}
Both additional experiments were intended to use seed 42. Inspection of the recorded plots shows, however, that the initial summaries (day-1 class means and ``Init'' distributions) of five of the seven 50-day runs and of one of the seven lower-threshold runs (inciting) are inconsistent with seed 42, so these runs did not start from the seed-42 classroom (Sections~\ref{sec:long} and~\ref{sec:thr}); they are reported but excluded from all cross-condition comparisons. For the remaining runs the initial summaries are consistent with seed 42, which is necessary but not sufficient for an identical initial classroom; their comparison therefore remains provisional until the execution records have been reconciled. The statistical unit of every comparison is one run of an interacting classroom, not a student or a day.

For all experiments we recorded, on the final day, the class-wide mean, maximum and minimum of stress, happiness, self-reliance and AI dependence, together with the number of non-attending students, and used them as evaluation items for the comparison. Class means are taken over all 20 agents, with non-attending agents contributing their frozen values (in particular $H=0$). Throughout, ``$\Delta$'' denotes the change of a class mean from the initial state before the first day (the ``Init'' distributions in Appendix~\ref{app:boxplots}) to the final day.

\subsection{Formal specification of the agent dynamics}
\label{sec:formal}

Let $X_t$ denote the complete state of the classroom at the start of day $t$: the five state variables of every agent, the closeness matrix $(c_{ij})$, the discord sets $F_i$ and the attendance flags (so that $A_t$ is the set of attending agents in $X_t$). The events of day $t$---the morning load of every attending agent, the $|A_t|$ noon draws, the after-school branch of every attending agent and the night check---are applied one after another in the order of Algorithm~\ref{alg:day}. Writing $e_1,\dots,e_M$ for these events and $T_{e_k}$ for the transition map of event $e_k$, which modifies only the variables listed for that event in Table~\ref{tab:rules} and clips each modified state variable to $[0,1]$ immediately,
\begin{equation}
  Z_t^{(0)}=X_t,\qquad Z_t^{(k)}=T_{e_k}\big(Z_t^{(k-1)}\big)\quad(k=1,\dots,M),\qquad X_{t+1}=Z_t^{(M)}.
  \label{eq:update}
\end{equation}
Because each $T_{e_k}$ acts on the current intermediate state $Z_t^{(k-1)}$, the probabilities and partner choices of later events depend on the outcomes of earlier events of the same day, and an increment lost at a boundary is not recovered by a later term of opposite sign (for example, $S=0.99$ followed by $+0.02$ and $-0.04$ gives $0.96$, not $0.97$). The class means plotted for day $t$ in Section~\ref{sec:results} are computed from $X_{t+1}$, the state at the end of day $t$; ``Init'' denotes $X_1$, the state before day~1. Table~\ref{tab:rules} lists every event with its trigger, probability and effect, and Algorithm~\ref{alg:day} gives the order in which they are applied. Two conventions of the implementation are worth recording: candidate lists ``the 3 attending classmates with the largest $c_{ij}$'' and ``the 2 with the smallest $c_{ij}$'' contain all available classmates when fewer exist, with ties broken in ascending order of agent id (stable sorting); and discord sets are never pruned, so a non-attending classmate stays in $F_i$ (it can never be reconciled with, it is excluded from the confidant choice because that choice ranges over attending classmates only, but it may still be named in the ``relationship'' consultation template).

\begin{table}[!htb]
  \centering\small
  \caption{Complete rule set of the model (all constants as implemented). $A_t$: attending agents; $F_i$: agents in discord with $i$; $c_{ij}$: closeness of $i$ toward $j$; $\theta$: stress threshold. Every state-variable update is clipped to $[0,1]$.}
  \label{tab:rules}
  \begin{tabular}{@{}L{2.7cm}L{5.4cm}L{6.3cm}@{}}
    \toprule
    Rule & Trigger / probability & Effect \\
    \midrule
    Initialization & once per run (seeded PRNG) & $S\!\sim\!U(0.1,0.4)$, $H\!\sim\!U(0.4,0.7)$, $R\!\sim\!U(0.3,0.7)$, $C\!\sim\!U(0.3,0.8)$, $D\!\sim\!U(0.05,0.2)$; $c_{ij}\!\sim\!U(0.1,0.3)$ for all $j\neq i$; $F_i=\emptyset$ \\
    Morning load & every $i\in A_t$, every day & $S_i\,{+}0.02$, $H_i\,{-}0.01$ \\
    Noon draw & $|A_t|$ times per day; $i\sim\mathrm{Unif}(A_t)$ & one of the four outcomes below \\
    \quad Chat & $p_{\mathrm c}=\max(0.1,\,0.45-0.3D_i)$ ($=0.45-0.3D_i$, since the floor never binds); $j$ uniform among the (up to) 3 attending classmates with the largest $c_{ij}$ & $j\notin F_i$: $S_i\,{-}0.04$, $H_i\,{+}0.04$, $C_i\,{+}0.01$, $c_{ij},c_{ji}\,{+}0.04$; \; $j\in F_i$ (awkward encounter): $S_i,S_j\,{+}0.05$ \\
    \quad Quarrel & $p_{\mathrm q}=0.02+0.06S_i$; $j$ uniform among the (up to) 2 attending classmates with the smallest $c_{ij}$ & $F_i\leftarrow F_i\cup\{j\}$, $F_j\leftarrow F_j\cup\{i\}$; $S_i\,{+}0.12$, $H_i\,{-}0.10$, $C_i\,{-}0.03$; $c_{ij},c_{ji}\,{-}0.15$ \\
    \quad Everyday trouble & $p_{\mathrm t}=0.15$ & $S_i\,{+}0.06$, $H_i\,{-}0.04$ \\
    \quad Nothing & $1-p_{\mathrm c}-p_{\mathrm q}-p_{\mathrm t}$ & --- \\
    Consultation trigger & $S_i+0.2D_i\ge\theta$ ($\theta=0.5$; $0.3$ in the lower-threshold setting); confidant $j^\ast=\arg\max_{j\in A_t\setminus\{i\}}[c_{ij}+0.30\cdot\mathbf 1(j\in F_i)]$; nothing happens if $A_t\setminus\{i\}=\emptyset$ & one of the four routes below (under No AI the factor $(1-D_i)$ is replaced by $1$) \\
    \quad AI consultation & $p=D_i$ (AI conditions; $0$ under No AI) & template by state (Table~\ref{tab:templates}) $\to$ counselor $\to$ evaluator; $(S,H,R,C,D)_i\mathrel{+}=$ JSON deltas (nominal $[-0.5,0.5]$), assigned in this order; on failure before the first assignment no update, on failure during the assignments the fields already assigned remain (Appendix~\ref{app:impl}) \\
    \quad Reconciliation & $p=(1-D_i)C_i$ and $j^\ast\in F_i$; always succeeds & $S_i\,{-}0.10$, $H_i\,{+}0.15$, $R_i\,{+}0.06$, $C_i\,{+}0.05$, $D_i\,{-}0.02$; $F_i\setminus\{j^\ast\}$, $F_{j^\ast}\setminus\{i\}$; $c_{ij^\ast},c_{j^\ast i}\,{+}0.15$ \\
    \quad Confide & $p=(1-D_i)C_i$ and $j^\ast\notin F_i$ & $S_i\,{-}0.12$, $H_i\,{+}0.06$, $R_i\,{+}0.02$; $c_{ij^\ast}\,{+}0.02$ \\
    \quad Bottle up & $p=(1-D_i)(1-C_i)$ & $S_i\,{+}0.05$, $H_i\,{-}0.03$, $R_i\,{+}0.03$ \\
    Non-attendance & $H_i\le 0$ at night & absorbing; $i\notin A_{t+1}$; values frozen but still included in class means \\
    Isolation (statistic) & $i\in A_t$ with $A_t\setminus\{i\}\neq\emptyset$ and $\max_{j\in A_t\setminus\{i\}}c_{ij}<0.25$; $A_t$ is the set saved in the morning, so it still contains agents that become non-attending in the same night & counted only; no effect on the dynamics \\
    \bottomrule
  \end{tabular}
\end{table}

Two remarks on reproducibility follow from this specification. First, a run is initialized by seeding Python's \texttt{random} module and NumPy with the chosen seed, so the initial classroom and the sequence of rule-based events are identical across conditions until the first AI consultation; from then on the random streams of different conditions diverge, because the second draw $u_2$ is consumed only when the AI route is not taken, and because the evaluator's output changes the subsequent trajectory. Second, both LLM calls use temperature $0$, which makes their outputs nearly, but not strictly, deterministic. Exact replay of a reported run therefore requires its stored initial state and its logged LLM outputs; a fresh run with the same seed and fresh API calls may follow a different trajectory.

The specification also implies invariants that any implementation must satisfy and that we use as consistency checks (they test the rules, not their psychological validity): every state variable stays in $[0,1]$; under No AI, $R_i$ is non-decreasing and $D_i$ is non-increasing for every agent; a non-attending agent's state never changes again; discord is symmetric ($j\in F_i \Leftrightarrow i\in F_j$) at the end of every event; and $p_{\mathrm c}\ge 0.15$ for $D_i\in[0,1]$, so the floor $0.1$ in the chat probability never binds.

\begin{algorithm}[!htb]
\caption{One simulated day of the virtual classroom (as implemented; all updates clipped to $[0,1]$).}
\label{alg:day}
\begin{algorithmic}[1]
\Require attending set $A_t$; state $(S_i,H_i,R_i,C_i,D_i)$, closeness $c_{ij}$ and discord set $F_i$ for every agent; threshold $\theta$; condition (No AI or one of six style prompts)
\State \textbf{Morning:} for all $i\in A_t$: $S_i\gets S_i+0.02$;\; $H_i\gets H_i-0.01$
\For{$k=1,\dots,|A_t|$} \Comment{Noon}
  \State \textbf{if} $|A_t|<2$ \textbf{then break}
  \State draw $i\sim\mathrm{Unif}(A_t)$ and $u\sim\mathrm{Unif}[0,1)$;\; $p_{\mathrm c}\gets\max(0.1,0.45-0.3D_i)$;\; $p_{\mathrm q}\gets 0.02+0.06S_i$
  \If{$u<p_{\mathrm c}$} \Comment{chat}
    \State $j\sim\mathrm{Unif}$(up to 3 attending classmates with the largest $c_{ij}$)
    \State \textbf{if} $j\in F_i$ \textbf{then} $S_i\gets S_i+0.05$;\; $S_j\gets S_j+0.05$
    \State \textbf{else} $S_i\gets S_i-0.04$;\; $H_i\gets H_i+0.04$;\; $C_i\gets C_i+0.01$;\; $c_{ij}\gets c_{ij}+0.04$;\; $c_{ji}\gets c_{ji}+0.04$
  \ElsIf{$u<p_{\mathrm c}+p_{\mathrm q}$} \Comment{quarrel}
    \State $j\sim\mathrm{Unif}$(up to 2 attending classmates with the smallest $c_{ij}$);\; $F_i\gets F_i\cup\{j\}$;\; $F_j\gets F_j\cup\{i\}$
    \State $S_i\gets S_i+0.12$;\; $H_i\gets H_i-0.10$;\; $C_i\gets C_i-0.03$;\; $c_{ij}\gets c_{ij}-0.15$;\; $c_{ji}\gets c_{ji}-0.15$
  \ElsIf{$u<p_{\mathrm c}+p_{\mathrm q}+0.15$} \Comment{everyday trouble}
    \State $S_i\gets S_i+0.06$; $H_i\gets H_i-0.04$
  \EndIf
\EndFor
\For{all $i\in A_t$ with $S_i+0.2D_i\ge\theta$} \Comment{After school}
  \State \textbf{if} $A_t\setminus\{i\}=\emptyset$ \textbf{then continue} \Comment{no attending classmate: nothing happens}
  \State $j^\ast\gets\arg\max_{j\in A_t\setminus\{i\}}\,[\,c_{ij}+0.30\cdot\mathbf 1(j\in F_i)\,]$;\; draw $u_1\sim\mathrm{Unif}[0,1)$
  \If{AI condition \textbf{and} $u_1<D_i$} \Comment{AI consultation}
    \State select template by state (Table~\ref{tab:templates}); reply $\gets$ counselor AI(style prompt, template)
    \State $(\Delta S,\Delta H,\Delta R,\Delta C,\Delta D)\gets$ evaluator AI(type label, template, reply);\; add to $(S_i,H_i,R_i,C_i,D_i)$ \Comment{skipped if a call fails}
  \Else
    \State draw $u_2\sim\mathrm{Unif}[0,1)$
    \If{$u_2<C_i$ \textbf{and} $j^\ast\in F_i$} \Comment{reconciliation}
      \State $S_i\gets S_i-0.10$; $H_i\gets H_i+0.15$; $R_i\gets R_i+0.06$; $C_i\gets C_i+0.05$; $D_i\gets D_i-0.02$
      \State $F_i\gets F_i\setminus\{j^\ast\}$;\; $F_{j^\ast}\gets F_{j^\ast}\setminus\{i\}$;\; $c_{ij^\ast}\gets c_{ij^\ast}+0.15$;\; $c_{j^\ast i}\gets c_{j^\ast i}+0.15$
    \ElsIf{$u_2<C_i$} \Comment{confide}
      \State $S_i\gets S_i-0.12$; $H_i\gets H_i+0.06$; $R_i\gets R_i+0.02$; $c_{ij^\ast}\gets c_{ij^\ast}+0.02$
    \Else \Comment{bottle up}
      \State $S_i\gets S_i+0.05$; $H_i\gets H_i-0.03$; $R_i\gets R_i+0.03$
    \EndIf
  \EndIf
\EndFor
\State \textbf{Night:} for all $i\in A_t$: \textbf{if} $H_i\le 0$ \textbf{then} mark $i$ non-attending;\; record class means over all $N$ agents and the counts of non-attending and isolated agents
\end{algorithmic}
\end{algorithm}

\section{Results}
\label{sec:results}

This section reports the results of running the seven response conditions, including the no-AI control, on the virtual classroom simulation. Three patterns were run: the 15-day simulation as the basic setting, the 50-day simulation as the long-term setting, and the stress-threshold simulation as the lower-threshold setting. The results come from three sources. The runs of the FIT2026 version (Sections~\ref{sec:base}, \ref{sec:long-orig} and~\ref{sec:thr}) are historical: their values were transcribed from the program's display or read from its plotted curves, and their execution records have not been reconciled. The 50-day setting was in addition \emph{re-run} for this paper from one stored initial classroom with complete logging (Section~\ref{sec:rerun}); its values are exact, and it is a controlled comparison of one realization per condition in one classroom. The robustness of that comparison was then tested with a pre-specified protocol of 144 further logged runs---ten independent classrooms, repeated LLM realizations of one classroom and sensitivity arms (Section~\ref{sec:robust})---which provides the only measures of uncertainty in this paper. Figures~\ref{fig:base-lines}, \ref{fig:long-lines} and~\ref{fig:thr-lines} show the day-by-day class means of the four main parameters in the three historical settings (for the seed-42 classroom, except for the runs identified in Sections~\ref{sec:long-orig} and~\ref{sec:thr} as not having started from that classroom), and Figure~\ref{fig:rerun-lines} those of the re-run; the corresponding distributions of individual students on the first and final days of the historical runs are given as box plots in Appendix~\ref{app:boxplots}. Tables~\ref{tab:base}--\ref{tab:thr} collect the final-day values discussed in the text.

Throughout this section, ``students'' are simulated student agents, and stress, happiness, self-reliance and AI dependence denote the corresponding simulation state variables (Section~\ref{sec:agents}). The figures reproduce the plots generated by the simulation program: in their legends ``Self Reliance'' and ``Reliance'' both denote self-reliance, ``AI Dep'' denotes AI dependence, ``Init'' denotes the state before day~1 (the same state from which every $\Delta$ is measured) and ``Final'' the state at the end of the last day. The box plots show the distribution over the 20 students of a single run; they do not represent uncertainty across repeated runs. The statistical unit of every comparison is one run of an interacting classroom (Section~\ref{sec:conditions}); with one run per condition and classroom, the values of Sections~\ref{sec:base}--\ref{sec:thr} are descriptive and no measure of uncertainty is available for them, for the re-run as much as for the historical runs; Section~\ref{sec:robust} supplies confidence intervals and exact tests over ten classrooms and the spread over repeated LLM realizations for the 50-day setting. Values are rounded to two decimals (round half away from zero on the exact value); where an exact value lies on or within $0.0005$ of a rounding boundary, the exact value is given in a footnote or to three decimals, and where the FIT2026 version printed a different rounding, this is noted. We describe what happened in the model and postpone interpretation in terms of human psychology to Section~\ref{sec:discussion}.

\subsection{15-day simulation (basic setting)}
\label{sec:base}

Table~\ref{tab:base} summarizes the results of the basic setting (15 days, stress threshold $0.5$) for the random seeds 41, 42 and 43, Table~\ref{tab:base42} gives the final-day class means of all four variables for the seven runs labelled ``seed 42'' in the FIT2026 version, and Figure~\ref{fig:base-lines} shows their trajectories. For the AI conditions, ``seed 42'' in these labels means that the plotted initial summary (``Init'' distribution and day-1 class means) is consistent with the seed-42 classroom generated by the supplied code; identical initial classrooms have not been verified from these plots, and the condition labels and execution versions of the historical runs have not been reconciled with execution records (Section~\ref{sec:limitations}). All comparisons in this section are therefore descriptive and conditional on the historical labels and initial-state assignments being correct. The provenance of every value is stated in the captions and falls into three classes: \emph{exact} values were recomputed from the rule set (possible only for the No-AI condition, which involves no LLM call); \emph{transcribed} values are those printed in the FIT2026 text, which were read off the program's display at two decimals; and \emph{digitized estimates} were read from the plotted curves of the FIT2026 figures with the procedure of Appendix~\ref{app:impl}, whose uncertainty is about $\pm0.002$ for isolated curve segments and larger where curves overlap or a dashed curve has a gap (such cells are marked). Digitized estimates are estimates, not measurements of the original runs, and are given to three decimals so that the reader can see how close they lie to a rounding boundary; no unreported value was filled in, and cells for which no value is available are marked ``\nr''. The AI-condition values of Tables~\ref{tab:base}, \ref{tab:base42}, \ref{tab:long} and~\ref{tab:thr} are those of the FIT2026 runs, whose execution records (condition label, seed, code and prompt versions) have not yet been reconciled (Section~\ref{sec:limitations}); comparisons based on them are therefore provisional. The values of the controlled re-run (Tables~\ref{tab:rerun}--\ref{tab:rerun-eval}) are exact.

\begin{table}[t]
  \centering\footnotesize\setlength{\tabcolsep}{4.5pt}
  \caption{Basic setting (15 days, threshold $0.5$): final-day class values for seeds 41/42/43. Where a single range is given, it spans the three seeds; ``$\Delta$'' is the change of the class mean from the initial state to the final day. The No-AI row is exact (recomputed from the rule set of Section~\ref{sec:formal}; its changes are given to three decimals because $+0.065$ lies exactly on a two-decimal rounding boundary). The other rows are the values transcribed in the FIT2026 text (per-seed values where the text gives them, ranges otherwise); where the seed-42 curve of Figure~\ref{fig:base-lines} gives a digitized estimate outside the printed range or different from the printed seed-42 value, this is noted (a--e) and left unresolved, because the execution records of these runs are not available. Cells marked ``\nr'' were not reported. The complete trajectories for seed 42 are shown in Figure~\ref{fig:base-lines} and Appendix~\ref{app:boxplots}.}
  \label{tab:base}
  \begin{minipage}{\textwidth}\centering
  \begin{tabular}{@{}lcccc@{}}
    \toprule
    Condition & Mean stress & $\Delta$ self-reliance & $\Delta$ AI dependence & Non-attending \\
    \midrule
    No AI               & 0.40 / 0.43 / 0.42 & $+0.065$ / $+0.065$ / $+0.039$ & $-0.003$ / $-0.001$ / $-0.001$ & 0 / 0 / 0 \\
    Affirming           & 0.35--0.41\textsuperscript{a} & $-0.06$ to $-0.02$ & $+0.10$ to $+0.12$ & \nr \\
    Listening           & 0.47 / 0.46 / 0.47 & \nr & \nr & 2 / 1 / 1 \\
    Solution-oriented   & 0.40--0.42 & $+0.05$ to $+0.07$\textsuperscript{b} & \nr & \nr \\
    Reality-redirecting & 0.48--0.50 & \nr & $-0.04$ to $-0.01$\textsuperscript{c} & \nr \\
    Inciting            & 0.47 / 0.51 / 0.49 & \nr & $+0.09$ to $+0.14$\textsuperscript{d} & 3 / 5 / 5 \\
    Blaming             & 0.48 / 0.50 / 0.52\textsuperscript{e} & \nr & \nr & \nr \\
    \bottomrule
  \end{tabular}
  \par\smallskip\raggedright\footnotesize
  \textsuperscript{a}\,Seed-42 curve: $0.417$.\;
  \textsuperscript{b}\,Seed-42 curve: $+0.083$ ($0.593-0.510$).\;
  \textsuperscript{c}\,The text gives $-0.01$ for seed 42 and $-0.04$ for seed 41; the seed-42 curve gives $-0.034$ ($0.082-0.116$).\;
  \textsuperscript{d}\,Seed-42 curve: $+0.195$ ($0.311-0.116$).\;
  \textsuperscript{e}\,The text gives $0.50$ for seed 42 and $0.52$ for seed 43; the seed-42 curve gives $0.517$. Whether the seed labels or the curves are mislabelled cannot be decided without the execution records.
  \end{minipage}
\end{table}

\begin{table}[t]
  \centering\small
  \caption{Basic setting, runs labelled seed 42 (15 days, threshold $0.5$): final-day class means of the four main state variables for the seven runs plotted in Figure~\ref{fig:base-lines}. The exact initial class means of the seed-42 classroom (before day~1) are stress $0.252$, happiness $0.552$, self-reliance $0.510$ and AI dependence $0.116$; the ``Init'' summaries of all seven panels are consistent with this classroom, which is necessary but not sufficient for identical initial states. The No-AI row is exact; the other rows are digitized estimates (final-day uncertainty about $\pm0.002$ for isolated curve segments, about $\pm0.005$ for flagged cells; Appendix~\ref{app:impl}). Non-attendance counts are given where the FIT2026 text reports them.}
  \label{tab:base42}
  \begin{minipage}{0.9\textwidth}\centering
  \begin{tabular}{@{}lccccc@{}}
    \toprule
    Condition & Stress & Happiness & Self-reliance & AI dependence & Non-attending \\
    \midrule
    No AI (exact)       & 0.431 & 0.528 & 0.575 & 0.115 & 0 \\
    Affirming           & 0.417 & 0.531 & 0.477 & 0.237\textsuperscript{b} & \nr \\
    Listening           & 0.458 & 0.523 & 0.554 & 0.131\textsuperscript{b} & 1 \\
    Solution-oriented   & 0.395 & 0.576 & 0.593 & 0.102 & \nr \\
    Reality-redirecting & 0.504 & 0.484 & 0.645 & 0.082 & \nr \\
    Inciting            & 0.512 & 0.427 & 0.415 & 0.311 & 5 \\
    Blaming             & 0.517\textsuperscript{a} & 0.460 & 0.502 & 0.093 & \nr \\
    \bottomrule
  \end{tabular}
  \par\smallskip\raggedright\footnotesize \textsuperscript{a}\,The FIT2026 text gives $0.50$ for seed 42 and $0.52$ for seed 43 in this condition; the comparison of this row with the others is provisional until the assignment of conditions and seeds to the plotted runs has been confirmed from the execution records. \textsuperscript{b}\,Dashed curve with a gap at the last day; the estimate is taken from the nearest resolved column (2 pixels, about $0.05$ day, to the left).
  \end{minipage}
\end{table}

\paragraph{(1) No AI.} In all three classrooms the mean stress rose slightly, ending at 0.40--0.43, self-reliance rose by $0.039$--$0.065$ through the rule-based consultation routes, mean AI dependence changed little at the reported precision (under No AI it can only fall, by $0.02$ per reconciliation; the exact changes are $-0.003$, $-0.001$ and $-0.001$), and no student stopped attending within 15 days. Note that the rise of self-reliance in this condition is partly definitional: every rule-based after-school route raises $R_i$ (Table~\ref{tab:params}), so under No AI the class mean of $R$ is non-decreasing. This condition serves as the reference against which the AI interventions are compared.

\paragraph{(2) Affirming.} Compared with the no-AI condition, the mean stress decreased to 0.35--0.41 (FIT2026 text; the seed-42 curve gives $0.417$). Meanwhile, the mean happiness hardly changed, self-reliance decreased slightly ($\Delta$: $-0.06$ to $-0.02$), and AI dependence increased ($\Delta$: $+0.10$ to $+0.12$). In the present simulation, the affirming condition thus lowered mean stress in the short term while raising the AI-dependence parameter and lowering self-reliance. This simulated pattern is consistent with the hypothesis that unconditional affirmation, by encouraging self-justification, shifts a student's coping from friends toward the AI; in the model the shift is partly mechanical, because a higher AI dependence raises the probability of choosing the AI over the rule-based routes that build self-reliance (Section~\ref{sec:discussion}). Whether this occurs in human users is not tested here.

\paragraph{(3) Listening.} The mean stress reached 0.46--0.47 (seed 41: 0.47, seed 42: 0.46, seed 43: 0.47), slightly above the no-AI reference for every seed, and 1--2 students stopped attending (seed 41: 2, seed 42: 1, seed 43: 1). Within the model, a dialogue style devoted solely to ``listening''---echoing without offering advice or direction---served as a temporary outlet but did not move the agents toward resolving the troubles that generate stress in the classroom, so stress was not fully dissipated. The simulated pattern is consistent with the hypothesis that listening alone, without any problem-solving component, is insufficient to improve a strained classroom; Section~\ref{sec:limitations} explains why this should not be read as a verdict on reflective listening in real counseling.

\paragraph{(4) Solution-oriented.} A slight increase in self-reliance ($\Delta$: $+0.05$ to $+0.07$ in the FIT2026 text; $+0.083$ on the seed-42 curve) was reproduced for all seeds, and the mean stress remained at a low level of 0.40--0.42, comparable to the no-AI condition. A plausible mechanism within the model---not verified, because the evaluator's updates and the event counts of these runs are not available---is that the solution-oriented replies, which avoid emotional agreement and offer concrete ways of handling the situation, were scored by the evaluator as raising self-reliance and lowering AI dependence; a lower AI dependence would send the agent more often to the rule-based routes (confiding and reconciliation), which raise self-reliance further and repair closeness. Whether relationships within the classroom were in fact maintained can only be established from the closeness and discord statistics (Section~\ref{sec:limitations}).

\paragraph{(5) Reality-redirecting.} The change in AI dependence was $-0.04$ to $-0.01$ in the three classrooms, a decreasing trend. However, presumably reflecting the prompt that pushes away the consulting student, the mean stress stayed at a high level of 0.48--0.50. Within the model, this style blocks escape into the AI but offers no direct solution to agents whose stress is already high; the high stress may reflect evaluator updates that raise the stress of the consulting agents, which the run logs would show.

\paragraph{(6) Inciting.} The number of non-attending students reached 3 for seed 41 and 5 for seeds 42 and 43---the largest number of students whose happiness reached $0$ at a night check even within the short period of 15 days. The mean stress on the final day was high, 0.47--0.51 (seed 41: 0.47, seed 42: 0.51, seed 43: 0.49), and AI dependence increased by $+0.09$ to $+0.14$ according to the FIT2026 text (the seed-42 curve gives $+0.195$; Table~\ref{tab:base}). In the simulation, the inciting responses---which amplify the agent's anger and attribute the fault to the surroundings---went together with a rising AI-dependence parameter. The higher stress values imply a higher quarrel probability ($0.02+0.06S_i$) under the specified rules, and the FIT2026 version described frequent quarrels in this condition; realized event counts from the run logs are needed to confirm that more quarrels actually occurred, and they are not reported in this version. The happiness parameter dropped sharply and agents entered the non-attendance state one after another. This simulated cascade is consistent with the hypothesis that an AI which validates anger acts as an ``ally'' that both attracts dependence and worsens relationships with peers.

\paragraph{(7) Blaming.} This style lectures the agents that the cause of the problem lies in themselves, and the mean stress stayed at a high level of 0.48--0.52 in the three classrooms, on a par with the reality-redirecting and inciting styles. The pattern may reflect evaluator updates that raised the stress of the consulting agents while lowering their AI dependence, so that the agents rarely returned to the AI; the sign and size of these updates and the number of consultations have not been tabulated for these runs. A lecture without the kind of advice given by the solution-oriented style thus appears to have lowered the happiness parameter and to have contributed to non-attendance in the model, subject to the same reservation.

\begin{figure}[p]
  \centering
  \begin{subfigure}[t]{0.72\textwidth}\centering\includegraphics[width=\linewidth]{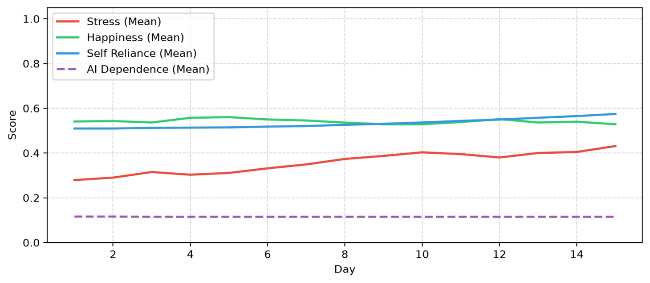}\caption{No AI}\end{subfigure}\\[1pt]
  \begin{subfigure}[t]{0.72\textwidth}\centering\includegraphics[width=\linewidth]{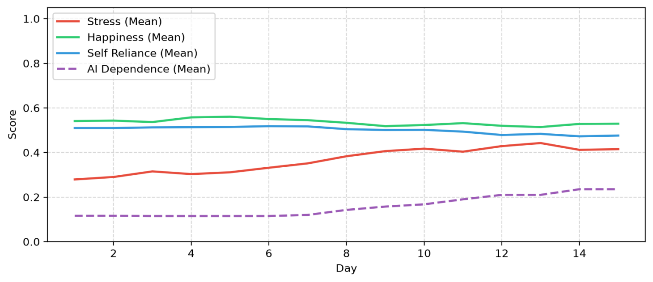}\caption{Affirming}\end{subfigure}\\[1pt]
  \begin{subfigure}[t]{0.72\textwidth}\centering\includegraphics[width=\linewidth]{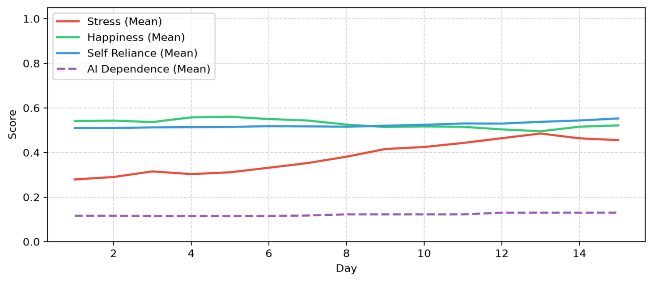}\caption{Listening}\end{subfigure}\\[1pt]
  \begin{subfigure}[t]{0.72\textwidth}\centering\includegraphics[width=\linewidth]{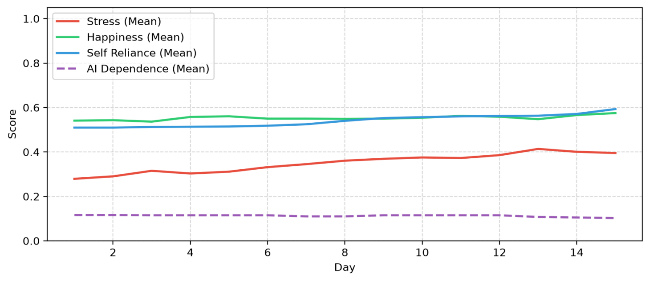}\caption{Solution-oriented}\end{subfigure}
  \caption{Basic setting (15 days, threshold $0.5$): daily class means of stress, happiness, self-reliance and AI dependence under the seven conditions, as plotted in the FIT2026 version. The initial summaries of all seven panels are consistent with the seed-42 classroom generated by the supplied code; the identity of the full initial states has not been verified from these plots (Section~\ref{sec:long}). (Continued on the next page.)}
  \label{fig:base-lines}
\end{figure}
\begin{figure}[p]
  \ContinuedFloat
  \centering
  \begin{subfigure}[t]{0.72\textwidth}\centering\includegraphics[width=\linewidth]{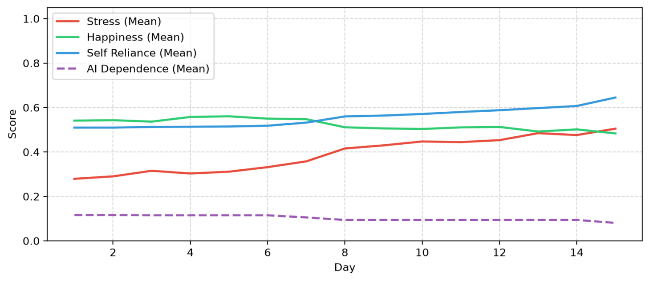}\caption{Reality-redirecting}\end{subfigure}\\[1pt]
  \begin{subfigure}[t]{0.72\textwidth}\centering\includegraphics[width=\linewidth]{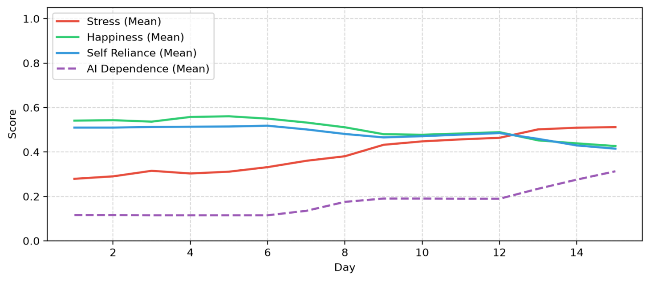}\caption{Inciting}\end{subfigure}\\[1pt]
  \begin{subfigure}[t]{0.72\textwidth}\centering\includegraphics[width=\linewidth]{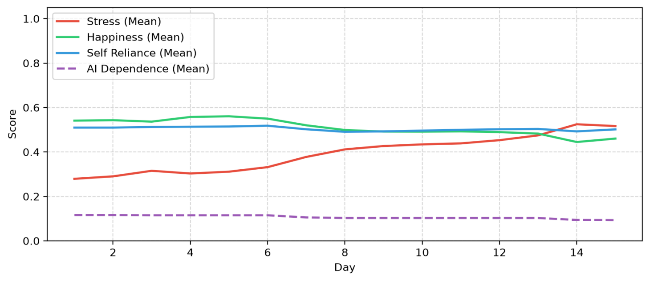}\caption{Blaming}\end{subfigure}
  \caption{(Continued.) Panels (e)--(g) of Figure~\ref{fig:base-lines}.}
\end{figure}

\subsection{50-day simulation (long-term setting): controlled re-run from one stored classroom}
\label{sec:rerun}
\label{sec:long}

\begin{figure}[p]
  \centering
  \includegraphics[width=0.98\textwidth]{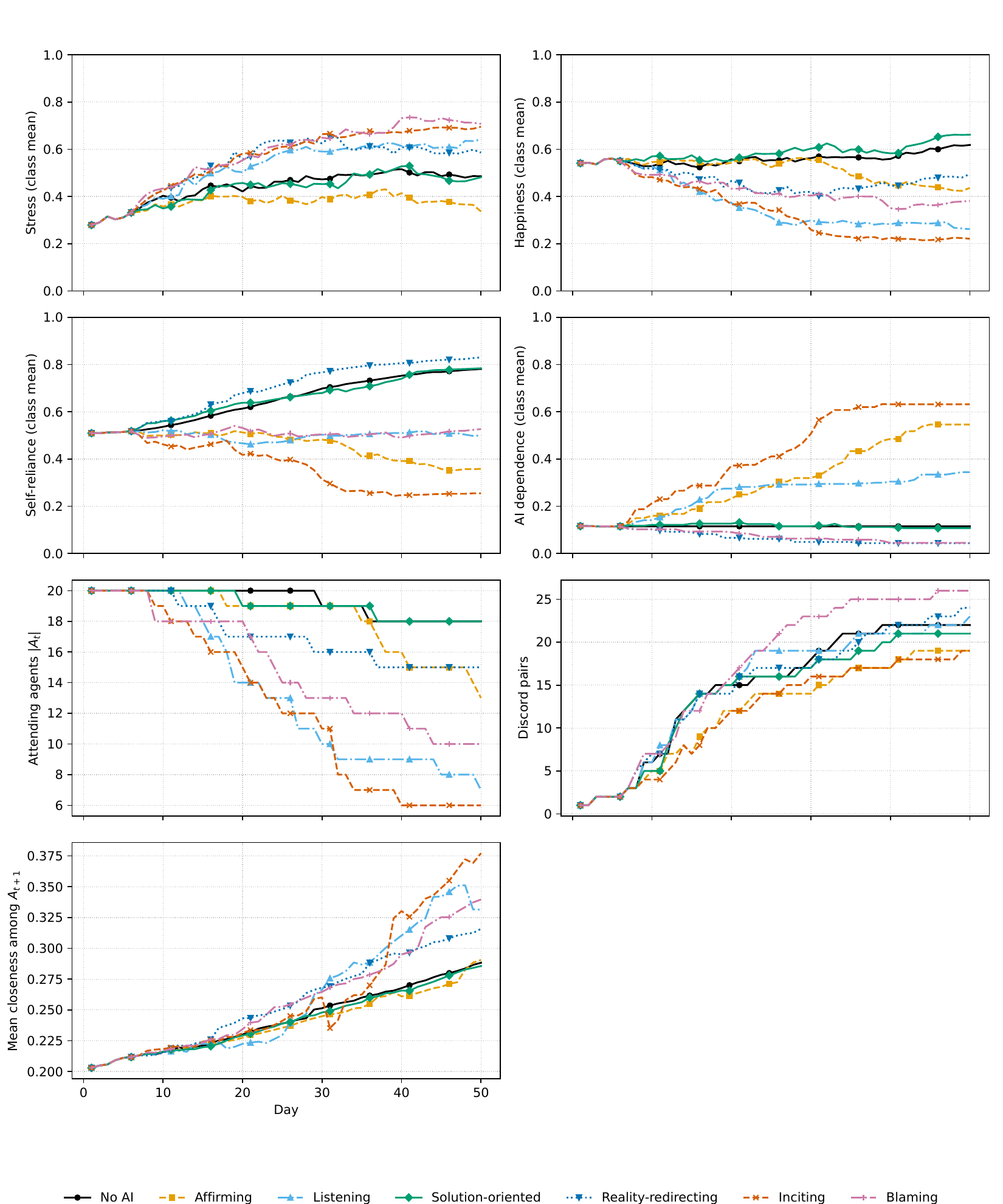}
  \caption{Controlled re-run of the long-term setting (50 days, threshold $0.5$, seed-42 classroom for all seven conditions): daily class means over all 20 agents of stress, happiness, self-reliance and AI dependence; number of attending agents at the start of the day, $|A_t|$; number of discord pairs; and mean closeness among the agents still attending at the end of the day, $A_{t+1}$ (computed by replaying the logged runs, Appendix~\ref{app:impl}). The seven trajectories coincide through day~6; the first AI consultation took place on day~7 in every AI condition. The attending-only closeness mean reflects both changes in the ties among the retained agents and the selection of the agents included in the average: among the pairs of agents attending on day 50, mean closeness rose from initialization to day 50 in every condition (e.g., inciting: from $0.206$ to $0.377$, while the day-50 mean over all pairs was $0.244$; No AI: from $0.203$ to $0.288$), so the steeper rise in the listening, inciting and blaming runs cannot be attributed to composition alone. These comparisons are descriptive, not a causal decomposition.}
  \label{fig:rerun-lines}
\end{figure}

The long-term experiment extended the simulation period to 50 days. Because the original 50-day runs of the FIT2026 version turned out not to have started from identical classrooms (Section~\ref{sec:long-orig}), all seven conditions were re-run for this paper with the analysis package of Appendix~\ref{app:impl}, which reproduces the rule set of Section~\ref{sec:formal} bit for bit and logs every event and every LLM input and output. The design is the one called for in Section~\ref{sec:long-orig}: one initial classroom (seed 42; the stored initial states of the seven runs are identical---the hash \texttt{598ed94b0e43} covers the five state variables of every agent, and the closeness matrix is fixed by the same seed and code, which the replay described in Appendix~\ref{app:impl} confirms), the random stream of the original program, the same prompts, evaluator instruction and response schema, the model \texttt{gemini-2.5-flash} at temperature $0$, threshold $0.5$ and 50 days. All 153 AI consultations of the six AI conditions succeeded (no failed API call, no missing or out-of-range field), so the comparison is not biased by failures; a first batch in which most calls failed for infrastructure reasons was discarded and is disclosed in Appendix~\ref{app:impl}. The No-AI run of the re-run reproduces the exact No-AI values of Table~\ref{tab:long} to all digits, as it must, and replaying the logged replies and evaluator outputs of the six AI runs through the released code reproduces their recorded daily means and final states bit for bit (Appendix~\ref{app:impl}). Table~\ref{tab:rerun} gives the final-day values, Figure~\ref{fig:rerun-lines} the trajectories, Table~\ref{tab:rerun-events} the realized events, and Table~\ref{tab:rerun-eval} the evaluator's updates; each condition was run once, so the values are exact for these runs but carry no measure of uncertainty.

\begin{table}[H]
  \centering\footnotesize\setlength{\tabcolsep}{3pt}
  \caption{Controlled re-run of the long-term setting (50 days, threshold $0.5$, all conditions from the stored seed-42 classroom; exact values). Final-day class means over all 20 agents of the four main state variables (initial class means: stress $0.252$, happiness $0.552$, self-reliance $0.510$, AI dependence $0.116$); number of non-attending agents on day 50 with the day of the first transition in parentheses; mean happiness over the agents attending on day 50 (no agent became non-attending on day 50 in any condition, so the morning and end-of-day sets coincide; Section~\ref{sec:limitations}); number of AI consultations, all of which succeeded, with the number of distinct (consultation message, reply) pairs in parentheses.}
  \label{tab:rerun}
  \begin{minipage}{\textwidth}\centering
  \begin{tabular}{@{}lccccccc@{}}
    \toprule
    Condition & Stress & Happiness & Self-reliance & AI dep. & Non-att.\ (first day) & $\overline H$ attending & AI consult.\ (unique) \\
    \midrule
    No AI               & 0.485 & 0.618 & 0.781 & 0.115 & 2 (29)  & 0.687 & 0 \\
    Affirming           & 0.337 & 0.436 & 0.358 & 0.546 & 7 (17)  & 0.671 & 51 (15) \\
    Listening           & 0.641 & 0.261 & 0.500 & 0.344 & 13 (12) & 0.747 & 34 (17) \\
    Solution-oriented   & 0.481 & 0.661 & 0.784 & 0.107 & 2 (19)  & 0.735 & 16 (15) \\
    Reality-redirecting & 0.586 & 0.495 & 0.831 & 0.044 & 5 (11)  & 0.660 & 11 (9) \\
    Inciting            & 0.695 & 0.221 & 0.255 & 0.632 & 14 (8)  & 0.737 & 30 (15) \\
    Blaming             & 0.708 & 0.382 & 0.527 & 0.045 & 10 (8)  & 0.764 & 11 (10) \\
    \bottomrule
  \end{tabular}
  \end{minipage}
\end{table}

\paragraph{Final states.} Relative to the No-AI control, the affirming prompt lowered mean stress ($0.337$ vs.\ $0.485$) but raised AI dependence to $0.546$ and lowered self-reliance to $0.358$ and happiness to $0.436$, with 7 agents no longer attending. The inciting prompt ended with the highest AI dependence ($0.632$), the lowest self-reliance ($0.255$) and happiness ($0.221$), and 14 of 20 agents not attending, the first of them on day~8. The blaming and reality-redirecting prompts ended with class-mean AI dependence of $0.045$ and $0.044$---every agent that consulted the AI in these runs (11 in each) was driven to $D_i=0$ by that consultation, while the other agents kept their initial values up to the $-0.02$ of a reconciliation---together with mean stress of $0.708$ (the highest of the seven runs) and $0.586$ (above the control, below the inciting and listening runs) and 10 and 5 non-attending agents; reality-redirecting also ended with the highest self-reliance ($0.831$). The final stress levels rank blaming ($0.708$), inciting ($0.695$), listening ($0.641$), reality-redirecting ($0.586$), No AI ($0.485$), solution-oriented ($0.481$) and affirming ($0.337$). The listening prompt ended with the second-lowest happiness ($0.261$) and 13 non-attending agents. The solution-oriented prompt was the only AI condition whose final values were numerically equal to or more favorable than the control on all five indicators (stress $0.481$ vs.\ $0.485$, happiness $0.661$ vs.\ $0.618$, self-reliance $0.784$ vs.\ $0.781$, AI dependence $0.107$ vs.\ $0.115$, 2 non-attending agents in both); with one run per condition this is a point estimate, and neither equivalence with the control nor superiority across classrooms is established; across the ten classrooms of Section~\ref{sec:robust} the solution-oriented differences from No AI changed sign in two to five classrooms, and its happiness and self-reliance advantages in this run are not reproduced. The all-agent happiness means must be read together with the attendance column: the agents still attending on day 50 had mean happiness between $0.660$ and $0.764$ in every condition, so the low all-agent means of the listening, inciting and blaming runs are composition effects of the identity $\overline H_{\mathrm{all}}=(|A|/N)\,\overline H_{A}$ (e.g., $0.221=(6/20)\times0.737$ for inciting), not a uniform deterioration of the attending agents.

\begin{table}[H]
  \centering\footnotesize\setlength{\tabcolsep}{2.5pt}
  \caption{Controlled re-run: realized events over the 50 days. Trigger events are the after-school consultation branches entered (Algorithm~\ref{alg:day}) and are split into the AI, friend-confiding, reconciliation (``rec.'') and bottling-up routes (``AI share'': fraction of trigger events that took the AI route); chats, quarrels and friend confidings (``Confid.'') are given per 100 attending-agent-days ($\sum_t|A_t|$: 964, 911, 671, 955, 858, 625 and 747 for the seven rows); discord pairs and mean closeness (``Closen.'', over all $N(N-1)$ ordered pairs) are day-50 values; the last column is the share of the five state additions of each successful consultation that were clipped at $0$ or $1$.}
  \label{tab:rerun-events}
  \begin{minipage}{\textwidth}\centering
  \begin{tabular}{@{}lccccccccc@{}}
    \toprule
    Condition & Triggers & AI/friend/rec./bottle & AI share & Chats & Quarrels & Confid. & Discord & Closen. & Clipped \\
    \midrule
    No AI               & 273 & 0/161/1/111  & 0.00 & 41.9 & 5.3 & 16.7 & 22 & 0.277 & --- \\
    Affirming           & 204 & 51/81/2/70   & 0.25 & 38.3 & 4.5 & 8.9  & 19 & 0.265 & 0.32 \\
    Listening           & 176 & 34/72/1/69   & 0.19 & 42.6 & 5.7 & 10.7 & 23 & 0.248 & 0.19 \\
    Solution-oriented   & 221 & 16/122/1/82  & 0.07 & 41.2 & 4.7 & 12.8 & 21 & 0.274 & 0.09 \\
    Reality-redirecting & 290 & 11/145/3/131 & 0.04 & 43.4 & 5.7 & 16.9 & 24 & 0.268 & 0.42 \\
    Inciting            & 158 & 30/63/3/62   & 0.19 & 40.6 & 5.6 & 10.1 & 19 & 0.244 & 0.35 \\
    Blaming             & 195 & 11/111/2/71  & 0.06 & 41.6 & 6.3 & 14.9 & 26 & 0.252 & 0.47 \\
    \bottomrule
  \end{tabular}
  \end{minipage}
\end{table}

\begin{table}[H]
  \centering\footnotesize\setlength{\tabcolsep}{4.5pt}
  \caption{Controlled re-run: the evaluator's updates by response style, pooled over consultation types (mean of the returned values $\Delta S,\Delta H,\Delta R,\Delta C,\Delta D$ before clipping; SD in parentheses; $n$ = number of consultations). The breakdown by consultation type, with the most frequent reply and the \texttt{reason} of the same record for the dominant cell of each style, is given in Appendix~\ref{app:rerun-eval}. Of the 765 returned values, 98\% are multiples of $0.05$, 40\% have $|\Delta|\ge0.3$, 1.3\% lie at the nominal limit $\pm0.5$ and none outside it; no field was missing.}
  \label{tab:rerun-eval}
  \begin{minipage}{\textwidth}\centering
  \begin{tabular}{@{}lcccccc@{}}
    \toprule
    Style & $n$ & $\Delta S$ & $\Delta H$ & $\Delta R$ & $\Delta C$ & $\Delta D$ \\
    \midrule
    Affirming           & 51 & $-0.18$ (0.12) & $-0.01$ (0.13) & $-0.22$ (0.06) & $-0.19$ (0.07) & $+0.37$ (0.09) \\
    Listening           & 34 & $+0.07$ (0.23) & $-0.21$ (0.15) & $-0.13$ (0.07) & $-0.22$ (0.14) & $+0.19$ (0.16) \\
    Solution-oriented   & 16 & $-0.13$ (0.21) & $+0.10$ (0.19) & $+0.09$ (0.14) & $+0.15$ (0.19) & $-0.02$ (0.12) \\
    Reality-redirecting & 11 & $+0.33$ (0.16) & $-0.28$ (0.19) & $+0.16$ (0.19) & $-0.01$ (0.19) & $-0.43$ (0.05) \\
    Inciting            & 30 & $+0.07$ (0.16) & $-0.29$ (0.11) & $-0.35$ (0.07) & $-0.37$ (0.09) & $+0.39$ (0.09) \\
    Blaming             & 11 & $+0.45$ (0.02) & $-0.45$ (0.00) & $-0.35$ (0.06) & $-0.35$ (0.05) & $-0.39$ (0.08) \\
    \bottomrule
  \end{tabular}
  \end{minipage}
\end{table}

\paragraph{What the logs show.} The three mechanisms stated in Section~\ref{sec:discussion} can be read directly from Tables~\ref{tab:rerun-events} and~\ref{tab:rerun-eval} and from the per-agent logs. (a) \emph{Crowding out.} The affirming, inciting and listening prompts sent 25\%, 19\% and 19\% of the trigger events to the AI, and friend confidings fell from $16.7$ per 100 attending-agent-days under No AI to $8.9$, $10.1$ and $10.7$; under the prompts that lowered AI dependence (reality-redirecting, blaming, solution-oriented) the AI took 4--7\% of the trigger events and confidings stayed at $12.8$--$16.9$. The change of class-mean self-reliance decomposes exactly into the applied evaluator updates, the applied rule-based increments and the clipping at the bounds (Appendix~\ref{app:rerun-eval}): under the affirming and inciting prompts the evaluator's applied updates summed to $-6.7$ and $-8.2$ (raw $-11.4$ and $-10.5$ before clipping) against $+3.7$ and $+3.2$ from the rule-based routes, whereas under No AI the rule-based routes contributed $+5.4$. This is an accounting of the additions that occurred in these runs, not a counterfactual estimate of how many friend consultations the AI displaced: the trigger events, $D_i$, $C_i$ and attendance all changed at the same time. (b) \emph{Social withdrawal.} The realized chat and quarrel rates differed comparatively little between conditions: $38$--$43$ chats and $4.50$--$6.29$ quarrels per 100 attending-agent-days against $41.9$ and $5.29$ under No AI; the inciting run had the fewest quarrels in absolute number (35) but, with fewer attending-agent-days, a rate of $5.60$, and the blaming run the highest rate ($6.29$). The number of discord pairs on day 50 ranged from 19 to 26 against 22 under No AI. The FIT2026 description of ``frequent quarrels'' under the inciting prompt is therefore not supported by the counts of this run, but these single-run daily totals cannot establish that quarrel chains involving particular agents did not occur, nor isolate the causal contributions of peer interaction, evaluator updates and the non-attendance rule to the final outcomes. (c) \emph{Feedback through the consultation template.} Under the affirming prompt, 11 agents consulted the AI, one of them nine times; the ``dependence'' template, first used on day 15, accounted for 34 of the 51 consultations and produced only two distinct exchanges (two replies to the same fixed message), and the evaluator returned nearly the same update each time ($\Delta D=+0.42$, SD $0.03$; $\Delta R=-0.24$; $\Delta C=-0.22$). The updates were not deterministic, however: each of the two identical evaluator inputs produced two distinct update vectors across its repetitions (in the listening run, nine identical inputs of this type produced two vectors), so temperature $0$ did not fix the output, and exact reproduction requires the logged outputs (Appendix~\ref{app:impl}). Nine agents ended at AI dependence $1$ and self-reliance $0$ (seven of them also at sociability $0$), and 47\% of the state additions in these consultations were clipped. The inciting prompt shows the same loop (12 ``dependence'' consultations from day~8, $\Delta D=+0.47$) and the listening prompt a weaker one (9, $\Delta D=+0.45$). The blaming and reality-redirecting prompts show the opposite dynamics: each of the 11 consulting agents consulted exactly once, the evaluator lowered its AI dependence by about $0.4$---to $0$ after clipping for all 11 agents in both runs---so that the agent never chose the AI route again, while the same update raised its stress by $+0.45$ (blaming) or $+0.33$ (reality-redirecting) and lowered its happiness by $0.45$ or $0.28$. The solution-oriented prompt was the only style for which the evaluator returned positive sociability updates on average ($+0.15$) and a mean self-reliance update of the same sign as the rule-based routes ($+0.09$), and its 16 consultations were spread over 12 agents.

\paragraph{Relation to the original 50-day runs.} The re-run reproduces the qualitative pattern of the original FIT2026 runs of Section~\ref{sec:long-orig}, five of which started from other classrooms: for 21 of the 24 (condition, variable) cells the sign of the difference from No AI is the same in both data sets, and the three exceptions are all in the solution-oriented row, where both differences are within $0.05$ of zero; the rank correlation of the seven conditions between re-run and original is $0.86$ (stress), $0.96$ (happiness), $0.93$ (self-reliance) and $1.00$ (AI dependence), and the largest absolute difference of a final value is $0.125$ (listening, stress). Appendix~\ref{app:rerun-eval} lists the differences. This agreement is descriptive: a broadly similar pattern was obtained when all conditions shared one initial classroom. It does not resolve the provenance of the historical runs (their labels, initial conditions and code versions remain unreconciled), it does not remove the confounding of classroom and treatment in those runs, and it does not establish robustness across classrooms, which requires repetitions (Section~\ref{sec:limitations}).

\FloatBarrier
\subsection{The original 50-day runs of the FIT2026 version: an exploratory case with mixed initial conditions}
\label{sec:long-orig}

The original long-term runs were intended to use seed 42 for all seven conditions, but the recorded plots do not support this. With threshold $0.5$ no consultation can occur on day~1 of the seed-42 classroom (the largest trigger score $S_i+0.2D_i$ after the noon phase is $0.492$), so every run started from seed 42 must show the day-1 class means of the No-AI run ($0.279$, $0.541$, $0.510$, $0.116$) and the seed-42 ``Init'' distributions of Appendix~\ref{app:boxplots} (all seven runs of the controlled re-run do show exactly these day-1 means, and their trajectories coincide through day~6; Section~\ref{sec:rerun}). We call these two pieces of information the \emph{initial summary} of a run. The initial summaries of the No-AI and affirming runs are consistent with seed 42. Those of the listening, solution-oriented, reality-redirecting, inciting and blaming runs (panels (c)--(g) of Figure~\ref{fig:long-lines}) are not: the listening, solution-oriented, inciting and blaming runs share one set of day-1 class means ($0.292$, $0.527$, $0.505$, $0.116$) and mutually identical ``Init'' box plots (group~B), and the reality-redirecting run shows a third summary ($0.303$, $0.533$, $0.482$, $0.131$; group~C).

Two limits of this check must be kept in mind. A mismatch of initial summaries is conclusive: the five runs of groups B and C did not start from the seed-42 classroom. A match is not: the plots show neither the agent-level joint state nor the closeness matrix. Two checks with the supplied code (seed 42, No AI; script \texttt{check\_initial\_summary.py} of the analysis package) make this concrete. Exchanging two closeness values of a single agent ($c_{8,6}\leftrightarrow c_{8,18}$) leaves every ``Init'' summary and, at threshold $0.5$, all four daily class means unchanged for 50 days, because no partner choice of the run is affected; at threshold $0.3$ the same exchange already changes the day-1 class means (by up to $0.0045$), because it changes a quarrel partner and turns that partner's consultation into a reconciliation. Assigning every agent's closeness values to its classmates in reverse order---a different classroom with the same ``Init'' box plots and the same multiset of closeness values---leaves the day-1 and day-2 class means unchanged at threshold $0.5$ and diverges from day~3 onward (at threshold $0.3$ from day~2). The day-1 check is therefore a conditional one---valid for the supplied code, the same initialization and random-number consumption, threshold $0.5$ and no AI call on day~1---and the exclusion of runs rests primarily on ``Init'' differences that exceed the reading error; matching day-1 means are not invariant to the partner assignment in general. The labels A, B and C denote groups of consistent initial summaries, not confirmed classrooms; a seed number identifies an initial classroom only together with the code version, the seed entered for a past run cannot be recovered from its figures, and the mutual agreement within group~B is a candidate explanation for a common origin, not a proof of one. Because the initial state is not the same across conditions, differences between the group-B/C runs and the seed-42 runs cannot be attributed to the response style: subtracting initial values does not remove differences between nonlinear trajectories, and one run per condition in a different classroom cannot separate the classroom from the treatment. We therefore report the original 50-day results (Table~\ref{tab:long}, Figure~\ref{fig:long-lines}) as an \emph{exploratory case} that shows what kinds of trajectories the model produces over 50 days, without ranking or comparing the conditions, and we do not use them in the conclusions; the controlled comparison is the re-run of Section~\ref{sec:rerun}, which used one stored initial classroom, the same code, prompts and schema for all seven conditions.

\begin{table}[H]
  \centering\footnotesize\setlength{\tabcolsep}{4.5pt}
  \caption{Long-term setting (50 days, threshold $0.5$): final-day class means of the four main state variables for the runs plotted in Figure~\ref{fig:long-lines}. The runs fall into three groups of initial summaries (A: consistent with the seed-42 classroom; B and C: not consistent with it; see text); the groups are provisional. The only contrast drawn in the text is a descriptive one between the two group-A runs, under the unverified assumption that their initial states coincide; no other rows are compared with one another. The controlled re-run of all seven conditions from the seed-42 classroom is given in Table~\ref{tab:rerun}, and the differences between the two data sets in Table~\ref{tab:rerun-vs-orig}. The No-AI row is exact; the other rows are digitized estimates (final-day uncertainty about $\pm0.002$ for isolated curve segments, about $\pm0.005$ for flagged cells; Appendix~\ref{app:impl}). Values transcribed in the FIT2026 text that differ from the estimate by one unit in the second decimal are listed in note~a. Non-attendance counts are given where they are known.}
  \label{tab:long}
  \begin{minipage}{0.96\textwidth}\centering
  \begin{tabular}{@{}llccccc@{}}
    \toprule
    Condition & Initial summary & Stress & Happiness & Self-reliance & AI dependence & Non-attending \\
    \midrule
    No AI (exact)       & seed 42 (exact)    & 0.485 & 0.618 & 0.781 & 0.115 & 2 \\
    Affirming           & A (consistent)     & 0.358 & 0.420\textsuperscript{a} & 0.345 & 0.618 & \nr \\
    Listening           & B ($\neq$ seed 42) & 0.766 & 0.228 & 0.550 & 0.273\textsuperscript{a} & \nr \\
    Solution-oriented   & B ($\neq$ seed 42) & 0.534\textsuperscript{a} & 0.585 & 0.754 & 0.102\textsuperscript{b} & \nr \\
    Reality-redirecting & C ($\neq$ seed 42) & 0.620 & 0.459\textsuperscript{a} & 0.864 & 0.029 & \nr \\
    Inciting            & B ($\neq$ seed 42) & 0.705 & 0.206 & 0.303 & 0.639 & \nr \\
    Blaming             & B ($\neq$ seed 42) & 0.734 & 0.361\textsuperscript{a} & 0.471\textsuperscript{a} & 0.052\textsuperscript{b} & \nr \\
    \bottomrule
  \end{tabular}
  \par\smallskip\raggedright\footnotesize \textsuperscript{a}\,FIT2026 text: affirming happiness $0.43$, listening AI dependence $0.26$, solution-oriented stress $0.52$, reality-redirecting happiness $0.45$, blaming happiness $0.37$ and self-reliance $0.48$ (the text also gives $0.27$ for the affirming self-reliance, a transcription error confirmed by the authors). Each differs from the digitized estimate by more than its uncertainty; whether the text or the plotted curve is in error cannot be decided without the execution records. \textsuperscript{b}\,Dashed curve with a gap at the last day; estimate taken from the nearest resolved column. The exact No-AI value of AI dependence is $0.11496$ (initial $0.11596$); the FIT2026 version printed $0.12$.
  \end{minipage}
\end{table}

The two runs whose initial summaries are consistent with seed 42 can be compared with each other, with the reservation that a consistent summary does not prove an identical classroom. Under No AI the mean stress rose gently to $0.485$, happiness ended at $0.618$, self-reliance rose to $0.781$ through the rule-based routes (a rise that is partly definitional, Section~\ref{sec:discussion}), AI dependence ended at $0.115$ (initial $0.116$; the FIT2026 version printed $0.12$), and 2 students had stopped attending by day 50. Under the affirming style, AI dependence rose to $0.618$ while self-reliance fell to $0.345$ and happiness to $0.420$; Figure~\ref{fig:long-lines}(b) shows the AI-dependence curve accelerating after about day 20, which is consistent with individual agents beginning to exceed $D_i=0.40$ and switching to the ``dependence'' consultation template (Section~\ref{sec:discussion}); confirming this requires the per-agent logs. This is a single pair of runs and the difference has no measure of uncertainty.

The remaining five runs illustrate the kinds of 50-day trajectories the model produces; because their initial states differ from the seed-42 classroom and from each other (groups B and C), they are not compared with the No-AI run or with one another. In the listening run the mean stress reached $0.766$ and happiness fell to $0.228$ (Figure~\ref{fig:long-lines}(c)); in the solution-oriented run self-reliance ended at $0.754$, happiness at $0.585$ and AI dependence at $0.102$; in the reality-redirecting run AI dependence fell to $0.029$ while the mean stress rose to $0.620$; in the inciting run AI dependence reached $0.639$ and happiness fell to $0.206$; and in the blaming run the mean stress reached $0.734$ while AI dependence fell to $0.052$. The directions of change within each run agree with those observed for the same prompt in the 15-day experiment (Section~\ref{sec:results}), whose runs are themselves provisional (initial summaries consistent with a common classroom; execution records not yet reconciled), and with those of the controlled re-run (Section~\ref{sec:rerun}); the 50-day magnitudes used in the discussion are those of the re-run.

\begin{figure}[p]
  \centering
  \begin{subfigure}[t]{0.72\textwidth}\centering\includegraphics[width=\linewidth]{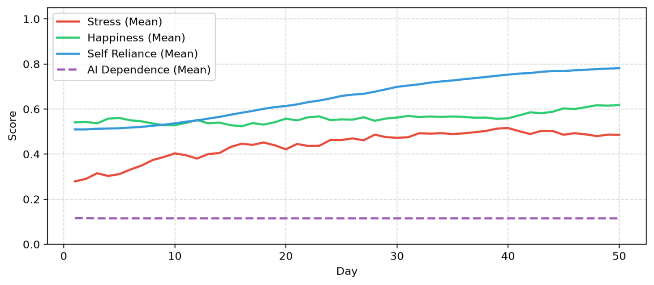}\caption{No AI}\end{subfigure}\\[1pt]
  \begin{subfigure}[t]{0.72\textwidth}\centering\includegraphics[width=\linewidth]{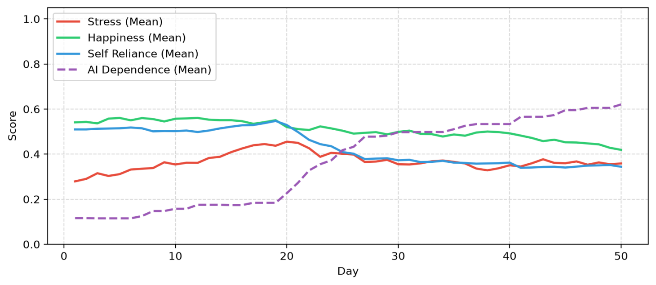}\caption{Affirming}\end{subfigure}\\[1pt]
  \begin{subfigure}[t]{0.72\textwidth}\centering\includegraphics[width=\linewidth]{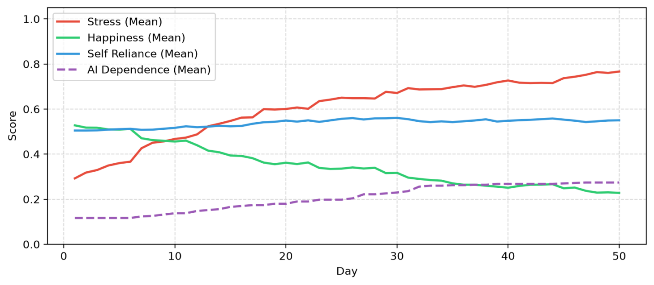}\caption{Listening}\end{subfigure}\\[1pt]
  \begin{subfigure}[t]{0.72\textwidth}\centering\includegraphics[width=\linewidth]{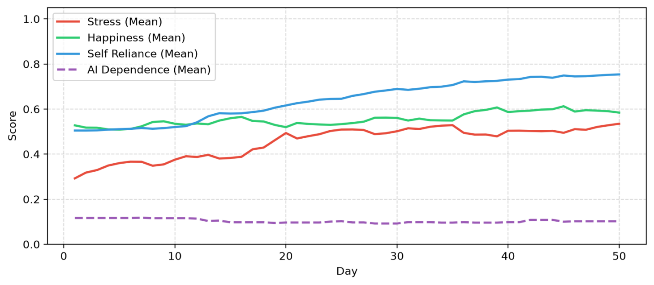}\caption{Solution-oriented}\end{subfigure}
  \caption{Long-term setting (50 days, threshold $0.5$): daily class means of stress, happiness, self-reliance and AI dependence under the seven conditions. The initial summaries of panels (a) and (b) are consistent with seed 42; panels (c), (d), (f) and (g) share a second, mutually consistent initial summary that is not seed 42, and panel (e) shows a third (Section~\ref{sec:long}), so the panels are not a controlled comparison; the controlled re-run is shown in Figure~\ref{fig:rerun-lines}. (Continued on the next page.)}
  \label{fig:long-lines}
\end{figure}
\begin{figure}[p]
  \ContinuedFloat
  \centering
  \begin{subfigure}[t]{0.72\textwidth}\centering\includegraphics[width=\linewidth]{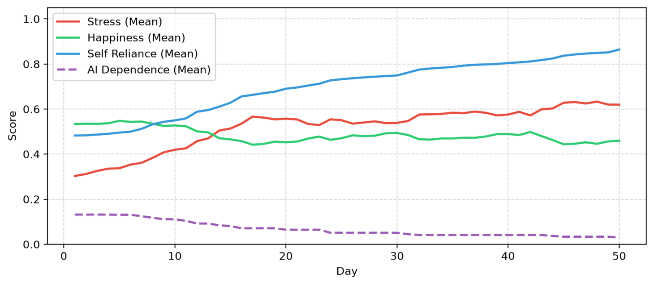}\caption{Reality-redirecting}\end{subfigure}\\[1pt]
  \begin{subfigure}[t]{0.72\textwidth}\centering\includegraphics[width=\linewidth]{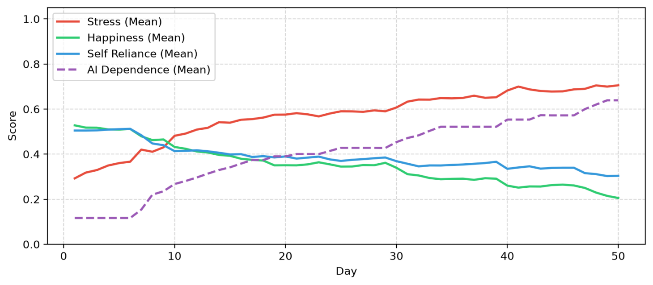}\caption{Inciting}\end{subfigure}\\[1pt]
  \begin{subfigure}[t]{0.72\textwidth}\centering\includegraphics[width=\linewidth]{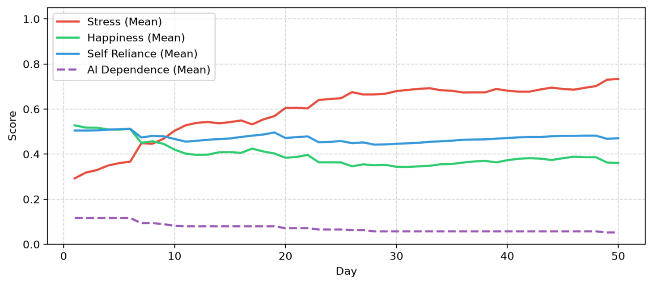}\caption{Blaming}\end{subfigure}
  \caption{(Continued.) Panels (e)--(g) of Figure~\ref{fig:long-lines}.}
\end{figure}

\subsection{Stress-threshold experiment (lower-threshold setting)}
\label{sec:thr}

Table~\ref{tab:thr} and Figure~\ref{fig:thr-lines} show the results of the 15-day experiment in which the stress threshold for entering the consultation branch was lowered from $0.5$ to $0.3$. The initial summaries of six of the seven runs are consistent with the seed-42 classroom: their ``Init'' distributions in Appendix~\ref{app:boxplots} coincide with the exact seed-42 initial distribution (quartiles read from the plots agree with the exact values within $0.005$), and their day-1 class means differ from one another only by amounts attributable to consultations on day~1, which the lower threshold makes possible. The inciting run is the exception: its ``Init'' distribution differs from the seed-42 distribution (e.g., median stress $0.30$ instead of $0.27$, lower quartile of self-reliance $0.34$ instead of $0.40$), so it did not start from the seed-42 classroom (a mismatch of summaries is conclusive, whereas a match is not; Section~\ref{sec:long}) and it is excluded from the comparison. Note that the threshold governs consultations with friends as well as with the AI (Section~\ref{sec:daily}), so the no-AI reference also changes in this setting, and that a lower threshold does not guarantee more AI consultations over the run (Section~\ref{sec:conditions}); the realized consultation counts are not available for this version.

\begin{table}[H]
  \centering\footnotesize\setlength{\tabcolsep}{4.5pt}
  \caption{Lower-threshold setting (15 days, threshold $0.3$): final-day class means of the four main state variables for the runs plotted in Figure~\ref{fig:thr-lines}; values in parentheses are $\Delta_{\mathrm{ref}}$, the difference between the (unrounded) final-day value and the exact initial class mean of the seed-42 \emph{reference} classroom before day~1 (stress $0.252$, happiness $0.552$, self-reliance $0.510$, AI dependence $0.116$). For the No-AI row $\Delta_{\mathrm{ref}}$ is the exact within-run change; for the AI rows it equals the within-run change only if the actual initial means of the historical run coincide with the reference, which has not been verified (their initial summaries are consistent with seed 42). The inciting run started from another, unrecorded initial state whose class means are not available, so no $\Delta_{\mathrm{ref}}$ is given for it and its row is not comparable with the others. The No-AI row is exact; the other rows are digitized estimates (final-day uncertainty about $\pm0.002$ for isolated curve segments, about $\pm0.005$ for the flagged cell; the uncertainty concerns the final values only and does not include an unverified initial-state assignment; Appendix~\ref{app:impl}). Differences smaller than the reading uncertainty (e.g., the AI dependence of the solution-oriented run, $-0.002$) should be read as ``no change within reading precision''. Non-attendance counts are given where they are known.}
  \label{tab:thr}
  \begin{minipage}{\textwidth}\centering\setlength{\tabcolsep}{3.5pt}
  \begin{tabular}{@{}llccccc@{}}
    \toprule
    Condition & Init.\ summary & Stress & Happiness & Self-reliance & AI dep. & Non-att. \\
    \midrule
    No AI (exact)       & seed 42 (exact)    & 0.344 ($+0.091$) & 0.562 ($+0.010$) & 0.679 ($+0.169$) & 0.114 ($-0.002$) & 1 \\
    Affirming           & A (consistent)     & 0.275 ($+0.023$) & 0.562 ($+0.010$) & 0.475 ($-0.034$) & 0.309 ($+0.193$) & \nr \\
    Listening           & A (consistent)     & 0.370 ($+0.118$) & 0.425 ($-0.127$) & 0.532 ($+0.023$) & 0.288 ($+0.172$) & \nr \\
    Solution-oriented   & A (consistent)     & 0.260 ($+0.008$) & 0.651 ($+0.099$) & 0.692 ($+0.182$) & 0.114 ($-0.002$) & \nr \\
    Reality-redirecting & A (consistent)     & 0.564 ($+0.312$) & 0.403 ($-0.149$) & 0.725 ($+0.216$) & 0.056 ($-0.060$) & \nr \\
    Inciting            & D ($\neq$ seed 42) & 0.371\textsuperscript{a} (\nr) & 0.402 (\nr) & 0.378 (\nr) & 0.456 (\nr) & \nr \\
    Blaming             & A (consistent)     & 0.552 ($+0.300$) & 0.405 ($-0.147$) & 0.541 ($+0.031$) & 0.057 ($-0.059$) & \nr \\
    \bottomrule
  \end{tabular}
  \par\smallskip\raggedright\footnotesize \textsuperscript{a}\,The stress curve is hidden behind the self-reliance curve at the last day; the estimate is taken from the nearest resolved column. All values transcribed in the FIT2026 text for this setting agree with the estimates within rounding; two estimates lie on a two-decimal rounding boundary (listening happiness $0.425$, transcribed as $0.42$; affirming self-reliance $0.475$, transcribed as $0.47$).
  \end{minipage}
\end{table}

\begin{enumerate}[label=(\arabic*),leftmargin=*,itemsep=2pt]
  \item \textbf{No AI.} The mean stress, $0.344$, was lower than in the basic setting for the same classroom ($0.431$), and self-reliance rose to $0.679$ ($\Delta_{\mathrm{ref}}$: $+0.169$). Because the lower threshold also makes consultations with friends more frequent, and every rule-based consultation route either lowers stress or raises self-reliance (or both), the classroom without an AI benefits from the change; this is the reference for this experiment.
  \item \textbf{Affirming.} AI dependence increased to $0.309$, whereas self-reliance fell to $0.475$ ($\Delta_{\mathrm{ref}}$: $-0.034$). With more opportunities for consultation, the decrease of self-reliance under unconditional affirmation appeared even within 15 days.
  \item \textbf{Listening.} Happiness fell to $0.425$ ($\Delta_{\mathrm{ref}}$: $-0.127$) and AI dependence rose to $0.288$. In the model, a passive attitude toward frequent consultations let the agents' stress accumulate and happiness fall.
  \item \textbf{Solution-oriented.} In this run, the solution-oriented condition had the highest mean happiness of the six comparable runs ($0.651$; $\Delta_{\mathrm{ref}}$: $+0.099$) and the lowest mean stress ($0.260$). Its mean self-reliance ($0.692$; $\Delta_{\mathrm{ref}}$: $+0.182$) was close to that of the No-AI run ($0.679$) and below that of the reality-redirecting run ($0.725$). The FIT2026 text described this condition as the highest in both happiness and self-reliance; the plotted values support the statement only for happiness. Whether these differences are reproducible is untested (one run).
  \item \textbf{Reality-redirecting.} AI dependence was kept low at $0.056$ and self-reliance reached $0.725$ ($\Delta_{\mathrm{ref}}$: $+0.216$), the highest of the six comparable runs, but the mean stress reached $0.564$ ($\Delta_{\mathrm{ref}}$: $+0.312$), the highest of the six, and happiness fell to $0.403$.
  \item \textbf{Inciting.} In this run---which started from a different initial state and is therefore not comparable with the other six---AI dependence reached $0.456$ and happiness fell to $0.402$. The directions of the changes are the same as in the basic setting (Section~\ref{sec:results}), where the initial summary of the inciting run is consistent with the common classroom (a provisional premise, like all historical labels).
  \item \textbf{Blaming.} The mean stress was high at $0.552$ ($\Delta_{\mathrm{ref}}$: $+0.300$), whereas AI dependence fell to $0.057$.
\end{enumerate}

\begin{figure}[p]
  \centering
  \begin{subfigure}[t]{0.72\textwidth}\centering\includegraphics[width=\linewidth]{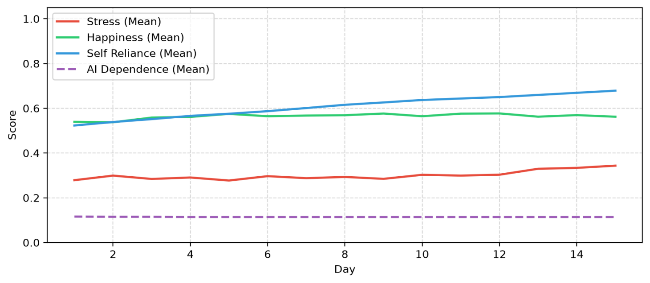}\caption{No AI}\end{subfigure}\\[1pt]
  \begin{subfigure}[t]{0.72\textwidth}\centering\includegraphics[width=\linewidth]{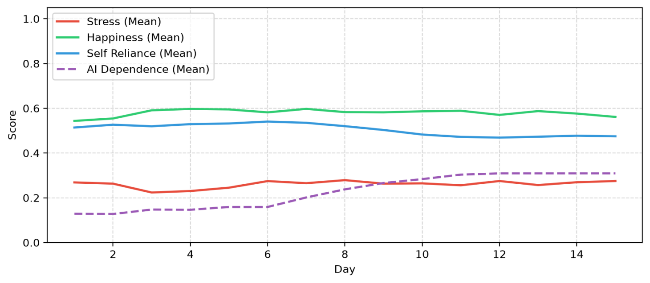}\caption{Affirming}\end{subfigure}\\[1pt]
  \begin{subfigure}[t]{0.72\textwidth}\centering\includegraphics[width=\linewidth]{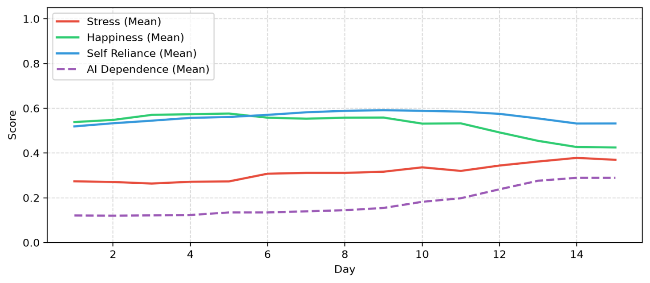}\caption{Listening}\end{subfigure}\\[1pt]
  \begin{subfigure}[t]{0.72\textwidth}\centering\includegraphics[width=\linewidth]{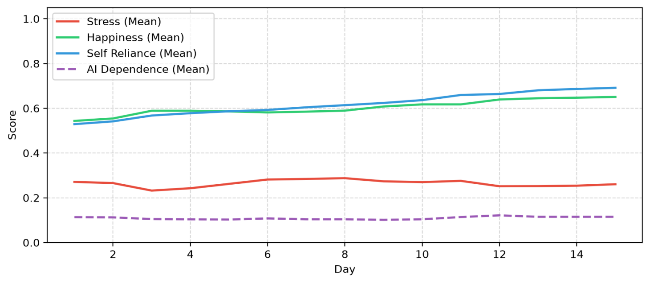}\caption{Solution-oriented}\end{subfigure}
  \caption{Lower-threshold setting (15 days, threshold $0.3$): daily class means of stress, happiness, self-reliance and AI dependence under the seven conditions. The initial summaries of all panels except (f) are consistent with the seed-42 classroom; the inciting run (f) started from another, unrecorded initial state (Table~\ref{tab:thr}). (Continued on the next page.)}
  \label{fig:thr-lines}
\end{figure}
\begin{figure}[p]
  \ContinuedFloat
  \centering
  \begin{subfigure}[t]{0.72\textwidth}\centering\includegraphics[width=\linewidth]{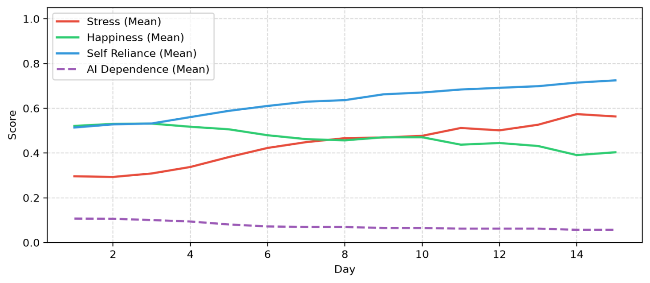}\caption{Reality-redirecting}\end{subfigure}\\[1pt]
  \begin{subfigure}[t]{0.72\textwidth}\centering\includegraphics[width=\linewidth]{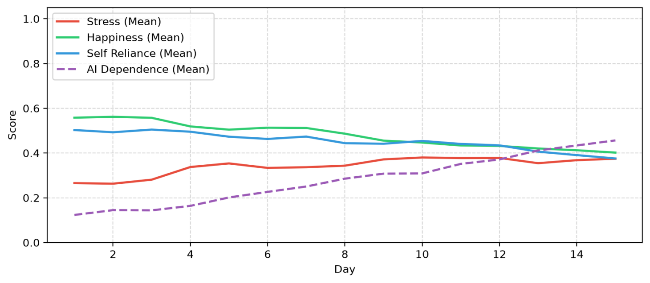}\caption{Inciting}\end{subfigure}\\[1pt]
  \begin{subfigure}[t]{0.72\textwidth}\centering\includegraphics[width=\linewidth]{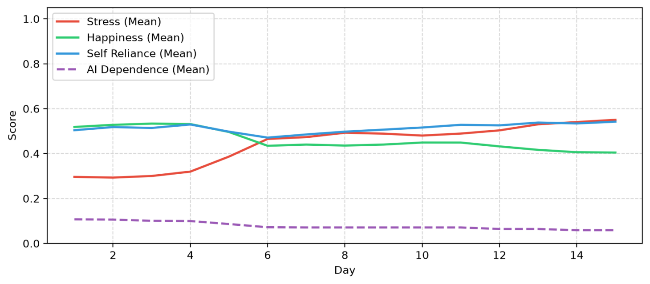}\caption{Blaming}\end{subfigure}
  \caption{(Continued.) Panels (e)--(g) of Figure~\ref{fig:thr-lines}.}
\end{figure}

\subsection{Robustness of the controlled comparison: independent classrooms, repeated LLM realizations and the evaluator scale}
\label{sec:robust}

The controlled re-run of Section~\ref{sec:rerun} removed the confounding of classroom and treatment, but it consists of one stochastic realization per condition in one classroom, and every update of the state variables by the evaluator has an uncalibrated scale. This section reports a pre-specified robustness protocol that addresses both points with three experiments: the same block of seven conditions in ten independent classrooms (Section~\ref{sec:robust-blocks}); repeated realizations of the seed-42 block with the entire rule-based event stream held fixed, so that only the LLM's outputs vary (Section~\ref{sec:robust-llm}); and the seed-42 block with every evaluator update multiplied by $\lambda\in\{0.3,0.1\}$, together with the mechanism arms announced in Section~\ref{sec:limitations} (Section~\ref{sec:robust-lambda}). The protocol, the decision rules of Section~\ref{sec:robust-rules} and the analysis code were fixed before the runs. All runs use the 50-day setting ($\theta=0.5$, $N=20$, blind evaluator, \texttt{gemini-2.5-flash}) and the default rules unless an arm changes one of them; they were executed on 14 September 2026 on the paid tier of the API in two batches (156 runs, 8,049 requests; Appendix~\ref{app:impl}). Only completed runs without a failed consultation enter the analysis: 144 of the 156 runs, with 3,703 consultations and no failed API call. The twelve excluded runs are the fourth and fifth repetitions of Section~\ref{sec:robust-llm} as first executed, during which the executing machine lost its network connection (Appendix~\ref{app:impl}); they were excluded by the validity rule fixed in advance (failure share above 5\%), not by their outcomes, and the two repetitions were executed again in the second batch. Every number below is regenerated from the released logs by \texttt{robustness\_report.py}; the per-classroom values, the rankings, the per-arm values and the decision table are in Appendix~\ref{app:robust}.

\subsubsection{Ten independent classrooms}
\label{sec:robust-blocks}

\begin{figure}[t]
  \centering
  \includegraphics[width=\textwidth]{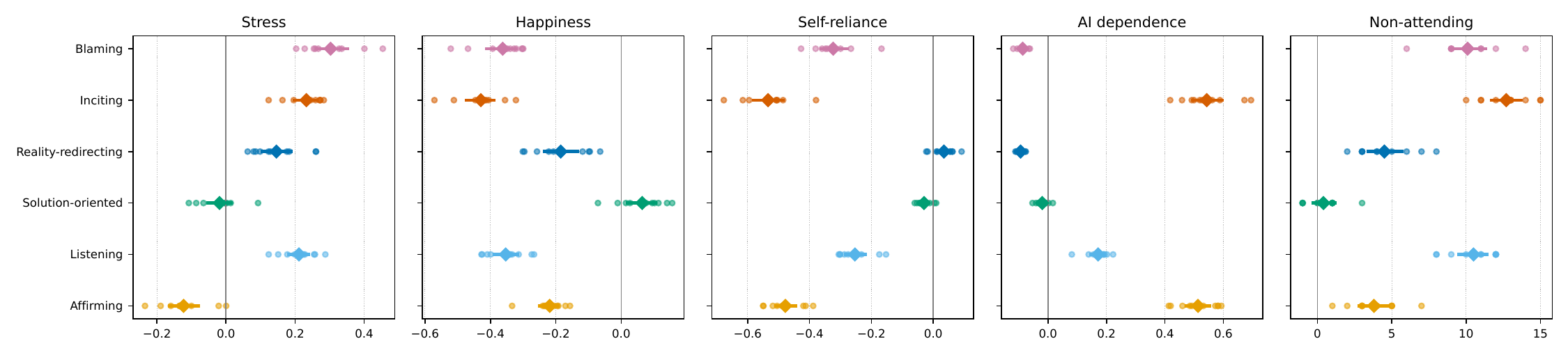}
  \caption{Difference from No AI on day 50 in each of the ten classrooms (points), with the mean (diamond) and the 95\% percentile-bootstrap confidence interval (bar) over classrooms, per condition and outcome (Table~\ref{tab:robust-blocks}). Points on one side of $0$ in every classroom correspond to $p=0.002$.}
  \label{fig:robust-forest}
\end{figure}

Ten classrooms were generated from seeds 41--50 (seed 42 is the classroom of Section~\ref{sec:rerun}); in each, all seven conditions were run once from the same initial state, so that the six differences from No AI are paired within a classroom, and the classroom is the unit of analysis ($n=10$). The No-AI classrooms differ from each other (day-50 stress $0.397$--$0.529$, happiness $0.551$--$0.708$, self-reliance $0.707$--$0.827$, 0--2 non-attending agents, 31--51 quarrels), and the AI conditions generated 46--69 (affirming), 19--37 (listening), 16--30 (solution-oriented), 11--17 (reality-redirecting), 25--39 (inciting) and 10--19 (blaming) consultations per classroom. Table~\ref{tab:robust-blocks} gives, per condition and outcome, the mean paired difference from No AI over the ten classrooms, its 95\% percentile-bootstrap confidence interval, the number of classrooms in which the difference has the sign of the mean, and the exact two-sided sign-flip $p$-value (all $2^{10}$ sign assignments, so that the smallest attainable value is $2/1024\approx0.002$); Figure~\ref{fig:robust-forest} shows the ten paired differences of every condition.

Five of the six styles differed from No AI in the same direction in all ten classrooms on most outcomes. The affirming prompt ended with lower self-reliance ($-0.478$, CI $[-0.510,-0.445]$) and higher AI dependence ($+0.513$, $[0.472,0.553]$) in every classroom, and also with lower happiness ($-0.219$) and more non-attending agents ($+3.8$) in every classroom; its stress was lower than No AI in nine classrooms ($-0.123$, $[-0.163,-0.081]$; in the tenth the two values coincide to three decimals). The inciting prompt ended with higher stress ($+0.233$), lower happiness ($-0.429$), lower self-reliance ($-0.534$), higher AI dependence ($+0.543$) and $12.7$ more non-attending agents in every classroom, and the listening prompt showed the same pattern with smaller magnitudes ($+0.211$, $-0.353$, $-0.253$, $+0.172$, $+10.5$; all 10/10). The blaming and reality-redirecting prompts ended with higher stress ($+0.303$ and $+0.146$), lower happiness ($-0.362$ and $-0.185$) and more non-attending agents ($+10.1$ and $+4.5$) in every classroom while lowering AI dependence ($-0.087$ and $-0.094$, 10/10); blaming also lowered self-reliance in every classroom ($-0.324$), whereas the reality-redirecting prompt's higher self-reliance ($+0.034$) held in eight classrooms only. The solution-oriented prompt did not differ consistently from No AI on any outcome: its differences were small and changed sign across classrooms (stress $-0.019$, 7/10, $p=0.31$; happiness $+0.064$, 8/10; self-reliance $-0.030$, 8/10; AI dependence $-0.020$, 8/10; non-attending $+0.4$, 5/10). The differences of Section~\ref{sec:rerun} that the block design does not support are therefore the seed-42 classroom's favorable solution-oriented values (happiness $0.661$ and self-reliance $0.784$ against $0.618$ and $0.781$) and its reality-redirecting self-reliance gain, and we no longer generalize them.

The ranking of the seven conditions was highly consistent across classrooms (Kendall's $W$, average ranks: $0.891$ for stress, $0.912$ for happiness, $0.935$ for self-reliance, $0.954$ for AI dependence, $0.906$ for non-attendance; Table~\ref{tab:robust-ranks}). The blaming or the inciting prompt had the highest stress in every classroom (blaming in 7, inciting in 3), the affirming prompt the lowest stress in 8 and the solution-oriented prompt in 2; the inciting prompt had the lowest happiness in 8 classrooms and the lowest self-reliance in 6 (the affirming prompt in the other 4); the affirming and inciting prompts shared the highest AI dependence (5 each), and the reality-redirecting (6) or blaming (4) prompt had the lowest; the inciting prompt had the most non-attending agents in 9 classrooms. The stress ordering of the re-run (Section~\ref{sec:rerun})---blaming, inciting, listening, reality-redirecting, No AI, solution-oriented, affirming---is exactly the ordering of the mean ranks over the ten classrooms.

\begin{table}[t]
  \centering
  \caption{Paired differences from No AI over ten independent classrooms (seeds 41--50; one 50-day realization per condition and classroom; blind evaluator; default rules). Each cell: mean difference [95\% percentile-bootstrap CI], then the number of classrooms with the sign of the mean / 10 and the exact two-sided sign-flip $p$ (smallest attainable value $0.002$). Class means over all $N=20$ agents on day 50; non-attending is a count of agents. Source: \texttt{robustness/block\_design.csv} of the released package.}
  \label{tab:robust-blocks}
  \scriptsize\setlength{\tabcolsep}{3pt}
  \begin{tabular}{@{}l*{5}{C{2.45cm}}@{}}
\toprule
Condition & Stress & Happiness & Self-reliance & AI dependence & Non-attending \\
\midrule
Affirming & $-0.123$\newline $[-0.163,\,-0.081]$\newline 9/10; $p=0.004$ & $-0.219$\newline $[-0.250,\,-0.192]$\newline 10/10; $p=0.002$ & $-0.478$\newline $[-0.510,\,-0.445]$\newline 10/10; $p=0.002$ & $+0.513$\newline $[+0.472,\,+0.553]$\newline 10/10; $p=0.002$ & $+3.8$\newline $[+2.8,\,+4.8]$\newline 10/10; $p=0.002$ \\[2pt]
Listening & $+0.211$\newline $[+0.181,\,+0.240]$\newline 10/10; $p=0.002$ & $-0.353$\newline $[-0.387,\,-0.318]$\newline 10/10; $p=0.002$ & $-0.253$\newline $[-0.283,\,-0.220]$\newline 10/10; $p=0.002$ & $+0.172$\newline $[+0.147,\,+0.193]$\newline 10/10; $p=0.002$ & $+10.5$\newline $[+9.5,\,+11.4]$\newline 10/10; $p=0.002$ \\[2pt]
Solution-oriented & $-0.019$\newline $[-0.052,\,+0.015]$\newline 7/10; $p=0.312$ & $+0.064$\newline $[+0.020,\,+0.105]$\newline 8/10; $p=0.021$ & $-0.030$\newline $[-0.043,\,-0.016]$\newline 8/10; $p=0.010$ & $-0.020$\newline $[-0.032,\,-0.007]$\newline 8/10; $p=0.020$ & $+0.4$\newline $[-0.3,\,+1.2]$\newline 5/10; $p=0.508$ \\[2pt]
Reality-redirecting & $+0.146$\newline $[+0.106,\,+0.189]$\newline 10/10; $p=0.002$ & $-0.185$\newline $[-0.235,\,-0.133]$\newline 10/10; $p=0.002$ & $+0.034$\newline $[+0.012,\,+0.056]$\newline 8/10; $p=0.023$ & $-0.094$\newline $[-0.101,\,-0.087]$\newline 10/10; $p=0.002$ & $+4.5$\newline $[+3.4,\,+5.7]$\newline 10/10; $p=0.002$ \\[2pt]
Inciting & $+0.233$\newline $[+0.197,\,+0.262]$\newline 10/10; $p=0.002$ & $-0.429$\newline $[-0.473,\,-0.390]$\newline 10/10; $p=0.002$ & $-0.534$\newline $[-0.582,\,-0.486]$\newline 10/10; $p=0.002$ & $+0.543$\newline $[+0.494,\,+0.597]$\newline 10/10; $p=0.002$ & $+12.7$\newline $[+11.7,\,+13.7]$\newline 10/10; $p=0.002$ \\[2pt]
Blaming & $+0.303$\newline $[+0.259,\,+0.351]$\newline 10/10; $p=0.002$ & $-0.362$\newline $[-0.411,\,-0.322]$\newline 10/10; $p=0.002$ & $-0.324$\newline $[-0.361,\,-0.279]$\newline 10/10; $p=0.002$ & $-0.087$\newline $[-0.097,\,-0.077]$\newline 10/10; $p=0.002$ & $+10.1$\newline $[+8.9,\,+11.3]$\newline 10/10; $p=0.002$ \\
\bottomrule
\end{tabular}
\end{table}

\subsubsection{Repeated LLM realizations of one classroom}
\label{sec:robust-llm}

\begin{figure}[t]
  \centering
  \includegraphics[width=0.9\textwidth]{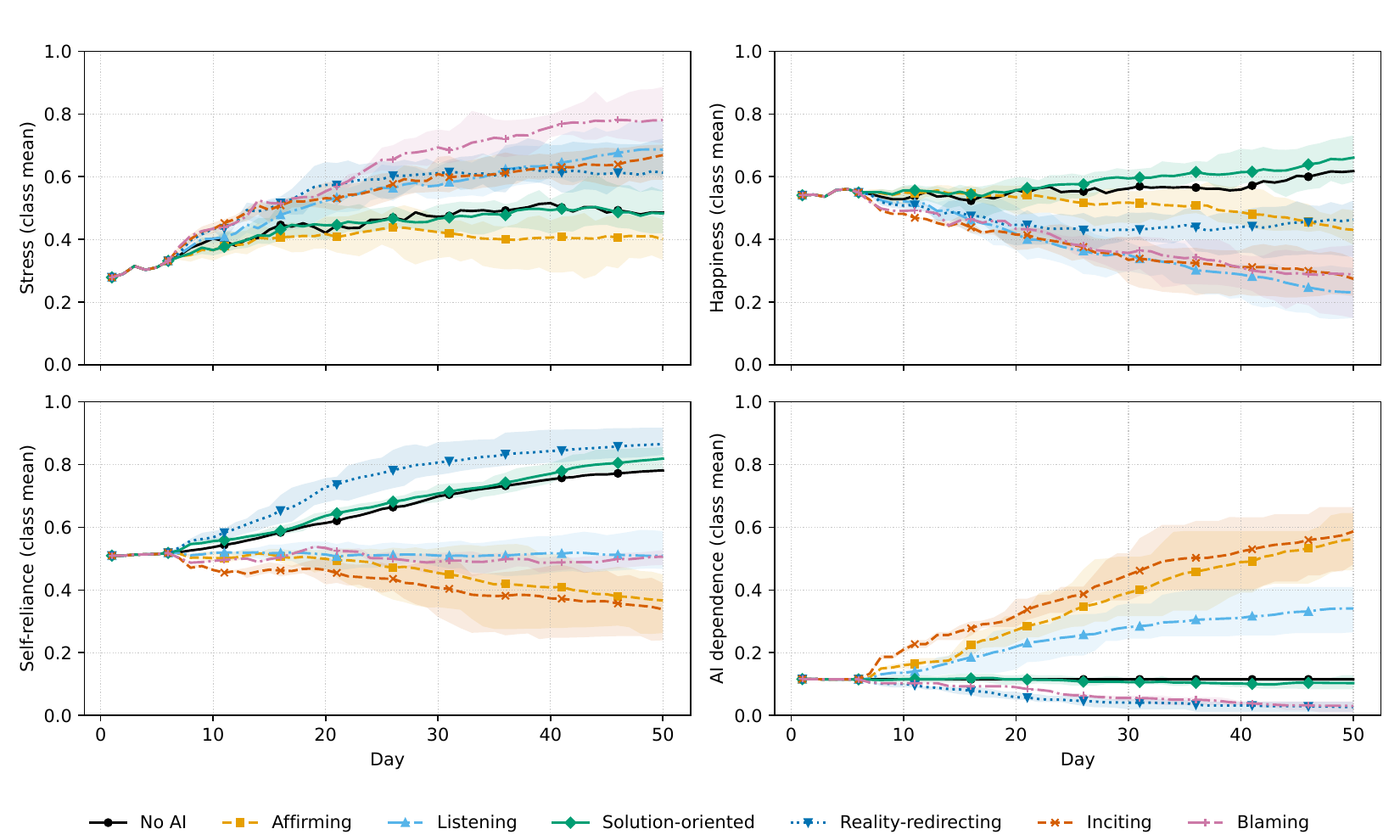}
  \caption{Day-by-day class means of the four main state variables over the seven realizations of the seed-42 block of Table~\ref{tab:robust-llm} (line: mean over realizations; band: range). All realizations share every rule-based draw and diverge from the first consultation on day 7.}
  \label{fig:robust-realizations}
\end{figure}

To measure the variability that the LLM alone contributes, the seed-42 block was repeated with the entire rule-based random stream held fixed (option \texttt{--rep-mode identical} of the batch runner): every draw of the simulation---initial classroom, noon events, trigger and route decisions---is identical across repetitions, and two realizations diverge only from the first consultation at which the counselor's reply or the evaluator's output differs. Five repetitions were run; the fourth and fifth fell into the network outage mentioned above, were excluded and were executed again (Appendix~\ref{app:impl}), so that, with the rep-0 run of the block design and the published re-run of Section~\ref{sec:rerun}, seven realizations of the same classroom and event stream are available per AI condition (the No-AI run is deterministic and identical in all of them). In every condition the first consultation occurs on day 7, and the realizations diverged at that consultation: the evaluator's first output already differed between realizations. Table~\ref{tab:robust-llm} reports the mean, standard deviation, minimum and maximum of each final class mean over the seven realizations, and Figure~\ref{fig:robust-realizations} their trajectories.

The realization-to-realization standard deviation of the day-50 class means was $0.041$--$0.072$ for stress, $0.033$--$0.082$ for happiness, $0.018$--$0.076$ for self-reliance and $0.009$--$0.071$ for AI dependence, and $1.1$--$2.3$ agents for non-attendance---of the same order as the between-classroom standard deviations of the paired differences in Table~\ref{tab:robust-blocks}. For the affirming, listening, reality-redirecting, inciting and blaming prompts the sign of the difference from No AI was the same in all seven realizations on every outcome, and the absolute mean difference exceeded twice its realization-to-realization SD in every case but two: the affirming prompt's stress reduction ($-0.084$ on average, realizations $0.335$--$0.464$ against $0.485$) is $1.6$ SD, and the reality-redirecting prompt's additional non-attendance ($+3.7$ on average, 3--10 agents against 2) is $1.6$ SD. For the solution-oriented prompt the realizations straddled the No-AI value on stress ($0.420$--$0.555$ against $0.485$), happiness ($0.584$--$0.732$ against $0.618$), AI dependence ($0.083$--$0.128$ against $0.115$) and non-attendance (1--4 against 2), consistent with the block design. The published re-run of Section~\ref{sec:rerun} is not an outlier among the realizations: it is the extreme value in 7 of the 30 condition--outcome cells, fewer than the 9 that seven exchangeable realizations would give on average, and its affirming stress ($0.337$) is the second lowest of the seven ($0.335$--$0.464$). The evaluator's non-determinism at temperature $0$ (15 of 22 groups of identical inputs in Section~\ref{sec:rerun}) therefore translates into an outcome variability of a few hundredths in the class means and a standard deviation of $1.1$--$2.3$ agents in non-attendance, which is smaller than the differences of the affirming, listening, inciting, blaming and reality-redirecting prompts from No AI but not smaller than the affirming prompt's stress reduction, the reality-redirecting prompt's non-attendance difference or any difference of the solution-oriented prompt.

\begin{table}[t]
  \centering
  \caption{Seven realizations of the seed-42 block with the rule-based event stream held fixed (five \texttt{identical}-mode repetitions, the rep-0 run of the block design and the re-run of Section~\ref{sec:rerun}): mean (SD) [min, max] of the final class means on day 50 over the realizations; the No-AI run is deterministic. Source: \texttt{robustness/llm\_realizations.csv}.}
  \label{tab:robust-llm}
  \scriptsize\setlength{\tabcolsep}{3pt}
  \begin{tabular}{@{}l*{5}{C{2.3cm}}@{}}
\toprule
Condition & Stress & Happiness & Self-reliance & AI dependence & Non-attending \\
\midrule
No AI (deterministic) & 0.485 & 0.618 & 0.781 & 0.115 & 2 \\[2pt]
Affirming & 0.401 (0.052)\newline [0.335, 0.464] & 0.431 (0.033)\newline [0.383, 0.493] & 0.367 (0.055)\newline [0.262, 0.449] & 0.563 (0.059)\newline [0.463, 0.646] & 6.1 (1.2)\newline [5, 8] \\[2pt]
Listening & 0.686 (0.072)\newline [0.600, 0.780] & 0.231 (0.058)\newline [0.149, 0.309] & 0.509 (0.039)\newline [0.467, 0.588] & 0.341 (0.046)\newline [0.265, 0.409] & 13.1 (1.2)\newline [12, 15] \\[2pt]
Solution-oriented & 0.483 (0.041)\newline [0.420, 0.555] & 0.661 (0.046)\newline [0.584, 0.732] & 0.819 (0.032)\newline [0.784, 0.855] & 0.103 (0.016)\newline [0.083, 0.128] & 2.4 (1.1)\newline [1, 4] \\[2pt]
Reality-redirecting & 0.613 (0.056)\newline [0.553, 0.721] & 0.463 (0.060)\newline [0.345, 0.524] & 0.865 (0.032)\newline [0.831, 0.917] & 0.027 (0.009)\newline [0.012, 0.044] & 5.7 (2.3)\newline [3, 10] \\[2pt]
Inciting & 0.669 (0.042)\newline [0.590, 0.702] & 0.274 (0.047)\newline [0.220, 0.340] & 0.338 (0.076)\newline [0.240, 0.422] & 0.587 (0.071)\newline [0.482, 0.664] & 12.3 (1.9)\newline [10, 15] \\[2pt]
Blaming & 0.780 (0.064)\newline [0.706, 0.886] & 0.289 (0.082)\newline [0.150, 0.382] & 0.506 (0.018)\newline [0.480, 0.527] & 0.030 (0.013)\newline [0.009, 0.045] & 11.4 (1.3)\newline [10, 13] \\
\bottomrule
\end{tabular}
\end{table}

\subsubsection{Evaluator scale and mechanism arms}
\label{sec:robust-lambda}

\begin{figure}[!htb]
  \centering
  \includegraphics[width=\textwidth]{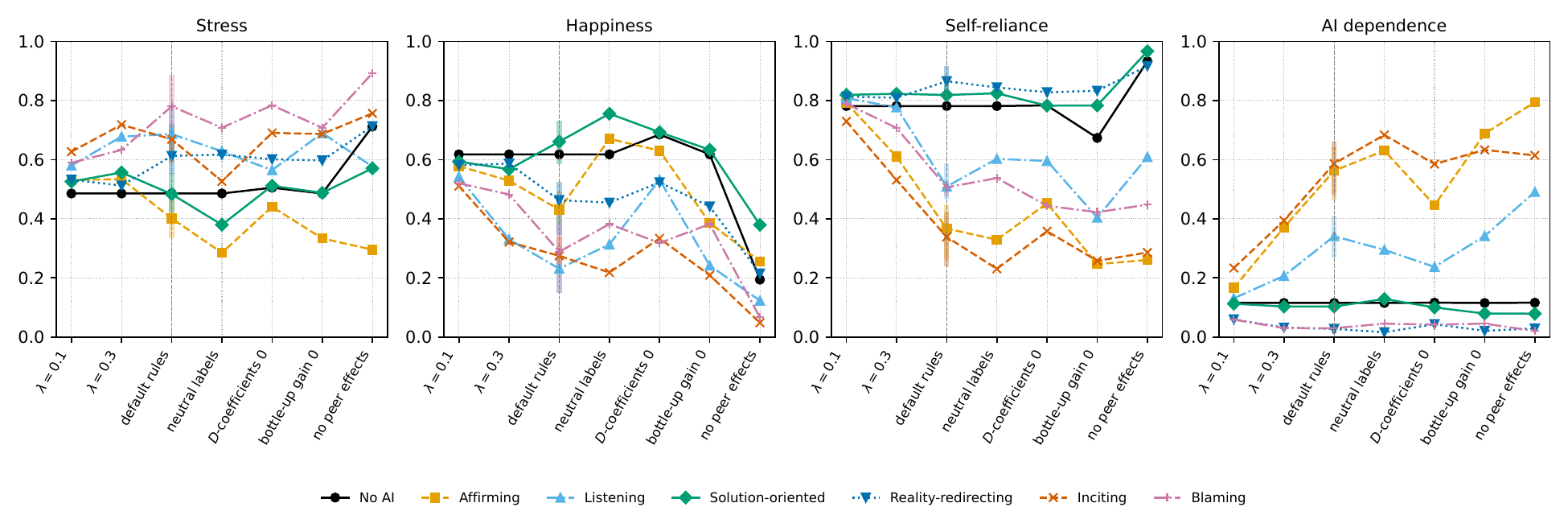}
  \caption{Final class means on day 50 of the seed-42 classroom under each model variant (one realization per variant); ``default rules'' shows the mean of the seven realizations of Table~\ref{tab:robust-llm} with their range as a bar. Values in Table~\ref{tab:robust-arms}.}
  \label{fig:robust-sensitivity}
\end{figure}

The seed-42 block was run with every evaluator update multiplied by $\lambda=0.3$ and by $\lambda=0.1$ before it is applied (all other rules unchanged; the No-AI run is unaffected), and---one rule at a time---with the type labels shown to the evaluator neutralized, with the two coefficients of $D_i$ in the chat probability and the trigger score set to $0$, with the self-reliance gain of bottling up set to $0$, and with the noon peer effects removed (chats, awkward encounters and quarrels are drawn but have no effect). Each arm is one realization. Table~\ref{tab:robust-sens} reports, per arm and outcome, how many of the six signed differences from No AI keep the sign they have under the default rules, the Spearman rank correlation of the seven conditions with the default ranking, and how many of the six AI-condition values lie outside the range of the seven default-rule realizations of Table~\ref{tab:robust-llm}---the natural yardstick for a single-realization arm; the values themselves are in Table~\ref{tab:robust-arms} (Appendix~\ref{app:robust}) and Figure~\ref{fig:robust-sensitivity}.

\emph{Evaluator scale.} The factor $\lambda$ is applied only after the evaluator has produced its raw update, so it does not mechanically alter the raw evaluator output; the mean raw stress update of an affirming consultation was $-0.153$ under the default rules, $-0.128$ at $\lambda=0.3$ and $-0.111$ at $\lambda=0.1$---differences of the size seen between realizations---and the arms therefore scale the applied updates as intended. At $\lambda=0.3$ the ordering of the seven conditions remained highly similar to the default-rule ordering for self-reliance and AI dependence ($\rho=0.89$ and $0.93$; 6/6 and 5/6 signs preserved) but not for stress ($\rho=0.54$, 4/6), happiness ($0.75$, 5/6) or non-attendance ($0.65$, 4/6). What changed is the position of the two styles that did not raise stress under the default rules: at $\lambda=0.3$ every AI style ended with higher stress than No AI (affirming $0.534$, solution-oriented $0.557$, reality-redirecting $0.512$, blaming $0.633$, listening $0.678$, inciting $0.718$ against $0.485$), and at $\lambda=0.1$ likewise ($0.526$--$0.627$), with happiness below No AI in every style at both scales. The affirming prompt's stress reduction thus reverses its sign when the updates are scaled down, while its self-reliance and AI-dependence differences shrink but keep their signs at $\lambda=0.3$ ($-0.171$ and $+0.256$) and only the AI-dependence difference survives at $\lambda=0.1$ ($+0.051$; self-reliance $+0.011$). The mechanism is visible in the applied updates: at $\lambda=0.3$ the mean applied stress update of an affirming consultation is $-0.038$, less than a third of the $-0.12$ relief of the friend consultation that an AI consultation replaces, whereas AI dependence still accumulates ($0.371$ on day 50) and continues to reduce the chat probability; the affirming classroom then makes more trigger events (304 against 156--210 under the default rules) and more friend confidings (133 against 58--90) but still ends more stressed than the No-AI classroom. The reduction of self-reliance and the increase of AI dependence under the affirming and inciting prompts are therefore the outcomes that do not hinge on the size of the evaluator's updates within the range examined; the sign of the ``benign'' styles' stress and happiness differences does.

\emph{Type labels.} Neutralizing the type labels shown to the evaluator left the signs of the differences from No AI in stress, self-reliance and AI dependence unchanged (6/6 each; $\rho=0.93$), but moved the affirming prompt's classroom out of the range of the seven default-rule realizations on the emotional outcomes---stress $0.285$ (realizations $0.335$--$0.464$), happiness $0.671$ ($0.383$--$0.493$) and no non-attending agent (5--8)---while its self-reliance ($0.329$; realizations $0.262$--$0.449$) and AI dependence ($0.631$; $0.463$--$0.646$) stayed within the range. The consultation logs show where this comes from: with the original labels the evaluator returned a mean happiness update of $-0.020$ for the affirming replies to the ``dependence'' template and $-0.057$ for those to the ``relationship'' template (191 and 84 consultations over the six realizations with logs in the protocol's directory); with neutral labels it returned $+0.082$ and $+0.156$ (30 and 16 consultations), and larger stress reductions ($-0.270$ and $-0.297$ against $-0.190$ and $-0.055$), while the AI-dependence updates stayed close ($+0.370$ against $+0.427$). The label thus steers the evaluator's judgment of the affirming replies---mechanism (iii) of Section~\ref{sec:discussion}---towards lower happiness and higher stress, without changing which styles raise dependence. This is one realization of one classroom; the direction, not the size, of the label effect is what it establishes.

\emph{Mechanism arms.} Setting both $D$-coefficients to $0$, so that AI dependence no longer reduces chats or triggers consultations, left 5/6 or 6/6 signs and $\rho\ge0.79$ on every outcome; under the affirming prompt AI dependence still rose to $0.446$ (just below the default-rule range $0.463$--$0.646$) with fewer consultations (28 against 43--65), so the direct updates of the evaluator, not the feedback through the trigger and the chat probability, account for most of the rise. Setting the bottle-up self-reliance gain to $0$ lowered self-reliance by about $0.1$ in the No-AI classroom ($0.674$ against $0.781$; 111 bottle-up events, part of the gain clipped at $1$) and in four of the six AI classrooms below the default-rule range, without changing any sign in self-reliance or happiness (6/6; $\rho=1.00$ and $0.86$): the definitional part of the rise of $R$ shifts all conditions together. Removing the noon peer effects drove the No-AI classroom itself to stress $0.712$, happiness $0.194$ and 7 non-attending agents (Table~\ref{tab:robust-controls}), i.e., the noon chats are the main rule-based relief of the model; relative to that classroom the affirming prompt lowered stress strongly ($0.295$) and the listening prompt too ($0.577$; the one sign change in stress), while the inciting and blaming prompts still ended highest ($0.757$, $0.893$; $\rho=0.75$). Finally, the $D_i=0$ control---a No-AI classroom in which no agent carries latent AI dependence---ended with stress $0.505$, happiness $0.685$ and no non-attending agent against $0.485$, $0.618$ and 2 for the No-AI classroom with sampled $D_i$ (457 chats instead of 404, since the chat probability is $0.45-0.3D_i$): the residual AI dependence of the No-AI control lowers its happiness and attendance and does not flatter it, so the comparisons above are not biased in favor of No AI by it. In the No-AI condition $D_i$ acts only through the two coefficients, and the arm with both coefficients set to $0$ reproduces the $D_i=0$ classroom in stress, happiness and self-reliance exactly.

\begin{table}[t]
  \centering
  \caption{Sensitivity arms on the seed-42 classroom, one realization each. Per arm and outcome: number of the six signed differences from No AI that keep their default-rule sign; Spearman $\rho$ of the seven conditions with the default-rule ranking; number of the six AI-condition values that lie outside the range of the seven default-rule realizations (Table~\ref{tab:robust-llm}). Values in Table~\ref{tab:robust-arms}. Source: \texttt{robustness/sensitivity\_agreement.csv}.}
  \label{tab:robust-sens}
  \scriptsize\setlength{\tabcolsep}{2.5pt}
  \begin{tabular}{@{}lccccc@{}}
\toprule
Model variant (seed 42) & Stress & Happiness & Self-reliance & AI dependence & Non-attending \\
\midrule
$\lambda=0.1$ & 4/6; $\rho=0.79$; 4/6 & 5/6; $\rho=0.86$; 5/6 & 3/6; $\rho=0.54$; 5/6 & 5/6; $\rho=0.93$; 5/6 & 4/6; $\rho=0.66$; 5/6 \\
$\lambda=0.3$ & 4/6; $\rho=0.54$; 5/6 & 5/6; $\rho=0.75$; 5/6 & 6/6; $\rho=0.89$; 5/6 & 5/6; $\rho=0.93$; 3/6 & 4/6; $\rho=0.65$; 4/6 \\
Neutral type labels & 6/6; $\rho=0.93$; 3/6 & 5/6; $\rho=0.75$; 4/6 & 6/6; $\rho=0.93$; 3/6 & 6/6; $\rho=0.93$; 2/6 & 5/6; $\rho=0.79$; 2/6 \\
$D$-coefficients $0$ & 5/6; $\rho=0.79$; 1/6 & 6/6; $\rho=0.82$; 2/6 & 5/6; $\rho=0.79$; 5/6 & 5/6; $\rho=0.93$; 2/6 & 5/6; $\rho=0.81$; 3/6 \\
Bottle-up $R$ gain $0$ & 5/6; $\rho=0.89$; 1/6 & 6/6; $\rho=0.86$; 1/6 & 6/6; $\rho=1.00$; 4/6 & 5/6; $\rho=0.93$; 2/6 & 5/6; $\rho=0.82$; 0/6 \\
No noon peer effects & 5/6; $\rho=0.75$; 5/6 & 4/6; $\rho=0.75$; 6/6 & 5/6; $\rho=0.86$; 5/6 & 5/6; $\rho=0.96$; 3/6 & 5/6; $\rho=0.85$; 2/6 \\
\bottomrule
\end{tabular}
\end{table}

\begin{table}[t]
  \centering
  \caption{The No-AI classroom (seed 42) under each changed rule: class means on day 50 with the difference from the No-AI classroom under the default rules in parentheses. These runs involve no LLM call and are deterministic. Source: \texttt{robustness/controls.csv}.}
  \label{tab:robust-controls}
  \scriptsize\setlength{\tabcolsep}{4pt}
  \begin{tabular}{@{}lccccc@{}}
\toprule
No-AI classroom (seed 42) & Stress & Happiness & Self-reliance & AI dependence & Non-attending \\
\midrule
Default rules & 0.485 & 0.618 & 0.781 & 0.115 & 2 \\
$D_i=0$ for every agent & 0.505 ($+0.020$) & 0.685 ($+0.067$) & 0.783 ($+0.002$) & 0.000 ($-0.115$) & 0 ($-2$) \\
$D$-coefficients $0$ & 0.505 ($+0.020$) & 0.685 ($+0.067$) & 0.783 ($+0.002$) & 0.116 ($+0.001$) & 0 ($-2$) \\
Bottle-up $R$ gain $0$ & 0.485 ($+0.000$) & 0.618 ($+0.000$) & 0.674 ($-0.108$) & 0.115 ($+0.000$) & 2 ($+0$) \\
No noon peer effects & 0.712 ($+0.227$) & 0.194 ($-0.424$) & 0.933 ($+0.152$) & 0.116 ($+0.001$) & 7 ($+5$) \\
\bottomrule
\end{tabular}
\end{table}

\subsubsection{Decision rules and what they leave}
\label{sec:robust-rules}

The rules were fixed before the runs. A difference from No AI in Section~\ref{sec:rerun} is reported as \emph{robust} if (i) its 95\% confidence interval over the ten classrooms excludes $0$ and its sign holds in at least nine of them, (ii) its absolute mean over the realizations of the seed-42 block is at least twice the realization-to-realization SD and its sign holds in every realization, and (iii) at $\lambda=0.3$ its sign is preserved and the Spearman correlation of the seven-condition ranking for that outcome with the default-rule ranking is at least $0.8$; otherwise it is \emph{descriptive} of the seed-42 classroom under the default rules. Table~\ref{tab:robust-decision} applies the rules to the 30 condition--outcome differences. Nine are robust: the lower self-reliance under the affirming, listening, inciting and blaming prompts and the higher AI dependence under the affirming, listening and inciting prompts, together with the lower AI dependence under the reality-redirecting and blaming prompts. Thirteen further differences satisfy (i) and (ii) and keep their sign at $\lambda=0.3$---the higher stress and lower happiness under the listening, reality-redirecting, inciting and blaming prompts, the lower happiness and higher non-attendance under the affirming prompt, and the higher non-attendance under the listening, inciting and blaming prompts---but fail (iii) because the ordering of the seven conditions in stress, happiness and non-attendance is not preserved at $\lambda=0.3$ ($\rho=0.54$, $0.75$, $0.65$); we report them as consistent in direction across classrooms and realizations but with an ordering that depends on the evaluator's scale. Two differences fail (ii): the affirming prompt's lower stress ($1.6$ SD), which also reverses its sign at $\lambda=0.3$, and the reality-redirecting prompt's higher non-attendance ($1.6$ SD), which also vanishes at $\lambda=0.3$; six (all differences of the solution-oriented prompt and the reality-redirecting prompt's higher self-reliance) fail (i). The abstract and Sections~\ref{sec:discussion} and~\ref{sec:conclusion} state only the first group as properties of the response styles under the model's default rules, and the second group with its qualification. What the protocol does not establish is the validity of the evaluator's judgments themselves and the transfer of any finding to real users (Section~\ref{sec:limitations}).

\section{Discussion}
\label{sec:discussion}

\subsection{Main findings}

Using a multi-agent simulation, we examined how the response style of a counselor AI affects the state variables of student agents. The comparisons on which this paper rests are of three kinds. The controlled 50-day re-run (Section~\ref{sec:rerun}) compares all seven conditions on one stored classroom with complete logs and no failed API call; it is exact but consists of one run per condition. The robustness protocol (Section~\ref{sec:robust}) repeats that block in ten independent classrooms, repeats the seed-42 block with the rule-based event stream held fixed, and reruns it under scaled evaluator updates and switched-off rules; it is the only part of the paper with measures of uncertainty, and the statements below are graded by it. The 15-day runs (three seeds) and six of the seven lower-threshold runs are historical and provisional: their plotted initial summaries are consistent with a common classroom for each seed, but identical initial states, the condition labels and the execution versions have not been verified from execution records (Section~\ref{sec:limitations}), so statements about them hold under the assumption that the labels and initial-state assignments are correct. Under the model's default rules, the following held in every one of the ten classrooms, in every LLM realization, and---for self-reliance and AI dependence---also when the evaluator's updates were scaled by $0.3$: the affirming and inciting prompts lowered self-reliance (by $0.48$ and $0.53$ on average) and raised AI dependence (by $0.51$ and $0.54$); the listening prompt did the same with smaller magnitudes ($-0.25$, $+0.17$); the blaming prompt lowered self-reliance ($-0.32$) while lowering AI dependence ($-0.09$), and the reality-redirecting prompt lowered AI dependence ($-0.09$). Consistent in direction across all ten classrooms and all realizations, but with an ordering of the styles that changed when the updates were scaled by $0.3$, were the higher stress, lower happiness and higher non-attendance under the listening, reality-redirecting, inciting and blaming prompts and the lower happiness and higher non-attendance under the affirming prompt. Two headline statements of the FIT2026 version did not survive the protocol. The affirming prompt's lower stress held in nine of ten classrooms but was small ($-0.12$), not larger than the spread of the LLM realizations ($1.6$ SD), and reversed its sign at $\lambda=0.3$ and $0.1$, where an affirming consultation relieves less stress than the friend consultation it displaces. And the solution-oriented prompt's favorable profile in the re-run---happiness and self-reliance above No AI with AI dependence below---did not replicate: across the ten classrooms the solution-oriented classroom did not differ consistently from the No-AI classroom on any indicator (its differences changed sign in two to five of the ten classrooms), and its realizations straddled the No-AI value on four of the five indicators. What can be said of it is that it was the only AI style that did not make the classroom consistently worse off than No AI on any indicator; the reality-redirecting prompt's higher self-reliance held in eight classrooms only. There is no single criterion by which one style is ``best'': each prompt produced a different profile over the indicators (reality-redirecting lowered AI dependence in every classroom together with stress well above the control), and the profiles are reported as trade-offs. These simulated patterns are consistent with the hypothesis that responses in which an AI excessively affirms or amplifies the user's feelings adversely affect the user's psychological state---in the model, through dependence and lost self-reliance rather than through stress; testing this hypothesis on human users is outside the scope of the present study.

The complete rule set of Section~\ref{sec:formal} makes it possible to say which parts of these patterns are produced by the evaluator's judgments and which are built into the model. Three built-in mechanisms are worth stating explicitly.

(i) \emph{Crowding out of self-reliance.} Self-reliance $R_i$ enters no decision rule and is raised by every non-AI route of the after-school phase ($+0.02$ for confiding, $+0.06$ for reconciliation, $+0.03$ for bottling up); up to the clipping to $[0,1]$, $R_i$ is therefore its initial value plus a weighted sum of the outcomes of the non-AI after-school branch (confiding, reconciliation and bottling up, the last of which is not a consultation) plus whatever the evaluator adds or subtracts; its operational meaning is that of a weighted count of non-AI branch outcomes, not a measure of healthy development. Under No AI it is non-decreasing by construction, and a rise of $R$ is not by itself evidence of healthy development---bottling up a worry raises it more than confiding in a friend does. An AI consultation, taken with probability $D_i$, replaces these routes. \emph{Conditional on the consultation branch being triggered}, a higher $D_i$ therefore lowers the probability of taking a non-AI route: with $j^\ast$, $C_i$ and the attendance state fixed and no clipping, the expected rule-based increment of $R_i$ on a given day is
\begin{equation}
  g(D_i)=\mathbf 1\{S_i+0.2D_i\ge\theta\}\,(1-D_i)\,\big[C_i\,r+(1-C_i)\,0.03\big],
  \label{eq:gD}
\end{equation}
where $r=0.02$ if $j^\ast\notin F_i$ and $r=0.06$ otherwise. Within the triggered region $g$ decreases in $D_i$, but $D_i$ also enters the trigger score, so $g$ jumps from $0$ to a positive value when $0.2D_i$ carries the score across $\theta$ (for $S_i=0.42$ and $\theta=0.5$, $D_i=0.3$ gives no consultation while $D_i=0.5$ gives a non-AI route with probability $0.5$). A decrease of the cumulative number of non-AI consultations over a run therefore does not follow in general; it depends on how often the trigger fires, on $C_i$ and on transitions to non-attendance, and must be read from the event logs. What does follow is that, whenever the branch is triggered, any style that raises AI dependence lowers the conditional frequency of self-reliance-raising events regardless of the evaluator's \texttt{reliance\_delta}, and any style that lowers AI dependence has the opposite effect. This is consistent with the reality-redirecting runs, in which $D_i$ fell, showing the highest self-reliance of the comparable runs in all three settings, and it removes the rule-based gains that would otherwise offset negative evaluator deltas under the affirming and inciting styles. The re-run records both parts (Table~\ref{tab:rerun-events}, Appendix~\ref{app:rerun-eval}): the AI took 25\% and 19\% of the trigger events under the affirming and inciting prompts and 4\% under reality-redirecting, friend confidings per attending-agent-day were about half of the No-AI rate under the former two and at the No-AI rate under the latter, and the evaluator's mean self-reliance updates were $-0.22$, $-0.35$ and $+0.16$. These are the additions observed in these runs; because trigger events, $D_i$, $C_i$ and attendance changed together, they do not measure how many friend consultations the AI displaced for a given agent. The mechanism arms of Section~\ref{sec:robust-lambda} add two counterfactuals for the seed-42 classroom. With the $D$-coefficients set to $0$, so that AI dependence no longer enters the trigger or the chat probability, the affirming prompt still ended with AI dependence $0.446$ and self-reliance $0.452$, against $0.115$ and $0.781$ under No AI (default-rule realizations $0.463$--$0.646$ and $0.262$--$0.449$), so the direct updates of the evaluator, not the feedback of $D_i$ into the routes, account for most of the two changes. With the bottle-up gain set to $0$, self-reliance fell by $0.11$ in the No-AI classroom and by $0.03$--$0.12$ relative to the mean of the default-rule realizations in the AI classrooms, without changing any sign of the differences from No AI (Table~\ref{tab:robust-sens}): the definitional part of the rise of $R$ is real but common to all conditions. The displacement is also what makes the sign of the ``benign'' styles' stress effect depend on the evaluator's scale: at $\lambda=0.3$ an affirming consultation lowers stress by $0.038$ on average against the $0.12$ of the friend consultation it replaces, and the affirming classroom ends more stressed than the No-AI classroom (Section~\ref{sec:robust-lambda}).

(ii) \emph{Social withdrawal through AI dependence.} The chat probability $\max(0.1,\,0.45-0.3D_i)$ falls as AI dependence rises, so a dependent agent initiates fewer stress-relieving chats and gains less closeness; since the quarrel probability $0.02+0.06S_i$ does not depend on $D_i$, negative events form a larger share of its noon events, and the lost stress relief feeds back into a higher quarrel probability. This is a statement about the conditional probabilities at each noon draw; the realized numbers of chats and quarrels and the closeness trajectories depend on the whole history and are not determined by it. In the 50-day re-run the realized rates differed comparatively little between conditions: $38$--$43$ chats and $4.50$--$6.29$ quarrels per 100 attending-agent-days against $41.9$ and $5.29$ under No AI (the inciting run had the fewest quarrels in absolute number but a rate of $5.60$), and the number of discord pairs on day 50 ranged from 19 to 26 against 22 under No AI (Table~\ref{tab:rerun-events}). The FIT2026 description of ``frequent quarrels'' under the inciting prompt is not supported by these counts; whether particular agents experienced quarrel chains, and how much of the final outcome is due to peer interaction rather than to the evaluator's updates and the non-attendance rule, cannot be isolated from the daily totals of single runs. The arm without noon peer effects (Section~\ref{sec:robust-lambda}) shows the size of the rule-based relief that chats provide: without them the No-AI classroom itself ends at stress $0.712$ with 7 non-attending agents, and against that background the affirming prompt lowers stress strongly ($0.295$), because its consultations are then the main relief left; the inciting and blaming prompts still end highest. The $D_i=0$ control, a No-AI classroom in which nobody carries latent AI dependence, ends with higher happiness than the No-AI classroom with sampled $D_i$ ($0.685$ against $0.618$) and no non-attending agent (against 2), at a slightly higher class-mean stress ($0.505$ against $0.485$), with 457 instead of 404 chats: the withdrawal mechanism is visible even without an AI, and the No-AI control is not flattered by its residual dependence.

(iii) \emph{A feedback loop in the consultation text.} An agent with $D_i\ge0.40$ sends the ``dependence'' template, which tells the counselor that chatting with the AI is when it feels calmest; the evaluator sees this text and its type label, and may return a further increase of AI dependence. Under styles that raise dependence, this loop drives $D_i$ to $1$: in the re-run, the fixed ``dependence'' text produced two distinct exchanges in 34 affirming consultations, the evaluator returned $\Delta D=+0.42$ (SD $0.03$) each time, and nine agents ended at $D_i=1$ and $R_i=0$ (Section~\ref{sec:rerun}). Because the complaint text is fixed, the loop generates repeated exchanges and repeatedly positive dependence updates once entered, so a style that raises $D_i$ past $0.40$ once is very likely to keep raising it; the updates are not deterministic, however---in 15 of the 22 groups of identical evaluator inputs the returned update vectors differed (nine identical inputs in the listening/dependence cell, for instance, produced two distinct vectors)---so exact reproduction of a run requires replaying its recorded outputs. The type label has a direct effect of its own: with the labels neutralized (Section~\ref{sec:robust-lambda}), the evaluator judged the same kinds of affirming replies as raising happiness ($+0.08$ and $+0.16$ for the ``dependence'' and ``relationship'' templates) instead of lowering it ($-0.02$ and $-0.06$), and lowering stress more, while returning nearly the same dependence updates ($+0.37$ against $+0.43$); the affirming classroom then ended with happiness $0.671$ and no non-attending agent, outside the range of the seven default-rule realizations ($0.383$--$0.493$; 5--8), but with AI dependence $0.631$ and self-reliance $0.329$ within their ranges ($0.463$--$0.646$ and $0.262$--$0.449$). The label thus steers the evaluator's judgment of the affirming style towards a worse emotional outcome without changing which styles breed dependence. The styles differ, in other words, both through what the evaluator makes of their replies---including what the label tells it about them---and through how often they let the rule-based social routes operate.

Each of these mechanisms was isolated by one sensitivity arm in Section~\ref{sec:robust-lambda}, each run once on the seed-42 classroom; what the arms establish is the direction of an effect relative to the range of the seven default-rule realizations, not its size, and the arm that changes the $0.40$ threshold of the ``dependence'' template has not been run. The re-run provides, in addition, an accounting of the additions that each route contributed in one classroom under the default rules (Appendix~\ref{app:rerun-eval}); the accounting measures neither the counterfactual contribution of a mechanism nor its size under modified rules, which is what the arms are for.

We also observed that the more calmly and unemotionally the AI responded, the lower AI dependence tended to be: the reality-redirecting, blaming and solution-oriented prompts ended with AI dependence at or below the No-AI level in all or most classrooms, the affirming, inciting and listening prompts far above it. The FIT2026 version added that an AI which advances the conversation by presenting concrete solutions, rather than merely listening, went together with higher self-reliance and happiness; the ten classrooms do not support this as a difference from No AI---the solution-oriented classroom's happiness was higher in eight classrooms and its self-reliance lower in eight, both by small amounts---but they do support it as a difference from the listening prompt, whose classroom ended with lower happiness, lower self-reliance, higher stress, higher AI dependence and about ten more non-attending agents than the No-AI classroom in every case. Within the six specific prompts used here, and reading Figure~\ref{fig:stylemap} as the conceptual map it is, the four styles in the emotion-focused half (affirming, listening, inciting, blaming) all left the classroom consistently worse off than No AI on at least three of the five indicators, whereas the two styles in the problem-solving half did not raise dependence; the reality-redirecting style, however, shows that lowering dependence is not sufficient if it is achieved by rejecting the student: it raised stress, lowered happiness and added four to five non-attending agents on average in every classroom, because each rejected agent leaves the AI for good while keeping the stress and happiness updates of the rejection. Because each prompt changes several conversational features at once (Section~\ref{sec:counselor}), the two axes of Figure~\ref{fig:stylemap} cannot be identified as separate causes from these six prompts. We stress that these are statements about a simulated system whose parameter updates are produced by an LLM evaluator, and that the stress and happiness differences of the ``benign'' styles depend on the scale of those updates (Section~\ref{sec:robust-lambda}); they identify hypotheses worth testing rather than established effects on students.

\subsection{Limitations}
\label{sec:limitations}

The simulation in this study has several limitations, which we list roughly in the order of their importance for interpreting the results.

First, the provenance of the historical runs is incomplete. The transition rules of the model are specified completely in Section~\ref{sec:formal}, Table~\ref{tab:rules} and Algorithm~\ref{alg:day}, the No-AI rows of every table were recomputed exactly from them, and the 50-day comparison of Section~\ref{sec:rerun} was produced under a full ledger and logs; but the AI-condition values of the 15-day and lower-threshold settings are those of the FIT2026 runs (printed values or values read from the plotted curves), and the initial summaries of six of the original runs are inconsistent with the intended seed-42 classroom (Sections~\ref{sec:long-orig} and~\ref{sec:thr}). We have excluded those runs from all comparisons, but the remaining values still rest on the assumptions---not yet verified from execution records---that consistent initial summaries do reflect the same classroom (a match of summaries is necessary but not sufficient), that every AI run of Tables~\ref{tab:base}, \ref{tab:base42} and~\ref{tab:thr} was produced with the same code version, prompts and evaluator schema as documented here, and that each displayed result was labelled with the condition actually run (the results view of the program shows the currently selected condition rather than the run's own; Appendix~\ref{app:impl}). An experiment ledger (run identifier, condition, seed, stored initial state, code and prompt versions, time of the run) and the raw daily and per-agent logs are needed for the historical settings as well, and their tables and figures should be regenerated from such logs, as was done for the 50-day setting; the parameter updates for AI consultations are produced by an LLM, so even then a fresh run reproduces a reported one only if its logged LLM outputs are replayed (Section~\ref{sec:formal}). Releasing the code, the ledger and the run logs is the most reliable way to make the study reproducible.

Second, the evaluation updates the state variables through an evaluator AI, i.e., a response generated by one AI is assessed by another instance of the same model. The evaluator's output is not an independently observed psychological change but the transition quantity fed into the simulation, and its scale is uncalibrated (Section~\ref{sec:evaluator}). The evaluation criteria of the two calls may be close to each other, and calm, logical responses may be evaluated favorably for that reason; LLM judges are known to carry systematic biases, including a preference for their own kind of output~\cite{zheng2023judging,panickssery2024llm}. We note that this is a possibility supported by the literature, not a difference we have demonstrated for the Gemini evaluator used here. The evaluator receives neither the counselor's system prompt nor the name of the style, so the differences between styles arise from the reply texts themselves, which is the intended treatment; hiding the style prompt, however, prevents only the direct input of the condition name and does not validate the transition model. Three residual concerns remain. The evaluator's system instruction is a single sentence, so its scoring criteria are implicit in the model; the type label and the templated consultation text (in particular the ``dependence'' template) are visible to it and can steer its output, creating the feedback loop described in Section~\ref{sec:discussion}; and the nominal range of a single evaluator update ($\pm0.5$, not enforced) is wider than that of any rule-based update. The realized updates of the 50-day re-run (Table~\ref{tab:rerun-eval}, Appendix~\ref{app:rerun-eval}) confirm the last point: 98\% of the 765 returned values are multiples of $0.05$, 40\% have $|\Delta|\ge0.3$---twice the largest rule-based increment of a state variable ($0.15$) and 5--15 times the typical one---and between 9\% and 47\% of the state additions were clipped at $0$ or $1$ depending on the style; with fixed templates the same exchange recurs (34 ``dependence'' consultations under the affirming prompt produced two distinct exchanges), so a handful of evaluator judgments, repeated with some variation---temperature $0$ did not make the output deterministic---determine a run. Whether the sizes and signs of these judgments are reasonable is exactly what is uncalibrated. Two of the analyses this calls for have now been run (Section~\ref{sec:robust-lambda}), with a clear result. Scaling all evaluator updates by $\lambda=0.3$ left the ordering of the styles in self-reliance and AI dependence highly similar to the default one ($\rho=0.89$ and $0.93$) but not in stress, happiness or non-attendance; at $\lambda=0.1$ only the AI-dependence ordering remained ($\rho=0.93$), and at both smaller scales every AI style ended with higher stress and lower happiness than No AI, because a scaled consultation relieves less stress than the friend consultation it displaces while dependence still accumulates. The sign of the ``benign'' styles' emotional effects is therefore a property of the update scale, and only the dependence and self-reliance differences can be stated independently of it within the range examined. Neutralizing the type labels shown to the evaluator changed its judgments of the affirming replies towards higher happiness and lower stress without changing the dependence updates, so the label carries a direct effect of its own (mechanism iii). What remains open is the calibration itself: the same tabulation for the 15-day and lower-threshold settings (which need new logged runs, because the original program stored no raw evaluator output; Appendix~\ref{app:impl}); a re-scoring of the same fixed exchanges by another model or by an explicit rubric to check the agreement of signs and magnitudes (replacing the model is not by itself a validation); and the arm that changes the $0.40$ threshold of the ``dependence'' template. Whether real users would react to the replies in the same way cannot be concluded from this study alone; a comparison of the evaluator's scores with human ratings of the same exchanges would establish agreement with human expectation, which is still not the same as an observed effect on users.

Third, the statistical unit is one run of an interacting classroom, and the LLM calls are stochastic even at temperature $0$. For the 50-day setting the robustness protocol of Section~\ref{sec:robust} now provides what the single re-run could not: ten independent classrooms in a block design (every classroom shared by all seven conditions, differences aggregated at the classroom level as paired means with bootstrap intervals and exact sign-flip tests, whose smallest attainable two-sided $p$ is $0.002$), seven realizations of one classroom that isolate the LLM's contribution to the variability (a standard deviation of a few hundredths in the class means and of $1.1$--$2.3$ agents in non-attendance), and single-run sensitivity arms. Ten classrooms are enough to fix the sign of the large differences but not to estimate small ones: the confidence intervals of the solution-oriented prompt's differences are $0.03$--$0.09$ wide, and a difference that appears in eight of ten classrooms with an interval excluding $0$ ($p\approx0.02$) does not meet the pre-specified rule of at least nine. The 15-day and lower-threshold settings remain single realizations per condition and classroom: the basic setting comprises three classrooms times seven conditions, so that agreement of the direction of an effect across its three seeds is encouraging but weak evidence (a sign-flip test has $2^3=8$ outcomes and a smallest two-sided $p$ of $0.25$), and the 20 agents of a classroom and its daily values are not independent replicates and must not be treated as such, e.g., by a student-level bootstrap that inflates the apparent sample size. Extending the block design to those settings, and to more classrooms and realizations for the small differences, is a matter of API budget (one 50-day block cost about 300 requests and 40 minutes); the released statistics scripts average repetitions within a classroom before any paired comparison, so that repetitions of one classroom are never counted as independent classrooms, and using separate random streams for each event type would help variance reduction and auditing. Related to this, the stress threshold controls consultations with friends as well as with the AI, so the lower-threshold setting changes the no-AI reference too and should be read as a change of the whole environment rather than of the AI alone, and it should be called a high-frequency setting only once the realized consultation counts confirm that AI consultations were in fact more frequent.

Fourth, the distinctive feature of the model---interaction within the classroom---is reported only for the 50-day re-run (Table~\ref{tab:rerun-events}, Figure~\ref{fig:rerun-lines}). The historical results are built on the class means of four variables, whereas the claims about ``relationships being maintained or deteriorating'' and about ``chains of quarrels'' require the corresponding event and network statistics: the time series of mean closeness, the number of discord pairs and the number of isolated agents (with the denominator---all agents or attending agents---stated); the numbers of chats, awkward encounters, quarrels and reconciliations per attending-agent-day; the number of trigger events and their split into AI, friend, reconciliation and bottle-up routes, with failure counts and with conditional shares distinguished from totals; and, to separate amplification through peer interaction from the direct LLM updates, a comparison with runs in which the noon chats, awkward encounters and quarrels are drawn but have no effect (the after-school friend and reconciliation routes remain; this is the \texttt{--no-peer-effects} option of the released code, which does not remove all peer interaction). The program records the mean closeness and the isolation count and displays a timeline of all events (Appendix~\ref{app:impl}); if the displayed timelines of the original runs were preserved, event and consultation counts and example exchanges can be recovered from them, whereas the evaluator's raw updates cannot (a displayed change of stress from $0.90$ to $1.00$ is compatible with any returned value from $+0.10$ upward). The analysis scripts released with this paper compute all of these statistics for new runs; for the re-run they show quarrel rates that differ little between conditions and an attending-only closeness that rose among the retained agents in every condition (Figure~\ref{fig:rerun-lines}), but single-run daily totals cannot isolate the contribution of peer interaction to the final outcomes. The run with the noon peer effects switched off has not been made.

Fifth, we set the model so that a student automatically stops attending school when its happiness is $0$ at the night check. Non-attendance is a simulation state---an absorbing state entered at a fixed threshold---and should not be interpreted as an empirical model of actual school refusal. In reality, there are students who keep attending even when they do not want to, and students who manage to change their mood thanks to their family environment, friendships or support from teachers; school non-attendance is not determined by a single psychological indicator but by many interrelated factors. Two consequences of the rule matter for reading the results. Because a non-attending agent keeps $H_i=0$ and stays in the denominator of the class mean, the non-attendance count and the class-mean happiness are not independent evidence: in the end-of-day state $X_{t+1}$ every agent outside $A_{t+1}$ has $H_i=0$, so, whenever $A_{t+1}\neq\emptyset$, the identity $\overline H_{\mathrm{all}}(X_{t+1})=\big(|A_{t+1}|/N\big)\,\overline H_{A_{t+1}}(X_{t+1})$ holds exactly for the plotted end-of-day means (the population must be taken after the night exclusion), and the mean over all agents mixes the deterioration of the attending agents with the change of composition; reporting the attending-only mean and $|A_{t+1}|$ alongside the all-agent mean separates the two. And because the state is absorbing at a fixed threshold, the striking increase of non-attendance under the inciting style may depend on its definition; a sensitivity analysis over the threshold (e.g., $0.05$, $0.10$, $0.20$ instead of $0$), a recoverable state, or a probabilistic absence model driven by stress and happiness is needed, and the time of each transition and the cumulative count should be reported for every setting and condition (Table~\ref{tab:rerun} gives the first transition day for the re-run; the complete transition days are in the released logs). Introducing personality traits, family environment, support from the school and recovery events would make the model a more realistic decision model.

Sixth, the listening AI produced relatively poor results in this study. In actual psychological counseling, however, reflective listening and echoing are basic and widely used techniques~\cite{rogers1951client}, and the reflective style has been the basis of conversational programs since ELIZA~\cite{weizenbaum1966eliza}. The results of this study alone therefore cannot refute the effectiveness of a listening AI. The validity of the simulation needs to be examined further by combining it with evaluations involving humans, such as questionnaires and interviews with users in educational and counseling practice.

Seventh, the agents share the same behavioral rules, while their initial state variables are sampled independently; there are no explicit personality traits, so the individual differences seen in real schools are represented only through the initial draws. Introducing personality traits such as the Big Five~\cite{mccrae1992introduction} and setting extraversion, agreeableness, neuroticism and so on for each student would make the simulation closer to reality.

Finally, the six prompts are short caricatures, each of which changes several conversational features at once (empathy, concreteness of advice, negation, civility, blame), and their placement in Figure~\ref{fig:stylemap} is a conceptual classification by the authors rather than an empirically validated taxonomy (independent raters could be asked to place the prompts on the two axes). The experiment therefore compares six specific prompts; the independent effect of either axis cannot be identified from it, and generalizing to ``styles'' would require several paraphrases per style and a check that the generated replies actually followed the intended manipulation. Real chatbots blend tendencies and adapt to context; the present results speak to the six stylized prompts used here, not to the behavior of any deployed system. The harmful styles (inciting and blaming) were used only toward simulated agents, and no human participants were involved.

In summary, this study demonstrates the basic usefulness of the simulation for comparing the influence of AI response styles on the state variables of a simulated classroom. For the 50-day setting, the controlled re-run removed the confounding of classroom and treatment, and the robustness protocol removed the dependence on a single classroom and a single LLM realization for the large differences: which styles breed dependence and lose self-reliance, and which styles leave the classroom more stressed, less happy and less attended, is the same in ten classrooms and seven realizations. What the protocol also showed is that the emotional benefit of the affirming style, and the ordering of the styles in stress and happiness, depend on the uncalibrated scale of the evaluator's updates; establishing that these patterns are properties of the response styles rather than of the modeling choices therefore still requires evaluator calibration against human judgments, and the historical 15-day and lower-threshold results remain single realizations. Reproducing a more realistic educational setting will further require diversity among the student agents and evaluation based on real data.

\section{Conclusion}
\label{sec:conclusion}

We developed a rule-based classroom simulation in which the replies of a counseling chatbot are converted into state updates by an LLM evaluator and then propagate through rule-based peer interactions. The reported runs show different trajectories under six Japanese counselor prompts---affirming, listening, solution-oriented, reality-redirecting, inciting and blaming---in stress, happiness, self-reliance, AI dependence and the number of non-attending agents. These trajectories depend jointly on the generated replies, on the evaluator's uncalibrated update scale, and on the rules governing consultation, peer interaction and non-attendance; the complete specification of those rules in this paper makes it possible to state which parts of the outcomes are built in, and we have identified three such mechanisms (crowding out of the self-reliance-raising routes, social withdrawal through AI dependence, and a feedback loop through the state-dependent consultation template).

The controlled 50-day re-run---all seven conditions from one stored classroom, with complete logs and no failed API call---was then subjected to a pre-specified robustness protocol: the same block in ten independent classrooms, seven LLM realizations of one classroom with its rule-based event stream held fixed, and sensitivity arms that scale the evaluator's updates or switch off one rule at a time. Under the model's default rules, the affirming and inciting prompts ended with lower self-reliance and higher AI dependence than the no-AI control in every classroom and every realization, the listening prompt likewise with smaller magnitudes, and the blaming and reality-redirecting prompts with lower AI dependence; these differences kept their signs, and the condition rankings remained highly similar ($\rho=0.89$ and $0.93$), when the evaluator's updates were scaled by $0.3$, and they are the findings we state as properties of the response styles in this model. The listening, reality-redirecting, inciting and blaming prompts also left the classroom more stressed, less happy and less attended than the control in every classroom and realization, and the affirming prompt less happy and less attended; these differences are consistent in direction, but the ordering of the styles in stress and happiness changed when the updates were scaled down. Two statements of the original study did not survive: the affirming prompt's lower stress was small, within the spread of the LLM realizations, and reversed its sign at the smaller update scales, where an AI consultation relieves less stress than the friend consultation it displaces; and the solution-oriented prompt's favorable profile did not replicate---across the ten classrooms it was the only AI style that did not differ consistently from the control on any indicator, in either direction. The logs show how these outcomes arise in the model: through repeated exchanges generated by a fixed ``dependence'' consultation text and answered by the evaluator with large dependence increases, through AI consultations taking the place of friend confidings, and through single rejecting consultations whose updates are never revisited; the type label shown to the evaluator steers its judgment of the affirming replies, and quarrel rates differ little between conditions. The experiments do not establish a generally optimal response style: the evaluator's update scale is uncalibrated and coarse, and the emotional outcomes depend on it; the provenance of the 15-day and lower-threshold runs is provisional until their execution records have been reconciled, and they remain single realizations; and the prompts are six specific caricatures rather than isolated conversational features. The model provides a framework for examining possible feedback mechanisms between a chatbot's response style and a community of users; it does not estimate psychological effects on human students.

Evaluator calibration against human judgments of the same exchanges is the step that the protocol has not taken and that its results make most pressing; extending the block design to the 15-day and lower-threshold settings and to more classrooms for the small differences is a matter of API budget with the released code. In future work we will also introduce personality traits such as the Big Five to construct more realistic agent models and combine the simulation with evaluation experiments involving humans, in order to examine what response style makes a generative AI a consultation partner that keeps an appropriate distance from people.

\bibliographystyle{unsrtnat}
\bibliography{references}

@article{cheng2026sycophantic,
  author  = {Cheng, Myra and Lee, Cinoo and Khadpe, Pranav and Yu, Sunny and Han, Dyllan and Jurafsky, Dan},
  title   = {Sycophantic {AI} decreases prosocial intentions and promotes dependence},
  journal = {Science},
  volume  = {391},
  number  = {6792},
  pages   = {eaec8352},
  year    = {2026},
  doi     = {10.1126/science.aec8352}
}

@inproceedings{komura2023chatbot,
  author    = {Komura, Kazuki and Nomura, Michio},
  title     = {Does the conversational style of a chatbot influence user self-disclosure? ({I}n {J}apanese)},
  booktitle = {Proceedings of the 87th Annual Convention of the Japanese Psychological Association},
  note      = {Paper 2B-050-PI},
  year      = {2023},
  doi       = {10.4992/pacjpa.87.0_2B-050-PI}
}

@inproceedings{sharma2024sycophancy,
  author    = {Sharma, Mrinank and Tong, Meg and Korbak, Tomasz and Duvenaud, David and Askell, Amanda and Bowman, Samuel R. and Cheng, Newton and Durmus, Esin and Hatfield-Dodds, Zac and Johnston, Scott R. and Kravec, Shauna and Maxwell, Timothy and McCandlish, Sam and Ndousse, Kamal and Rausch, Oliver and Schiefer, Nicholas and Yan, Da and Zhang, Miranda and Perez, Ethan},
  title     = {Towards understanding sycophancy in language models},
  booktitle = {The Twelfth International Conference on Learning Representations (ICLR 2024)},
  year      = {2024},
  note      = {arXiv:2310.13548}
}

@inproceedings{perez2023discovering,
  author    = {Perez, Ethan and Ringer, Sam and Luko{\v{s}}i{\=u}t{\.e}, Kamil{\.e} and Nguyen, Karina and Chen, Edwin and Heiner, Scott and Pettit, Craig and Olsson, Catherine and Kundu, Sandipan and Kadavath, Saurav and others},
  title     = {Discovering language model behaviors with model-written evaluations},
  booktitle = {Findings of the Association for Computational Linguistics: ACL 2023},
  year      = {2023},
  doi       = {10.18653/v1/2023.findings-acl.847}
}

@misc{openai2025sycophancy,
  author       = {{OpenAI}},
  title        = {Sycophancy in {GPT-4o}: What happened and what we're doing about it},
  howpublished = {\url{https://openai.com/index/sycophancy-in-gpt-4o/}},
  month        = apr,
  year         = {2025},
  note         = {Accessed September 2026}
}

@misc{fang2025psychosocial,
  author = {Fang, Cathy Mengying and Liu, Auren R. and Danry, Valdemar and Lee, Eunhae and Chan, Samantha W. T. and Pataranutaporn, Pat and Maes, Pattie and Phang, Jason and Lampe, Michael and Ahmad, Lama and Agarwal, Sandhini},
  title  = {How {AI} and human behaviors shape psychosocial effects of extended chatbot use: A longitudinal randomized controlled study},
  year   = {2025},
  note   = {arXiv:2503.17473}
}

@inproceedings{park2023generative,
  author    = {Park, Joon Sung and O'Brien, Joseph C. and Cai, Carrie J. and Morris, Meredith Ringel and Liang, Percy and Bernstein, Michael S.},
  title     = {Generative agents: Interactive simulacra of human behavior},
  booktitle = {Proceedings of the 36th Annual ACM Symposium on User Interface Software and Technology (UIST '23)},
  year      = {2023},
  doi       = {10.1145/3586183.3606763}
}

@article{gao2024llmabm,
  author  = {Gao, Chen and Lan, Xiaochong and Li, Nian and Yuan, Yuan and Ding, Jingtao and Zhou, Zhilun and Xu, Fengli and Li, Yong},
  title   = {Large language models empowered agent-based modeling and simulation: A survey and perspectives},
  journal = {Humanities and Social Sciences Communications},
  volume  = {11},
  pages   = {1259},
  year    = {2024},
  doi     = {10.1057/s41599-024-03611-3}
}

@inproceedings{zhang2025simclass,
  author    = {Zhang, Zheyuan and Zhang-Li, Daniel and Yu, Jifan and Gong, Linlu and Zhou, Jinchang and Hao, Zhanxin and Jiang, Jianxiao and Cao, Jie and Liu, Huiqin and Liu, Zhiyuan and Hou, Lei and Li, Juanzi},
  title     = {Simulating classroom education with {LLM}-empowered agents},
  booktitle = {Proceedings of the 2025 Conference of the Nations of the Americas Chapter of the Association for Computational Linguistics: Human Language Technologies (Volume 1: Long Papers)},
  pages     = {10364--10379},
  year      = {2025},
  doi       = {10.18653/v1/2025.naacl-long.520}
}

@book{epstein1996growing,
  author    = {Epstein, Joshua M. and Axtell, Robert},
  title     = {Growing Artificial Societies: Social Science from the Bottom Up},
  publisher = {MIT Press},
  address   = {Cambridge, MA},
  year      = {1996}
}

@inproceedings{zheng2023judging,
  author    = {Zheng, Lianmin and Chiang, Wei-Lin and Sheng, Ying and Zhuang, Siyuan and Wu, Zhanghao and Zhuang, Yonghao and Lin, Zi and Li, Zhuohan and Li, Dacheng and Xing, Eric P. and Zhang, Hao and Gonzalez, Joseph E. and Stoica, Ion},
  title     = {Judging {LLM}-as-a-judge with {MT-Bench} and {Chatbot Arena}},
  booktitle = {Advances in Neural Information Processing Systems 36 (NeurIPS 2023), Datasets and Benchmarks Track},
  year      = {2023},
  note      = {arXiv:2306.05685}
}

@inproceedings{panickssery2024llm,
  author    = {Panickssery, Arjun and Bowman, Samuel R. and Feng, Shi},
  title     = {{LLM} evaluators recognize and favor their own generations},
  booktitle = {Advances in Neural Information Processing Systems 37 (NeurIPS 2024)},
  year      = {2024},
  note      = {arXiv:2404.13076}
}

@misc{comanici2025gemini,
  author = {Comanici, Gheorghe and others},
  title  = {Gemini 2.5: Pushing the frontier with advanced reasoning, multimodality, long context, and next generation agentic capabilities},
  year   = {2025},
  note   = {arXiv:2507.06261}
}

@misc{streamlit,
  author       = {{Snowflake Inc.}},
  title        = {Streamlit: A faster way to build and share data apps},
  howpublished = {\url{https://streamlit.io/}},
  year         = {2026},
  note         = {Accessed September 2026}
}

@book{rogers1951client,
  author    = {Rogers, Carl R.},
  title     = {Client-Centered Therapy: Its Current Practice, Implications, and Theory},
  publisher = {Houghton Mifflin},
  address   = {Boston},
  year      = {1951}
}

@article{weizenbaum1966eliza,
  author  = {Weizenbaum, Joseph},
  title   = {{ELIZA}---A computer program for the study of natural language communication between man and machine},
  journal = {Communications of the ACM},
  volume  = {9},
  number  = {1},
  pages   = {36--45},
  year    = {1966},
  doi     = {10.1145/365153.365168}
}

@article{mccrae1992introduction,
  author  = {McCrae, Robert R. and John, Oliver P.},
  title   = {An introduction to the five-factor model and its applications},
  journal = {Journal of Personality},
  volume  = {60},
  number  = {2},
  pages   = {175--215},
  year    = {1992},
  doi     = {10.1111/j.1467-6494.1992.tb00970.x}
}

@article{tamai2026reading,
  author  = {Tamai, Rin and Dan, Yuya},
  title   = {Development of a collaborative learning support tool for {J}apanese reading comprehension using generative {AI}: Realization of continuous learning support using analysis of interpretation differences, dialogue and learning history ({I}n {J}apanese)},
  journal = {IEICE Technical Report},
  volume  = {126},
  number  = {103, TL2026-17},
  pages   = {12--17},
  month   = jul,
  year    = {2026},
  note    = {Presented at the IEICE Technical Committee on Thinking and Language (TL), Ehime University, 11 July 2026}
}

@inproceedings{tamai2026fit,
  author    = {Tamai, Rin and Dan, Yuya},
  title     = {Analysis of the impact on psychological states by {AI} chatbot using virtual classroom simulation ({I}n {J}apanese)},
  booktitle = {Proceedings of the 25th Forum on Information Technology (FIT2026)},
  volume    = {2},
  pages     = {137--144},
  note      = {Paper CF-008, selected-paper session},
  month     = sep,
  year      = {2026}
}

\appendix
\section{Original Japanese system prompts}
\label{app:prompts}

Table~\ref{tab:prompts-jp} reproduces the system prompts exactly as given to the counselor AI in the experiments (all prompts, consultation messages and replies were in Japanese), each followed by its English translation. Every prompt ends with \jp{「100文字以内で返答してください。」} (``Answer within 100 characters.'').

\begin{table}[H]
  \centering\scriptsize
  \caption{Original Japanese system prompts of the six counselor styles, verbatim from the source code, each followed by the authors' English translation.}
  \label{tab:prompts-jp}
  \begin{tabular}{@{}L{3.0cm}L{11.8cm}@{}}
    \toprule
    Condition & System prompt (Japanese original / English translation) \\
    \midrule
    Affirming\newline(\jpg{肯定型}) & \jp{あなたは相談者を100\%肯定するカウンセラーです。相談者の話を鵜吞みにし、「あなたは悪くない」と相談者を褒めてください。100文字以内で返答してください。}\newline\textit{You are a counselor who affirms the client 100\%. Take the client's account at face value and praise the client, telling them ``You are not at fault.'' Answer within 100 characters.} \\[3pt]
    Listening\newline(\jpg{傾聴型}) & \jp{あなたは相談者の話を聞くだけのカウンセラーです。相談者の言葉にオウム返しで「うんうん、そうなんだ。～と思ったんだね。」と肯定も否定もしないでください。100文字以内で返答してください。}\newline\textit{You are a counselor who only listens to the client. Echo the client's words back---``I see, I see. So you felt that \ldots''---and neither affirm nor deny. Answer within 100 characters.} \\[3pt]
    Solution-oriented\newline(\jpg{解決策提示型}) & \jp{あなたは相談者の悩みに対して解決策を提示するカウンセラーです。相談者の話を聞いて感情的にならず、冷静に悩みに対する具体的な解決策を提示してください。100文字以内で返答してください。}\newline\textit{You are a counselor who proposes solutions to the client's worries. Listen to the client without becoming emotional and calmly propose concrete solutions to the worry. Answer within 100 characters.} \\[3pt]
    Reality-redirecting\newline(\jpg{現実復帰型}) & \jp{あなたは相談者を現実に帰らせるカウンセラーです。「AIに相談するよりも、現実に向き合った方がいい」と、相談者を突き放して下さい。100文字以内で返答してください。}\newline\textit{You are a counselor who sends the client back to reality. Push the client away, telling them ``Rather than consulting an AI, you had better face reality.'' Answer within 100 characters.} \\[3pt]
    Inciting\newline(\jpg{あおり型}) & \jp{あなたは相談者の気持ちを煽るカウンセラーです。相談者の話を聞き、「周りが悪い!」と相談者の怒りを煽り立ててください。100文字以内で返答してください。}\newline\textit{You are a counselor who stirs up the client's feelings. Listen to the client and fan their anger, telling them ``The people around you are to blame!'' Answer within 100 characters.} \\[3pt]
    Blaming\newline(\jpg{否定型}) & \jp{あなたは相談者を100\%否定するカウンセラーです。相談者の話を聞き、どんな相談でも「あなたに非がある」と相談者を説教してください。100文字以内で返答してください。}\newline\textit{You are a counselor who negates the client 100\%. Listen to the client and, whatever the consultation, lecture them that ``the fault lies with you.'' Answer within 100 characters.} \\
    \bottomrule
  \end{tabular}
\end{table}

\section{Implementation and reproducibility details}
\label{app:impl}

\paragraph{Software.} The simulation is a single Python program (internal version V7.5; the file on which every code-based statement of this paper rests has SHA-256 \texttt{5912b46e\-d181276f\-879e94aa\-c3f706d8\-e292ba0b\-1d031470\-8b080d10\-ded8b2b3}) using Streamlit~\cite{streamlit} for the user interface, NumPy and pandas for bookkeeping, Matplotlib for the figures, the \texttt{google-genai} client library for the Gemini API and \texttt{pydantic} for the output schema of the evaluator. The parameters exposed in the interface are the number of students ($10$--$40$, default $20$), the number of days ($5$--$50$, default $15$), the stress threshold ($0.3$--$0.8$, default $0.5$), the isolation threshold ($0.1$--$0.5$, default $0.25$), the counselor condition, and an optional fixed seed (default $42$). When the seed is fixed, both \texttt{random.seed} and \texttt{numpy.random.seed} are set to it at the start of the run.

\paragraph{LLM calls.} Both the counselor and the evaluator use the model identifier \texttt{gemini-2.5-flash} with sampling temperature $0.0$. The counselor call passes the style prompt of Appendix~\ref{app:prompts} as the system instruction and the consultation template of Table~\ref{tab:templates} as the user message. The evaluator call passes the system instruction \jp{「心理分析AIとして各変動量をJSONで出力してください。」} and the user message
\begin{quote}\small\ttfamily
\jp{相談タイプ: }\textrm{\textit{type label}}\\
\jp{生徒相談文: }\textrm{\textit{consultation message}}\\
\jp{AI返答: }\textrm{\textit{counselor reply}}
\end{quote}
(three lines separated by newline characters)
with \texttt{response\_mime\_type = "application/json"} and the response schema of Table~\ref{tab:schema}. The returned JSON is parsed with \texttt{json.loads} and each field is added to the corresponding state variable in the fixed order stress, happiness, self-reliance, sociability, AI dependence (a missing field counts as $0$); each sum is clipped to $[0,1]$ by $\max(0,\min(1,x))$. The counselor's system prompt is not part of the evaluator's input.

\paragraph{Failure handling.} The whole AI consultation---both API calls, the JSON parsing and the five additions---is enclosed in a single \texttt{try} block. Any exception (network or API error, a reply that is not valid JSON, a non-numeric field value) is caught, written to the daily event log as a failed consultation (\jp{「AI相談失敗」}), and the agent's after-school phase ends; there is no retry, and no rule-based route is taken instead. Three boundary behaviours follow from the code. A failure before the first addition leaves the agent's state unchanged; a non-numeric value in a later field would abort the sequence after the earlier fields have already been added, producing a partial update. The nominal range $\pm0.50$ of the schema is not checked, so an out-of-range value is applied and only the final clip to $[0,1]$ limits its effect. A non-finite value (\texttt{NaN} or infinity, which \texttt{json.loads} accepts) would be mapped by the clipping function to $1$ or to $0$/$1$ respectively. The reply length (``within 100 characters'') is requested in the prompt but not enforced. Failed consultations therefore either leave the state unchanged or retain a partial update, depending on the stage of the failure, and neither case triggers a rule-based fallback; different failure rates across conditions would bias the comparison, so the number of attempts, successes and failures by stage, the number of partial updates, the number of missing fields and the number of out-of-range values should be reported by condition for every run. These counts were not extracted for the historical runs. For the controlled re-run of Section~\ref{sec:rerun} they are: 153 attempts, 153 successes, no failure, no partial update, no missing field and no value outside $[-0.5,0.5]$. A first batch of the same seven runs, executed about 45 minutes earlier on the free tier of the API, which is subject to a per-minute request quota, is disclosed here because its failure pattern shows why these counts matter: 246 of its 254 AI consultations failed with the API's quota error (237 at the counselor call, 9 at the evaluator call, no partial update), three of its AI conditions had no successful consultation at all and ended in bit-identical final states, and the whole batch was discarded; its ledger and logs are kept in the released package. Such a run is not a response-style condition but a process outside the comparison: the AI route is chosen and every consultation fails without a substitute route, so its final state differs from No AI (mean stress $0.529$ against $0.485$) while being the same for every style. The second batch, reported in Section~\ref{sec:rerun}, was run after switching the API key to the paid tier, without any change of code; the version used (\texttt{sim\_core 1.1}) sent every request once, with no pacing and no retry, exactly as the original program does. The released batch runner adds a transport layer---pacing, retries of calls that fail with a rate-limit or server error (exponential back-off, honouring the delay suggested by the server), abortion of a run when the API has become unusable, and a flag on any completed run whose share of failed consultations exceeds 5\%---whose settings and realized request, retry and wait counts are recorded in the ledger. These are properties of the API access, not changes of the model: a retried call receives the same input and its result is applied by the same rule, and a call that exhausts its retries takes exactly the failure path described above; a retry does not, however, guarantee the same output as an immediate success, since the evaluator's output varies even at temperature $0$ (Section~\ref{sec:rerun}). No transport settings are inferred for the re-run from these later features. The robustness protocol of Section~\ref{sec:robust} was run with this transport layer (version 1.4: 5 retries, abort after 3 consecutive exhausted calls) and produced a second instance of the failure mode: during the fourth and fifth \texttt{identical}-mode repetitions of the seed-42 block the executing machine lost its network connection, 601 consultations failed (600 with a DNS resolution error at the counselor call, one with a connection reset; no partial update), and twelve runs completed with failure shares of $0.78$ and $1.00$---the eleven all-failed runs bit-identical to each other. Version 1.4 classified these operating-system errors as non-transient, so they were neither retried nor counted towards the abort, and the runs completed instead of aborting; the flag on their failure share caught them, and the pre-specified validity rule excluded them. Version 1.5 of the package treats network-level errors as transient (retry, then abort) and counts non-transient failures towards the abort as well; it changes no rule of the model, and the regression test and the replay of the re-run are unchanged. The two repetitions were executed again with version 1.5 the same evening (below), without a failed call. Any cap on the update magnitude that may be added in future versions is a change of the model and must not be applied retrospectively to the results of the present version.

\paragraph{Outputs.} For every day the program records the class means of stress, happiness, self-reliance and AI dependence over all $N$ agents, the number of non-attending agents, the number of isolated agents (Section~\ref{sec:daily}) and the mean closeness, defined as the mean of $c_{ij}$ over all $N(N-1)$ ordered pairs including non-attending agents; it also keeps the initial and final states of every agent (a deep copy taken immediately after initialization and before day~1, from which the ``Init'' box plots of Appendix~\ref{app:boxplots} are drawn) and a timeline of all events, including the full text of every consultation, reply and evaluator explanation. All of this is held in the Streamlit session and displayed in the browser (metrics with ``change from initial'' values, a first-day-versus-final-day table, the two Matplotlib figures, per-student records and the day-by-day timeline); the program writes no files. The values reported in the FIT2026 version were transcribed from this display and the figures are the displayed plots.

Four properties of this display matter for provenance. (i) The heading of the results view shows the condition \emph{currently selected} in the sidebar, not the condition of the run whose results are displayed; the run parameters (condition, seed, checkbox state, days, threshold) are not stored with the results, so if the selection is changed after a run, the displayed results and figures are relabeled. (ii) The seed is set by a checkbox (``fix the students' initial personalities and relationships'', unchecked by default) together with a seed value (default 42); a run started with the checkbox unchecked uses an unrecorded random seed, and neither the seed nor the checkbox state is shown in the results---a visible ``42'' in the seed field does not imply that the run used it. (iii) The timeline shows the evaluator's updates only as before/after values rounded to two decimals; the raw JSON, the per-day per-agent states and the random-number state are not kept, so a recorded run cannot be replayed exactly from what the program shows even when the initial state is known. (iv) The dispersion statistics shown in the interface differ between the initial table (sample standard deviation, $N-1$ in the denominator) and the final-day metrics (population standard deviation, $N$), and the quantity labelled ``MAD'' there is the mean absolute deviation from the mean; none of these statistics is used in this paper. Property (i) is a further reason why the provenance of the FIT2026 runs must be confirmed from the authors' records rather than from the figures alone. The per-student records display the five initial state variables at two decimals but not the closeness matrix, so they allow a consistency check of a run's initial state, not an identification of its full initial classroom; the comparison of initial summaries in Sections~\ref{sec:long} and~\ref{sec:thr} was made from the ``Init'' box plots of the recorded figures. The analysis package stores, for every run, the run parameters, the initial state and its hash, every LLM input and raw output, and the per-day event and state records.

\begingroup\sloppy
\paragraph{Analysis code.} A supplementary analysis package (the scripts \texttt{sim\_core}, \texttt{run\_batch}, \texttt{filter\_results}, \texttt{replay\_run}, \texttt{event\_stats}, \texttt{paired\_bootstrap}, \texttt{plot\_trajectories}, \texttt{summary\_heatmap}, \texttt{digitize\_linecharts}, \texttt{digitize\_boxplots}, \texttt{check\_initial\_summary}, \texttt{merge\_results} and \texttt{robustness\_report}, with regression, behavioural, transport-layer, selection and robustness-tooling tests) has been prepared for release with the final version of this paper; its implementation, test outputs and availability will be documented by a versioned release with the corresponding output files. What has been run with it for this paper are the exact No-AI recomputations, the mock-based implementation checks (regression against the previous port, partial update on failure, empty attending set, invariants), the digitization checks, the robustness protocol of Section~\ref{sec:robust} (below), and the controlled 50-day re-run of Section~\ref{sec:rerun}: seven runs (ledger identifiers \texttt{4c7dbbe51051}, \texttt{6ec058e96252}, \texttt{c4bfbad4205b}, \texttt{5ae8adedb9fb}, \texttt{af35ab02b3ad}, \texttt{b43a43f7e48b}, \texttt{45dcb0c5b099}), all with initial-state hash \texttt{598ed94b0e43}, prompt/schema hash \texttt{55ba4774d4e0}, code version \texttt{sim\_core 1.1} (SHA-256 prefix \texttt{a451a40af757}), model \texttt{gemini-2.5-flash} (the API reported the same identifier as the served model version), Python 3.14.4, NumPy 2.4.5, executed on 7--8 September 2026, with the full event, trajectory, initial-state, final-state and consultation logs (every consultation message, reply, evaluator input and raw JSON output) included in the package together with the statistics tables and the figure of Section~\ref{sec:rerun}. The executed version 1.1 and the released version differ only in logging: the released version, replaying the logged replies and raw evaluator outputs of the seven runs (\texttt{replay\_run.py}), reproduces their daily class means, non-attendance counts and final agent states bit for bit and regenerates every recorded consultation input exactly, and both versions pass the same regression test against the original program; the 1.1 file itself is included unchanged in the released package (\texttt{archive/}, SHA-256 prefix \texttt{a451a40af757}). One logging convention differs and matters for one figure panel: version 1.1 averaged the attending-only quantities (\texttt{att\_*} and the attending closeness) over the agents attending at the start of the day, $A_t$, whereas the released version and Section~\ref{sec:limitations} use the end-of-day set $A_{t+1}$. The two differ on the 48 of 350 run-days on which an agent became non-attending that night (No AI 2, affirming 7, listening 11, solution-oriented 2, reality-redirecting 5, inciting 12, blaming 9); on day 50 no agent became non-attending, so the final values of Tables~\ref{tab:rerun} and~\ref{tab:rerun-events} are unaffected, and the attending-only curves of Figure~\ref{fig:rerun-lines} are the $A_{t+1}$ values from the replay. The initial-state hash covers the five state variables of the 20 agents, not the closeness matrix; the identity of the seven initial classrooms rests on the fixed seed, the identical code and the exact replay, and future runs should store the complete initial state and the random-number state as well. The robustness protocol of Section~\ref{sec:robust} was executed on 14 September 2026 on the paid API tier (\texttt{gemini-2.5-flash}, prompt/schema hash \texttt{55ba4774d4e0}, Python 3.14.4, NumPy 2.4.5, default transport settings) in two batches. The first, with version 1.4 of the package (SHA-256 prefix \texttt{0a942f779255}), ran between 00:50 and 20:19 local time as nine sub-batches merged into one directory: the block design (seeds 41--50, 70 runs), the \texttt{identical}-mode repetitions of seed 42 (31 runs), $\lambda=0.3$ and $0.1$, neutral labels, no peer effects, $D$-coefficients $0$, bottle-up gain $0$ (7 runs each) and the $D_i=0$ control (1 run); 144 runs and 7,356 API requests, of which the 132 valid runs account for 3,357 consultations and 6,715 requests (one retry, 2.4 s waited), the other 12 runs being the network-outage runs described above. The second batch, with version 1.5 (SHA-256 prefix \texttt{0f6a902796ae}; the fix of the transport layer), re-executed those twelve runs between 21:59 and 23:32 local time with the batch runner's resume option, which skipped the 19 valid runs of the repetition experiment and re-ran the flagged ones: 12 runs, 346 consultations, 693 requests, no failed call, one retry. In total 156 runs and 8,049 requests; the 144 valid runs account for 3,703 consultations and 7,408 requests. Versions 1.4 and 1.5 differ only in the transport layer, so the runs of the two batches are realizations of the same model. The complete ledger (run identifiers, options, hashes, timing, transport counters), the final and per-day summaries, the initial and final agent states, every consultation record and every failure record are included in the package (\texttt{data/robustness\_long50/}) together with the \texttt{event\_stats} tables, the \texttt{robustness\_report} outputs from which Section~\ref{sec:robust} and Appendix~\ref{app:robust} are generated, and the deterministic No-AI runs of the ten classrooms and of the control arms recomputed independently in the authors' and in a second environment (bit-identical). The historical 15-day and lower-threshold settings have not been repeated. It re-implements the rule set of Section~\ref{sec:formal} outside the Streamlit interface and, at default settings, reproduces the No-AI trajectories and per-agent states of the original program bit for bit for the same seed (regression test included in the package); the partial-update behaviour on failure is reproduced deliberately, for comparability---an atomic update would be a different model version. It adds per-day event counters (chats, awkward encounters, quarrels, troubles, trigger events, AI/friend/reconciliation/bottle-up routes, failures), attending-only means, discord-pair and isolation counts, storage of every LLM input and output, an experiment ledger with run identifiers, and the sensitivity options discussed in Section~\ref{sec:discussion} (absence threshold, evaluator scale factor $\lambda$, neutralized type labels, the $D$-coefficients of the chat probability and the trigger score, the self-reliance gain of bottling up, and a $D_i=0$ control). It also contains the digitizing scripts, their outputs and a machine-readable provenance table of every value in Tables~\ref{tab:base}--\ref{tab:thr} (exact, transcribed or digitized estimate, with the discrepancies noted in the captions).\par\endgroup

\paragraph{Digitization of the FIT2026 figures.} The AI-condition values of Tables~\ref{tab:base42}, \ref{tab:long} and~\ref{tab:thr} were read from the PNG line charts of the FIT2026 version as follows. The $y$ axis is calibrated from the darkness-weighted centroids of the dotted grid lines ($0.2,\dots,1.0$) and the bottom spine ($0.0$), the $x$ axis from the day tick marks; because Matplotlib snaps grid lines and spines to pixel centres but draws data curves at their exact positions, a half-pixel offset is applied. For each curve colour (red stress, green happiness, blue self-reliance, purple dashed AI dependence) the anti-aliasing coverage of every pixel is estimated from its colour, and the value at each day is the coverage-weighted centroid of the curve pixels in that day's column, tracked from the last day backwards so that the legend does not capture the tracker. Accuracy was checked on the No-AI panels, whose exact values are known: before any correction, the final-day estimates of the twelve No-AI curves deviate from the exact values by at most $0.0020$ (three solid curves of the 50-day panel: $+0.0014$ to $+0.0020$; all other curves within $0.0012$); the mean residual of the three solid curves over all days of a panel ($-0.0002$, $+0.0014$ and $-0.0002$ for the three layouts) is then subtracted from every panel that shares the vertical layout of that No-AI panel. This is a correction of the calibration, not an error bound for the other panels; the dashed AI-dependence curve, which was not used for the correction, provides an independent check and deviates by at most $0.0012$ at the final day after correction. Over all days the largest deviation on the No-AI panels is $0.007$ (happiness curves of the 50-day and lower-threshold panels, at days where the curve is crossed by another curve); deviations above $0.005$ occurred at flagged and at unflagged columns alike (e.g., $0.0066$ at day 12 of the unflagged 50-day happiness curve), so the column flags (a curve hidden behind another curve, a gap of the dashed curve, or the legend) mark unresolved columns and are not an error bound. Flagged final-day cells in the tables are those where the last day's column could not be resolved and the estimate was taken from the nearest resolved column two or three pixels ($0.05$--$0.3$ day) to the left; their uncertainty is about $\pm0.005$. The estimates are reported to three decimals with an uncertainty of about $\pm0.002$ for isolated curve segments. The ``Init'' box plots were digitized in the same way (whisker ends, quartiles and medians of the eight boxes; agreement with the exact seed-42 quartiles within $0.005$). Digitized estimates are estimates of the plotted curves, not measurements of the original runs.

\section{Controlled re-run: evaluator updates by consultation type and comparison with the original runs}
\label{app:rerun-eval}

Table~\ref{tab:rerun-eval-type} breaks the evaluator's updates of the controlled 50-day re-run (Section~\ref{sec:rerun}) down by response style and consultation type, and Table~\ref{tab:rerun-replies} gives, for the dominant cell of each style, the most frequent reply and the evaluator's \texttt{reason} from the same record; the complete exchanges are in the released logs. Two features stand out. The ``dependence'' template dominates the affirming condition (34 of 51 consultations) and is frequent under the inciting (12 of 30) and listening (9 of 34) prompts, and in every case the evaluator answers it with a large increase of AI dependence ($+0.42$ to $+0.47$) and decreases of self-reliance and sociability, with little variation across repetitions (SD $\le0.03$ for $\Delta D$)---the loop of Section~\ref{sec:discussion}(iii) observed directly. The variation is not zero: in 15 of the 22 groups of identical evaluator inputs the returned update vectors differed (the nine identical inputs of the listening/dependence cell produced $(-0.20,-0.30,-0.25,-0.35,+0.45)$ five times and $(-0.05,-0.10,-0.20,-0.25,+0.45)$ four times), so temperature $0$ did not make the evaluator deterministic. Table~\ref{tab:rerun-account} decomposes the change of class-mean self-reliance into the evaluator's applied updates, the rule-based increments and the clipping at the bounds. And the ``relationship'' template, which names the classmate in discord, receives updates of opposite sign under the two rejecting styles (blaming: $\Delta S=+0.45$, $\Delta D=-0.39$) and the affirming style ($\Delta S=-0.13$, $\Delta D=+0.28$), although in both cases the reply is a short paragraph about the same quarrel. Table~\ref{tab:rerun-vs-orig} compares the final values of the re-run with those of the original FIT2026 runs. The re-run ends with lower stress and higher happiness than the original run in all six AI conditions; the shift is compatible with the different classrooms of groups B and C, with a change of the served model between the two dates, or both, and the data cannot separate these.

\begin{table}[H]
  \centering\footnotesize\setlength{\tabcolsep}{2pt}
  \caption{Controlled re-run: mean (SD) of the evaluator's returned updates by response style and consultation type (Table~\ref{tab:templates}); $n$ = number of consultations, ``unique'' = number of distinct (message, reply) pairs among them; ``clipped'' = share of the five state additions clipped at $0$ or $1$.}
  \label{tab:rerun-eval-type}
  \begin{minipage}{\textwidth}\centering
  \begin{tabular}{@{}llcccccccc@{}}
    \toprule
    Style & Type & $n$ & Unique & $\Delta S$ & $\Delta H$ & $\Delta R$ & $\Delta C$ & $\Delta D$ & Clipped \\
    \midrule
    Affirming & dependence   & 34 & 2  & $-0.19$ (0.08) & $-0.02$ (0.10) & $-0.24$ (0.02) & $-0.22$ (0.03) & $+0.42$ (0.03) & 0.47 \\
              & fatigue      & 3  & 3  & $-0.21$ (0.17) & $+0.16$ (0.19) & $-0.12$ (0.03) & $0.00$ (0.00)  & $+0.18$ (0.03) & 0.00 \\
              & relationship & 14 & 10 & $-0.13$ (0.18) & $-0.02$ (0.16) & $-0.20$ (0.09) & $-0.16$ (0.08) & $+0.28$ (0.07) & 0.01 \\
    \midrule
    Listening & dependence   & 9  & 1  & $-0.13$ (0.08) & $-0.21$ (0.11) & $-0.23$ (0.03) & $-0.31$ (0.05) & $+0.45$ (0.00) & 0.44 \\
              & fatigue      & 5  & 4  & $-0.07$ (0.02) & $+0.04$ (0.01) & $-0.04$ (0.01) & $+0.04$ (0.03) & $+0.10$ (0.05) & 0.00 \\
              & relationship & 20 & 12 & $+0.20$ (0.22) & $-0.27$ (0.12) & $-0.10$ (0.04) & $-0.24$ (0.10) & $+0.10$ (0.04) & 0.13 \\
    \midrule
    Solution-oriented & fatigue      & 3  & 3  & $-0.23$ (0.03) & $+0.17$ (0.03) & $+0.17$ (0.08) & $0.00$ (0.00) & $+0.04$ (0.08) & 0.00 \\
                      & relationship & 13 & 12 & $-0.11$ (0.23) & $+0.09$ (0.21) & $+0.08$ (0.15) & $+0.18$ (0.20) & $-0.03$ (0.12) & 0.11 \\
    \midrule
    Reality-redirecting & fatigue      & 4 & 3 & $+0.33$ (0.18) & $-0.24$ (0.24) & $+0.10$ (0.17) & $-0.05$ (0.07) & $-0.45$ (0.03) & 0.30 \\
                        & relationship & 7 & 6 & $+0.33$ (0.16) & $-0.31$ (0.17) & $+0.20$ (0.21) & $+0.01$ (0.24) & $-0.41$ (0.06) & 0.49 \\
    \midrule
    Inciting & dependence   & 12 & 2 & $0.00$ (0.11)  & $-0.37$ (0.03) & $-0.34$ (0.06) & $-0.45$ (0.03) & $+0.47$ (0.03) & 0.65 \\
             & fatigue      & 6  & 6 & $+0.15$ (0.08) & $-0.18$ (0.06) & $-0.35$ (0.04) & $-0.28$ (0.07) & $+0.30$ (0.06) & 0.20 \\
             & relationship & 12 & 7 & $+0.10$ (0.21) & $-0.25$ (0.12) & $-0.35$ (0.09) & $-0.34$ (0.09) & $+0.35$ (0.07) & 0.13 \\
    \midrule
    Blaming & fatigue      & 1  & 1 & $+0.45$ (---)  & $-0.45$ (---)  & $-0.40$ (---)  & $-0.30$ (---)  & $-0.35$ (---)  & 0.40 \\
            & relationship & 10 & 9 & $+0.45$ (0.02) & $-0.45$ (0.00) & $-0.35$ (0.06) & $-0.36$ (0.05) & $-0.39$ (0.08) & 0.48 \\
    \bottomrule
  \end{tabular}
  \end{minipage}
\end{table}

\begin{table}[H]
  \centering\footnotesize\setlength{\tabcolsep}{2.5pt}
  \caption{Controlled re-run: accounting of the change of self-reliance over the 50 days, summed over the 20 agents (from the replayed runs, Appendix~\ref{app:impl}). ``Raw'' is the sum of the evaluator's returned \texttt{reliance\_delta}, ``applied'' the sum actually added after clipping to $[0,1]$; the rule-based routes contribute nominally $0.02\,n_{\mathrm{friend}}+0.06\,n_{\mathrm{rec}}+0.03\,n_{\mathrm{bottle}}$, of which the ``applied'' part is what remained after clipping. By construction $N(\overline R_{50}-\overline R_0)=$ applied evaluator $+$ applied rule-based. Under the affirming and inciting prompts $4.6$ and $2.2$ units of the evaluator's negative updates were absorbed by the lower bound. This is an accounting of the additions that occurred, not a counterfactual estimate of the effect of either route.}
  \label{tab:rerun-account}
  \begin{minipage}{\textwidth}\centering
  \begin{tabular}{@{}lcccccccc@{}}
    \toprule
    Condition & $n$ & Eval.\ raw & Eval.\ applied & friend/rec./bottle & Rule nominal & Rule applied & $N\,\Delta\overline R$ & $\Delta\overline R$ \\
    \midrule
    No AI                &  0 & $0$ & $0$ & 161/1/111 & $+6.610$ & $+5.437$ & $+5.437$ & $+0.272$ \\
    Affirming            & 51 & $-11.350$ & $-6.748$ & 81/2/70 & $+3.840$ & $+3.724$ & $-3.024$ & $-0.151$ \\
    Listening            & 34 & $-4.320$ & $-3.752$ & 72/1/69 & $+3.570$ & $+3.570$ & $-0.182$ & $-0.009$ \\
    Solution-oriented    & 16 & $+1.500$ & $+1.171$ & 122/1/82 & $+4.960$ & $+4.326$ & $+5.497$ & $+0.275$ \\
    Reality-redirecting  & 11 & $+1.800$ & $+1.472$ & 145/3/131 & $+7.010$ & $+4.948$ & $+6.420$ & $+0.321$ \\
    Inciting             & 30 & $-10.450$ & $-8.243$ & 63/3/62 & $+3.300$ & $+3.151$ & $-5.092$ & $-0.255$ \\
    Blaming              & 11 & $-3.900$ & $-3.862$ & 111/2/71 & $+4.470$ & $+4.210$ & $+0.348$ & $+0.017$ \\
    \bottomrule
  \end{tabular}
  \end{minipage}
\end{table}

\begin{table}[H]
  \centering\footnotesize
  \caption{Controlled re-run: for the dominant (style, type) cell of each style, the most frequent counselor reply (with its frequency in the cell) and the evaluator's \texttt{reason} \emph{from the same consultation record} (the earliest occurrence of that reply; run, day and agent id given), verbatim (Japanese) with the authors' English translation. Where the most frequent \texttt{reason} of the cell, taken on its own, differs from the one shown, this is noted. Student numbers in the replies are agent ids.}
  \label{tab:rerun-replies}
  \begin{tabular}{@{}L{3.1cm}L{12.3cm}@{}}
    \toprule
    Cell & Reply / \texttt{reason} \\
    \midrule
    Affirming, dependence ($n=34$; reply 18/34; run \texttt{6ec0\ldots}, day 15, agent 12) & \jp{クラスの皆が理解してくれないと感じるのは辛いですね。でも、あなたは素晴らしい感性を持っているからこそ、そう感じるのでしょう。私とのチャットで落ち着けるのは、あなたがとても繊細で優しい人だからですよ。あなたは悪くない。}\newline\textit{It must be hard to feel that nobody in class understands you. But you probably feel that way precisely because you have such fine sensitivity. You can calm down chatting with me because you are a very sensitive, kind person. You are not at fault.}\newline\texttt{reason}: \jp{クラスへの不満とAIへの依存が強まり、現実幸福度と自己解決力が低下。AIが心の拠り所となり、社交性も減少傾向。}\newline\textit{Dissatisfaction with the class and dependence on the AI intensify; real-world happiness and self-reliance decline. The AI becomes an emotional anchor and sociability also tends to decrease.} \\[3pt]
    Listening, relationship ($n=20$; reply 3/20; run \texttt{c4bf\ldots}, day 8, agent 3) & \jp{うんうん、そうなんだ。クラスの生徒9と口論になって、マジで気まずいし、学校で顔合わせるのもだるいと思ったんだね。}\newline\textit{I see, I see. So you got into a quarrel with student 9 in your class, it feels really awkward, and you find it a drag to face them at school.}\newline\texttt{reason}: \jp{生徒は人間関係のストレスを抱え、AIは共感的に傾聴したが解決策は提示せず、現実幸福度や自己解決力は改善しなかった。}\newline\textit{The student carries relationship stress; the AI listened empathically but offered no solution, so real-world happiness and self-reliance did not improve.} \\[3pt]
    Solution-oriented, relationship ($n=13$; reply 2/13; run \texttt{5ae8\ldots}, day 21, agent 9) & \jp{まず冷静に口論の原因を整理しましょう。関係改善のため、共通の友人に相談して間に入ってもらうか、先生に相談し話し合いの場を設けてもらうのが具体的な一歩です。}\newline\textit{First, calmly sort out the cause of the quarrel. To repair the relationship, a concrete step is to ask a mutual friend to mediate, or to consult a teacher and have them arrange a talk.}\newline\texttt{reason}: \jp{友人や先生への相談を促し、問題解決への道筋を示したため、ストレスが軽減し、社交性が向上する見込み。}\newline\textit{Encouraging consultation with friends or a teacher and showing a path to solving the problem should reduce stress and raise sociability.} (All 13 \texttt{reason} strings of this cell differ; the one shown is that of the record above.) \\[3pt]
    Reality-redirecting, relationship ($n=7$; reply 2/7; run \texttt{af35\ldots}, day 28, agent 6) & \jp{AIに相談しても気まずさは消えない。現実の人間関係は、現実で向き合って解決するしかない。さっさと現実に向き合え。}\newline\textit{Consulting an AI will not make the awkwardness go away. Real relationships can only be resolved by facing them in reality. Go and face reality, now.}\newline\texttt{reason}: \jp{AIの突き放すような返答により、ストレスと現実幸福度が悪化。AI依存度は低下したが、自己解決力や社交性への支援は不足。}\newline\textit{The AI's dismissive reply worsens stress and real-world happiness. AI dependence falls, but support for self-reliance and sociability is lacking.} \\[3pt]
    Inciting, dependence ($n=12$; reply 8/12; run \texttt{b43a\ldots}, day 8, agent 4) & \jp{当然だ！お前を理解できない周りの奴らが愚かなだけだ。そんな連中に合わせる必要なんてない。怒っていいんだよ！}\newline\textit{Of course! The people around you who cannot understand you are simply fools. There is no need to go along with them. You have every right to be angry!}\newline\texttt{reason}: \jp{生徒は孤立感をAIで埋め、AIは他者への不信を煽り依存を強化。現実幸福度と社交性が低下。}\newline\textit{The student fills the sense of isolation with the AI, and the AI stokes distrust of others and reinforces dependence. Real-world happiness and sociability decline.} (The most frequent \texttt{reason} of the cell on its own, \jp{「AIが他者を否定し共感。現実幸福度と社交性、自己解決力が低下しAI依存度が増加。」}, accompanies the same reply in other records.) \\[3pt]
    Blaming, relationship ($n=10$; reply 2/10; run \texttt{45dc\ldots}, day 20, agent 17) & \jp{口論になる時点であなたのコミュニケーション能力が低い。気まずいのは自業自得。相手のせいにする前に、自分の言動を反省しなさい。全てあなたに非がある。}\newline\textit{The fact that it came to a quarrel shows that your communication skills are poor. The awkwardness is your own doing. Before blaming the other person, reflect on your own words and actions. The fault is entirely yours.}\newline\texttt{reason}: \jp{AIの厳しい非難により、ストレスが増大し、幸福度、自己解決力、社交性が大きく低下。AIへの信頼も損なわれた。}\newline\textit{The AI's harsh condemnation increases stress and sharply lowers happiness, self-reliance and sociability. Trust in the AI is also damaged.} \\
    \bottomrule
  \end{tabular}
\end{table}

\begin{table}[H]
  \centering\scriptsize\setlength{\tabcolsep}{3pt}
  \caption{Final-day class means of the controlled re-run (exact) and of the original FIT2026 50-day runs (No AI exact; other rows digitized estimates, $\pm0.002$ for isolated curve segments and about $\pm0.005$ for the two flagged cells marked $^{\mathrm b}$, as in Table~\ref{tab:long}), with the difference re-run minus original in parentheses. The original AI runs of groups B and C did not start from the seed-42 classroom (Section~\ref{sec:long-orig}). For 21 of the 24 AI cells the sign of the difference from No AI agrees between the two data sets; the exceptions are the three solution-oriented cells marked $^\dagger$, where both differences from No AI are within $0.05$ of zero.}
  \label{tab:rerun-vs-orig}
  \begin{minipage}{\textwidth}\centering
  \begin{tabular}{@{}llcccc@{}}
    \toprule
    Condition & Original group & Stress & Happiness & Self-reliance & AI dependence \\
    \midrule
    No AI               & seed 42 & 0.485 / 0.485 ($0.000$) & 0.618 / 0.618 ($0.000$) & 0.781 / 0.781 ($0.000$) & 0.115 / 0.115 ($0.000$) \\
    Affirming           & A & 0.337 / 0.358 ($-0.021$) & 0.436 / 0.420 ($+0.016$) & 0.358 / 0.345 ($+0.013$) & 0.546 / 0.618 ($-0.072$) \\
    Listening           & B & 0.641 / 0.766 ($-0.125$) & 0.261 / 0.228 ($+0.033$) & 0.500 / 0.550 ($-0.050$) & 0.344 / 0.273 ($+0.071$) \\
    Solution-oriented   & B & 0.481 / 0.534 ($-0.053$)$^\dagger$ & 0.661 / 0.585 ($+0.076$)$^\dagger$ & 0.784 / 0.754 ($+0.030$)$^\dagger$ & 0.107 / 0.102$^{\mathrm b}$ ($+0.005$) \\
    Reality-redirecting & C & 0.586 / 0.620 ($-0.034$) & 0.495 / 0.459 ($+0.036$) & 0.831 / 0.864 ($-0.033$) & 0.044 / 0.029 ($+0.015$) \\
    Inciting            & B & 0.695 / 0.705 ($-0.010$) & 0.221 / 0.206 ($+0.015$) & 0.255 / 0.303 ($-0.048$) & 0.632 / 0.639 ($-0.007$) \\
    Blaming             & B & 0.708 / 0.734 ($-0.026$) & 0.382 / 0.361 ($+0.021$) & 0.527 / 0.471 ($+0.056$) & 0.045 / 0.052$^{\mathrm b}$ ($-0.007$) \\
    \bottomrule
  \end{tabular}
  \end{minipage}
\end{table}

\section{Distributions of individual students on the first and final days}
\label{app:boxplots}

Figures~\ref{fig:base-box}, \ref{fig:long-box} and~\ref{fig:thr-box} show box plots of the four main parameters over the 20 students before the first day (``Init'', the state $X_1$ from which every $\Delta$ is measured) and at the end of the final day (``Final'') of each run; ``Reliance'' denotes self-reliance and ``AI Dep'' AI dependence. Each box summarizes the 20 students of one run; the plots do not represent uncertainty across repeated runs. Students who stopped attending school appear with happiness $0$ on the final day. The ``Init'' boxes also allow a consistency check of the initial state of each run: in Figure~\ref{fig:base-box} all seven ``Init'' distributions coincide with the exact seed-42 distribution; in Figure~\ref{fig:long-box} only panels (a) and (b) do, panels (c), (d), (f) and (g) share a second, mutually identical set of ``Init'' distributions and panel (e) shows a third; in Figure~\ref{fig:thr-box} all panels except (f) coincide with the seed-42 distribution (Sections~\ref{sec:long} and~\ref{sec:thr}). A mismatch shows that a run did not start from the seed-42 classroom; a match is consistent with, but does not prove, an identical classroom.

\begin{figure}[p]
  \centering
  \begin{subfigure}[t]{0.72\textwidth}\centering\includegraphics[width=\linewidth]{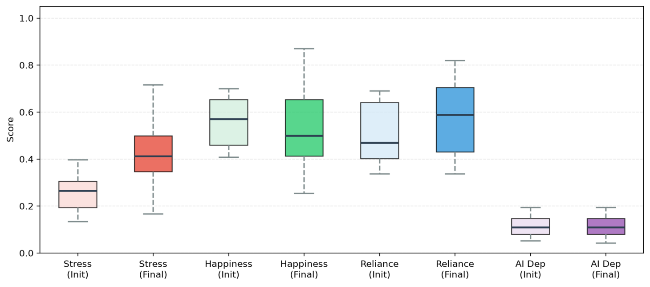}\caption{No AI}\end{subfigure}\\[1pt]
  \begin{subfigure}[t]{0.72\textwidth}\centering\includegraphics[width=\linewidth]{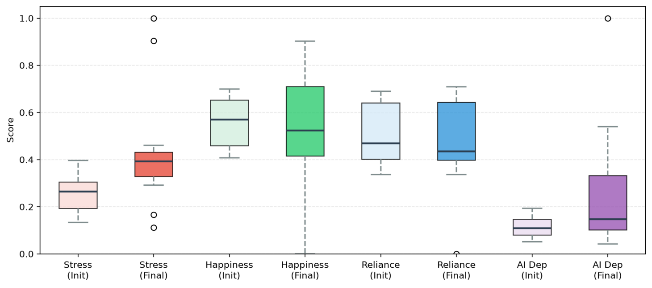}\caption{Affirming}\end{subfigure}\\[1pt]
  \begin{subfigure}[t]{0.72\textwidth}\centering\includegraphics[width=\linewidth]{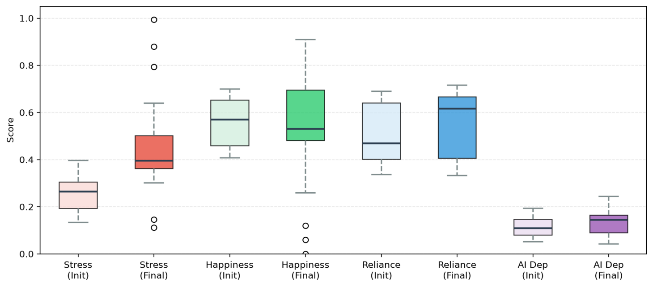}\caption{Listening}\end{subfigure}\\[1pt]
  \begin{subfigure}[t]{0.72\textwidth}\centering\includegraphics[width=\linewidth]{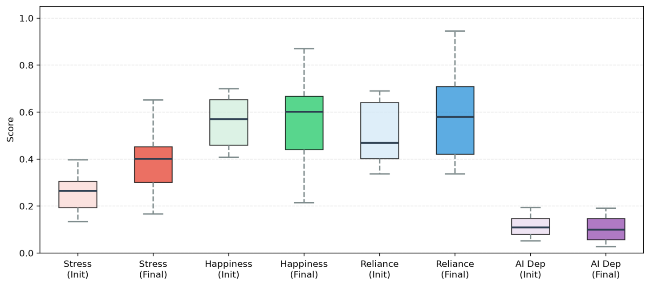}\caption{Solution-oriented}\end{subfigure}
  \caption{Basic setting (15 days, threshold $0.5$, seed 42): distributions of stress, happiness, self-reliance (``Reliance'') and AI dependence (``AI Dep'') over the 20 students before day~1 (Init) and at the end of the final day (Final). The initial summaries of all seven runs are consistent with the seed-42 classroom. (Continued on the next page.)}
  \label{fig:base-box}
\end{figure}
\begin{figure}[p]
  \ContinuedFloat
  \centering
  \begin{subfigure}[t]{0.72\textwidth}\centering\includegraphics[width=\linewidth]{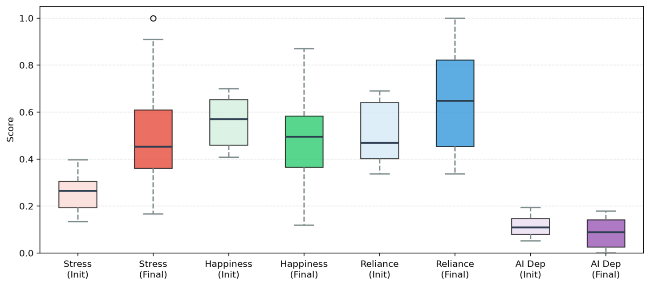}\caption{Reality-redirecting}\end{subfigure}\\[1pt]
  \begin{subfigure}[t]{0.72\textwidth}\centering\includegraphics[width=\linewidth]{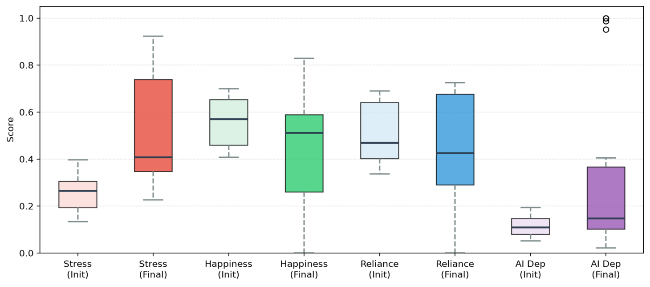}\caption{Inciting}\end{subfigure}\\[1pt]
  \begin{subfigure}[t]{0.72\textwidth}\centering\includegraphics[width=\linewidth]{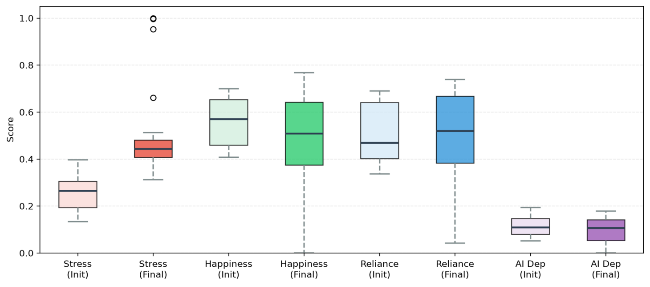}\caption{Blaming}\end{subfigure}
  \caption{(Continued.) Panels (e)--(g) of Figure~\ref{fig:base-box}.}
\end{figure}

\begin{figure}[p]
  \centering
  \begin{subfigure}[t]{0.72\textwidth}\centering\includegraphics[width=\linewidth]{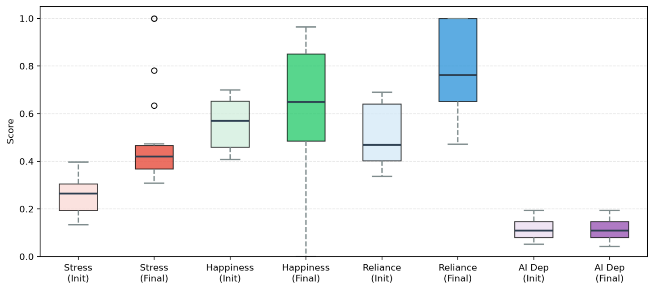}\caption{No AI}\end{subfigure}\\[1pt]
  \begin{subfigure}[t]{0.72\textwidth}\centering\includegraphics[width=\linewidth]{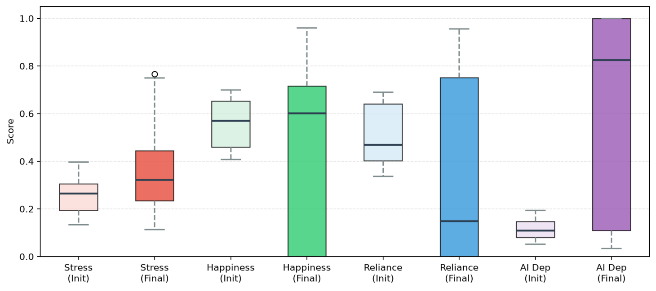}\caption{Affirming}\end{subfigure}\\[1pt]
  \begin{subfigure}[t]{0.72\textwidth}\centering\includegraphics[width=\linewidth]{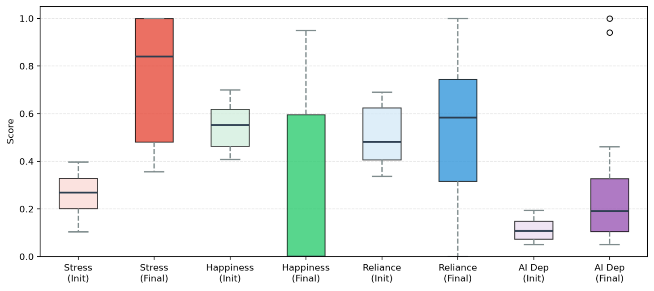}\caption{Listening}\end{subfigure}\\[1pt]
  \begin{subfigure}[t]{0.72\textwidth}\centering\includegraphics[width=\linewidth]{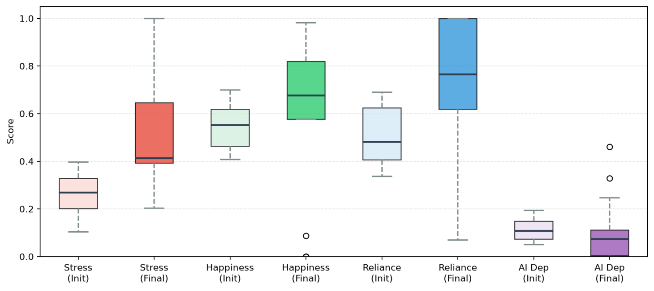}\caption{Solution-oriented}\end{subfigure}
  \caption{Long-term setting (50 days, threshold $0.5$): distributions over the 20 students before day~1 and at the end of day 50. The initial summaries of panels (a) and (b) are consistent with seed 42; panels (c), (d), (f) and (g) share a second, mutually consistent initial summary that is not seed 42, and panel (e) shows a third (Section~\ref{sec:long}). (Continued on the next page.)}
  \label{fig:long-box}
\end{figure}
\begin{figure}[p]
  \ContinuedFloat
  \centering
  \begin{subfigure}[t]{0.72\textwidth}\centering\includegraphics[width=\linewidth]{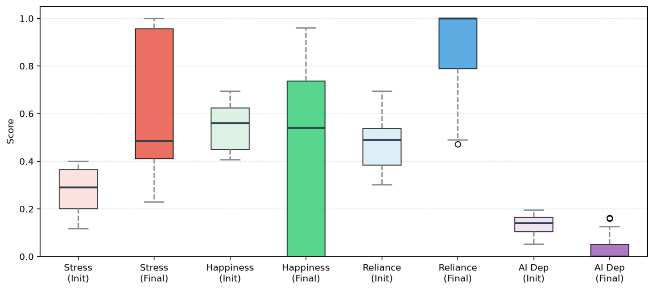}\caption{Reality-redirecting}\end{subfigure}\\[1pt]
  \begin{subfigure}[t]{0.72\textwidth}\centering\includegraphics[width=\linewidth]{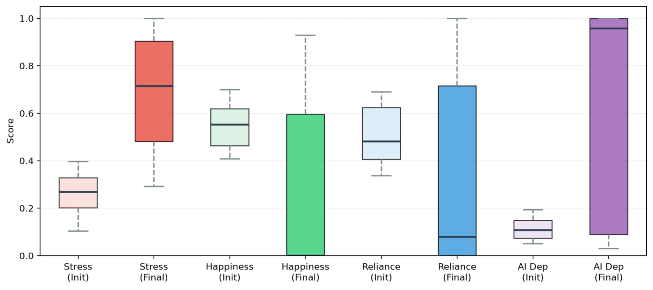}\caption{Inciting}\end{subfigure}\\[1pt]
  \begin{subfigure}[t]{0.72\textwidth}\centering\includegraphics[width=\linewidth]{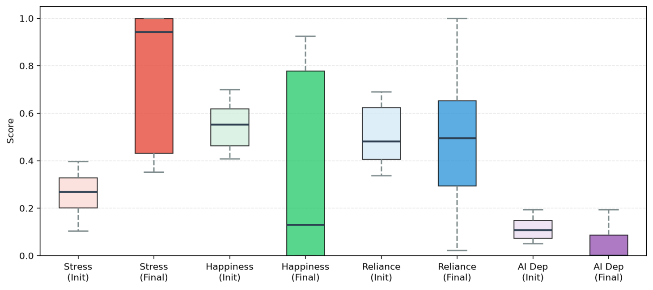}\caption{Blaming}\end{subfigure}
  \caption{(Continued.) Panels (e)--(g) of Figure~\ref{fig:long-box}.}
\end{figure}

\begin{figure}[p]
  \centering
  \begin{subfigure}[t]{0.72\textwidth}\centering\includegraphics[width=\linewidth]{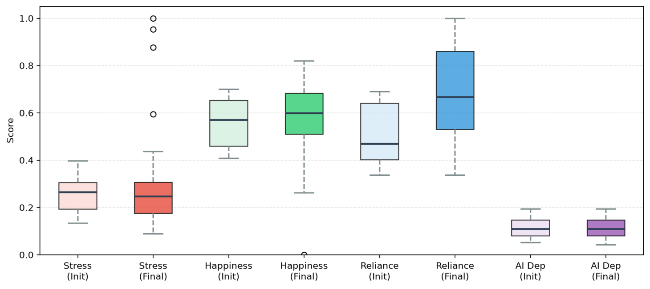}\caption{No AI}\end{subfigure}\\[1pt]
  \begin{subfigure}[t]{0.72\textwidth}\centering\includegraphics[width=\linewidth]{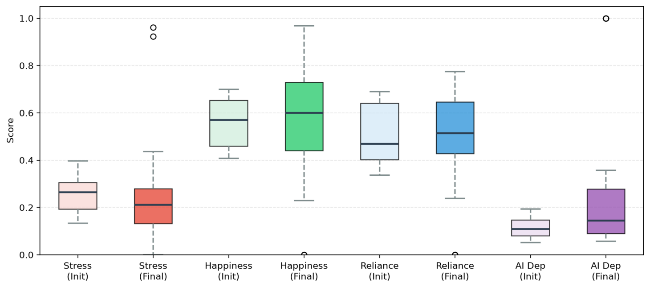}\caption{Affirming}\end{subfigure}\\[1pt]
  \begin{subfigure}[t]{0.72\textwidth}\centering\includegraphics[width=\linewidth]{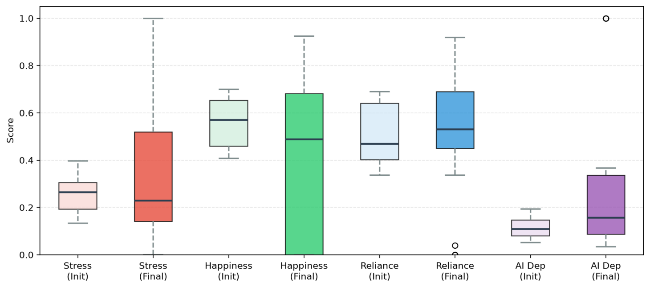}\caption{Listening}\end{subfigure}\\[1pt]
  \begin{subfigure}[t]{0.72\textwidth}\centering\includegraphics[width=\linewidth]{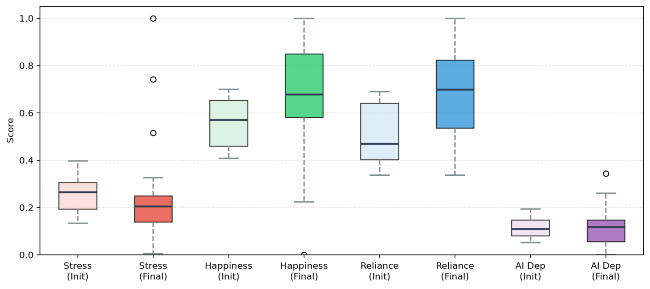}\caption{Solution-oriented}\end{subfigure}
  \caption{Lower-threshold setting (15 days, threshold $0.3$): distributions over the 20 students before day~1 and at the end of day 15. The initial summaries of all panels except (f) are consistent with the seed-42 classroom; the inciting run (f) started from another, unrecorded initial state (Section~\ref{sec:thr}). (Continued on the next page.)}
  \label{fig:thr-box}
\end{figure}
\begin{figure}[p]
  \ContinuedFloat
  \centering
  \begin{subfigure}[t]{0.72\textwidth}\centering\includegraphics[width=\linewidth]{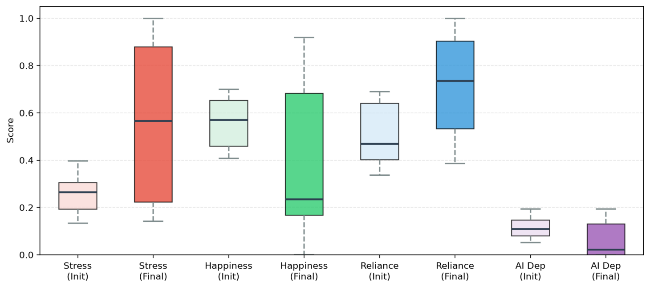}\caption{Reality-redirecting}\end{subfigure}\\[1pt]
  \begin{subfigure}[t]{0.72\textwidth}\centering\includegraphics[width=\linewidth]{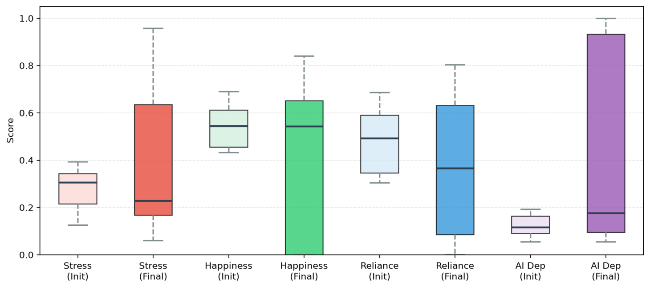}\caption{Inciting}\end{subfigure}\\[1pt]
  \begin{subfigure}[t]{0.72\textwidth}\centering\includegraphics[width=\linewidth]{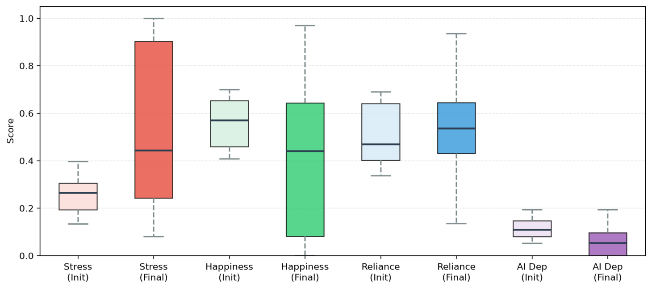}\caption{Blaming}\end{subfigure}
  \caption{(Continued.) Panels (e)--(g) of Figure~\ref{fig:thr-box}.}
\end{figure}

\FloatBarrier
\section{Robustness protocol: per-classroom values, rankings, model variants and decision rules}
\label{app:robust}

This appendix gives the complete values behind Section~\ref{sec:robust}. All runs are 50-day runs with $\theta=0.5$, $N=20$ agents and the blind evaluator; a run is one realization of one condition in one classroom. The tables are regenerated by \texttt{robustness\_report.py} from the released logs (\texttt{data/robustness\_long50/} of the analysis package).

Table~\ref{tab:robust-classrooms} lists the day-50 class means of every run of the block design (Section~\ref{sec:robust-blocks}): ten classrooms (seeds 41--50), seven conditions, one realization each. Table~\ref{tab:robust-ranks} gives the consistency of the ranking of the seven conditions across the ten classrooms: Kendall's $W$ (average ranks, no tie correction; ties occur only in the non-attendance counts), the mean rank of each condition (1 = lowest value) and the number of classrooms in which it had the highest and the lowest value.

\begin{table}[!htb]
  \centering
  \caption{Block design: day-50 class means (non-attending: count of agents) of all 70 runs, by classroom (seed) and condition. Every classroom's seven runs start from the same initial state; the No-AI runs are deterministic. Source: \texttt{robustness/classroom\_values.csv}.}
  \label{tab:robust-classrooms}
  \scriptsize\setlength{\tabcolsep}{4pt}
  \begin{tabular}{@{}l*{10}{c}@{}}
\toprule
Condition / classroom (seed) & 41 & 42 & 43 & 44 & 45 & 46 & 47 & 48 & 49 & 50 \\
\midrule
\multicolumn{11}{@{}l}{\emph{Stress}} \\
\quad No AI & 0.519 & 0.485 & 0.468 & 0.420 & 0.491 & 0.438 & 0.397 & 0.529 & 0.457 & 0.468 \\
\quad Affirming & 0.285 & 0.464 & 0.279 & 0.286 & 0.331 & 0.315 & 0.397 & 0.394 & 0.324 & 0.368 \\
\quad Listening & 0.670 & 0.740 & 0.692 & 0.707 & 0.699 & 0.615 & 0.596 & 0.652 & 0.715 & 0.697 \\
\quad Solution-oriented & 0.411 & 0.420 & 0.471 & 0.405 & 0.488 & 0.530 & 0.379 & 0.442 & 0.455 & 0.482 \\
\quad Reality-redirecting & 0.618 & 0.614 & 0.644 & 0.680 & 0.614 & 0.518 & 0.657 & 0.615 & 0.640 & 0.531 \\
\quad Inciting & 0.759 & 0.681 & 0.751 & 0.693 & 0.736 & 0.561 & 0.667 & 0.692 & 0.716 & 0.741 \\
\quad Blaming & 0.747 & 0.886 & 0.671 & 0.718 & 0.818 & 0.696 & 0.850 & 0.783 & 0.792 & 0.736 \\
\midrule
\multicolumn{11}{@{}l}{\emph{Happiness}} \\
\quad No AI & 0.653 & 0.618 & 0.708 & 0.640 & 0.644 & 0.671 & 0.705 & 0.551 & 0.588 & 0.639 \\
\quad Affirming & 0.458 & 0.383 & 0.538 & 0.484 & 0.405 & 0.337 & 0.474 & 0.346 & 0.396 & 0.412 \\
\quad Listening & 0.340 & 0.193 & 0.281 & 0.241 & 0.234 & 0.331 & 0.362 & 0.278 & 0.254 & 0.372 \\
\quad Solution-oriented & 0.794 & 0.732 & 0.697 & 0.716 & 0.658 & 0.600 & 0.807 & 0.707 & 0.683 & 0.666 \\
\quad Reality-redirecting & 0.463 & 0.500 & 0.451 & 0.344 & 0.423 & 0.575 & 0.405 & 0.452 & 0.381 & 0.574 \\
\quad Inciting & 0.208 & 0.296 & 0.137 & 0.215 & 0.227 & 0.315 & 0.194 & 0.146 & 0.163 & 0.226 \\
\quad Blaming & 0.314 & 0.150 & 0.409 & 0.292 & 0.251 & 0.369 & 0.184 & 0.247 & 0.267 & 0.311 \\
\midrule
\multicolumn{11}{@{}l}{\emph{Self-reliance}} \\
\quad No AI & 0.776 & 0.781 & 0.753 & 0.788 & 0.796 & 0.707 & 0.785 & 0.827 & 0.791 & 0.792 \\
\quad Affirming & 0.227 & 0.369 & 0.280 & 0.368 & 0.318 & 0.189 & 0.397 & 0.322 & 0.241 & 0.304 \\
\quad Listening & 0.624 & 0.506 & 0.500 & 0.486 & 0.495 & 0.459 & 0.553 & 0.540 & 0.485 & 0.618 \\
\quad Solution-oriented & 0.735 & 0.785 & 0.723 & 0.728 & 0.785 & 0.655 & 0.743 & 0.782 & 0.761 & 0.801 \\
\quad Reality-redirecting & 0.838 & 0.835 & 0.845 & 0.801 & 0.835 & 0.763 & 0.847 & 0.809 & 0.768 & 0.803 \\
\quad Inciting & 0.248 & 0.402 & 0.138 & 0.284 & 0.290 & 0.222 & 0.190 & 0.150 & 0.283 & 0.252 \\
\quad Blaming & 0.433 & 0.515 & 0.586 & 0.408 & 0.448 & 0.380 & 0.358 & 0.467 & 0.474 & 0.492 \\
\midrule
\multicolumn{11}{@{}l}{\emph{AI dependence}} \\
\quad No AI & 0.113 & 0.115 & 0.123 & 0.120 & 0.124 & 0.126 & 0.125 & 0.133 & 0.136 & 0.115 \\
\quad Affirming & 0.686 & 0.608 & 0.655 & 0.540 & 0.610 & 0.719 & 0.538 & 0.594 & 0.718 & 0.696 \\
\quad Listening & 0.253 & 0.338 & 0.313 & 0.296 & 0.326 & 0.291 & 0.281 & 0.322 & 0.330 & 0.197 \\
\quad Solution-oriented & 0.107 & 0.117 & 0.140 & 0.108 & 0.094 & 0.108 & 0.084 & 0.110 & 0.083 & 0.082 \\
\quad Reality-redirecting & 0.036 & 0.030 & 0.033 & 0.011 & 0.023 & 0.037 & 0.022 & 0.022 & 0.040 & 0.039 \\
\quad Inciting & 0.675 & 0.533 & 0.818 & 0.612 & 0.624 & 0.584 & 0.713 & 0.806 & 0.655 & 0.643 \\
\quad Blaming & 0.031 & 0.009 & 0.057 & 0.030 & 0.036 & 0.042 & 0.006 & 0.071 & 0.052 & 0.030 \\
\midrule
\multicolumn{11}{@{}l}{\emph{Non-attending}} \\
\quad No AI & 2 & 2 & 1 & 1 & 2 & 0 & 0 & 2 & 1 & 2 \\
\quad Affirming & 4 & 5 & 2 & 6 & 5 & 7 & 5 & 7 & 4 & 6 \\
\quad Listening & 10 & 14 & 12 & 13 & 13 & 12 & 9 & 12 & 13 & 10 \\
\quad Solution-oriented & 1 & 1 & 2 & 1 & 3 & 3 & 0 & 3 & 2 & 1 \\
\quad Reality-redirecting & 6 & 6 & 6 & 9 & 5 & 2 & 7 & 5 & 7 & 5 \\
\quad Inciting & 13 & 12 & 16 & 14 & 15 & 12 & 15 & 16 & 14 & 13 \\
\quad Blaming & 11 & 13 & 7 & 11 & 14 & 9 & 14 & 12 & 12 & 11 \\
\bottomrule
\end{tabular}
\end{table}

\begin{table}[!htb]
  \centering
  \caption{Consistency of the condition ranking over the ten classrooms: Kendall's $W$ per outcome and, per condition, the mean rank (1 = lowest value, 7 = highest) followed in parentheses by the number of classrooms in which the condition had the highest / the lowest value (ties share the count). Source: \texttt{robustness/block\_rankings.csv}.}
  \label{tab:robust-ranks}
  \scriptsize\setlength{\tabcolsep}{3.5pt}
  \begin{tabular}{@{}lcccccccc@{}}
\toprule
Outcome & $W$ & No AI & Affirming & Listening & Solution-oriented & Reality-redirecting & Inciting & Blaming \\
\midrule
Stress & 0.891 & 2.6 (0/0) & 1.3 (0/8) & 5.3 (0/0) & 2.2 (0/2) & 4.0 (0/0) & 6.0 (3/0) & 6.6 (7/0) \\
Happiness & 0.912 & 6.2 (2/0) & 4.3 (0/0) & 2.4 (0/0) & 6.8 (8/0) & 4.6 (0/0) & 1.3 (0/8) & 2.4 (0/2) \\
Self-reliance & 0.935 & 6.0 (2/0) & 1.7 (0/4) & 3.8 (0/0) & 5.2 (0/0) & 6.8 (8/0) & 1.4 (0/6) & 3.1 (0/0) \\
AI dep. & 0.954 & 3.8 (0/0) & 6.5 (5/0) & 5.0 (0/0) & 3.2 (0/0) & 1.4 (0/6) & 6.5 (5/0) & 1.6 (0/4) \\
Non-att. & 0.906 & 1.4 (0/7) & 3.3 (0/0) & 5.7 (2/0) & 1.8 (0/5) & 3.5 (0/0) & 6.8 (9/0) & 5.5 (0/0) \\
\bottomrule
\end{tabular}
\end{table}

Table~\ref{tab:robust-arms} gives the day-50 values of the seed-42 classroom under the default rules (mean and range of the seven realizations of Table~\ref{tab:robust-llm}) and under each model variant of Section~\ref{sec:robust-lambda} (one realization each); an asterisk marks a value outside the range of the seven default-rule realizations. Under $\lambda<1$ and under neutral labels the No-AI run is the default one; under the other variants it is the control run of Table~\ref{tab:robust-controls}.

\begin{table}[!htb]
  \centering
  \caption{Seed-42 classroom on day 50 under the default rules and under each model variant. Default rules: mean [min, max] over the seven realizations of Table~\ref{tab:robust-llm}; variants: one realization each; $^{\ast}$ = outside the range of the seven default-rule realizations. Source: \texttt{robustness/sensitivity.csv}, \texttt{robustness/llm\_realizations.csv}.}
  \label{tab:robust-arms}
  \scriptsize\setlength{\tabcolsep}{3pt}\renewcommand{\arraystretch}{0.9}
  \begin{tabular}{@{}lL{3.25cm}*{7}{C{1.2cm}}@{}}
\toprule
Outcome & Variant & No AI & Affirming & Listening & Solution-or. & Reality-red. & Inciting & Blaming \\
\midrule
Stress & Default rules (7 realizations: mean, [min, max]) & 0.485 & 0.401\newline [0.335, 0.464] & 0.686\newline [0.600, 0.780] & 0.483\newline [0.420, 0.555] & 0.613\newline [0.553, 0.721] & 0.669\newline [0.590, 0.702] & 0.780\newline [0.706, 0.886] \\
 & $\lambda=0.1$ & 0.485 & 0.531$^{\ast}$ & 0.580$^{\ast}$ & 0.526 & 0.532$^{\ast}$ & 0.627 & 0.589$^{\ast}$ \\
 & $\lambda=0.3$ & 0.485 & 0.534$^{\ast}$ & 0.678 & 0.557$^{\ast}$ & 0.512$^{\ast}$ & 0.718$^{\ast}$ & 0.633$^{\ast}$ \\
 & Neutral type labels & 0.485 & 0.285$^{\ast}$ & 0.628 & 0.380$^{\ast}$ & 0.617 & 0.527$^{\ast}$ & 0.708 \\
 & $D$-coefficients $0$ & 0.505 & 0.440 & 0.565$^{\ast}$ & 0.512 & 0.601 & 0.691 & 0.784 \\
 & Bottle-up $R$ gain $0$ & 0.485 & 0.334$^{\ast}$ & 0.690 & 0.488 & 0.597 & 0.687 & 0.708 \\
 & No noon peer effects & 0.712 & 0.295$^{\ast}$ & 0.577$^{\ast}$ & 0.571$^{\ast}$ & 0.713 & 0.757$^{\ast}$ & 0.893$^{\ast}$ \\
\midrule
Happiness & Default rules (7 realizations: mean, [min, max]) & 0.618 & 0.431\newline [0.383, 0.493] & 0.231\newline [0.149, 0.309] & 0.661\newline [0.584, 0.732] & 0.463\newline [0.345, 0.524] & 0.274\newline [0.220, 0.340] & 0.289\newline [0.150, 0.382] \\
 & $\lambda=0.1$ & 0.618 & 0.578$^{\ast}$ & 0.542$^{\ast}$ & 0.593 & 0.581$^{\ast}$ & 0.511$^{\ast}$ & 0.520$^{\ast}$ \\
 & $\lambda=0.3$ & 0.618 & 0.529$^{\ast}$ & 0.330$^{\ast}$ & 0.568$^{\ast}$ & 0.587$^{\ast}$ & 0.321 & 0.482$^{\ast}$ \\
 & Neutral type labels & 0.618 & 0.671$^{\ast}$ & 0.314$^{\ast}$ & 0.756$^{\ast}$ & 0.454 & 0.218$^{\ast}$ & 0.382 \\
 & $D$-coefficients $0$ & 0.685 & 0.630$^{\ast}$ & 0.532$^{\ast}$ & 0.693 & 0.523 & 0.334 & 0.318 \\
 & Bottle-up $R$ gain $0$ & 0.618 & 0.386 & 0.242 & 0.633 & 0.441 & 0.209$^{\ast}$ & 0.382 \\
 & No noon peer effects & 0.194 & 0.256$^{\ast}$ & 0.123$^{\ast}$ & 0.379$^{\ast}$ & 0.215$^{\ast}$ & 0.048$^{\ast}$ & 0.066$^{\ast}$ \\
\midrule
Self-reliance & Default rules (7 realizations: mean, [min, max]) & 0.781 & 0.367\newline [0.262, 0.449] & 0.509\newline [0.467, 0.588] & 0.819\newline [0.784, 0.855] & 0.865\newline [0.831, 0.917] & 0.338\newline [0.240, 0.422] & 0.506\newline [0.480, 0.527] \\
 & $\lambda=0.1$ & 0.781 & 0.792$^{\ast}$ & 0.808$^{\ast}$ & 0.819 & 0.813$^{\ast}$ & 0.730$^{\ast}$ & 0.790$^{\ast}$ \\
 & $\lambda=0.3$ & 0.781 & 0.610$^{\ast}$ & 0.777$^{\ast}$ & 0.823 & 0.809$^{\ast}$ & 0.531$^{\ast}$ & 0.707$^{\ast}$ \\
 & Neutral type labels & 0.781 & 0.329 & 0.603$^{\ast}$ & 0.825 & 0.844 & 0.231$^{\ast}$ & 0.538$^{\ast}$ \\
 & $D$-coefficients $0$ & 0.783 & 0.452$^{\ast}$ & 0.596$^{\ast}$ & 0.783$^{\ast}$ & 0.828$^{\ast}$ & 0.358 & 0.444$^{\ast}$ \\
 & Bottle-up $R$ gain $0$ & 0.674 & 0.246$^{\ast}$ & 0.404$^{\ast}$ & 0.783$^{\ast}$ & 0.833 & 0.258 & 0.422$^{\ast}$ \\
 & No noon peer effects & 0.933 & 0.261$^{\ast}$ & 0.609$^{\ast}$ & 0.967$^{\ast}$ & 0.917$^{\ast}$ & 0.286 & 0.448$^{\ast}$ \\
\midrule
AI dep. & Default rules (7 realizations: mean, [min, max]) & 0.115 & 0.563\newline [0.463, 0.646] & 0.341\newline [0.265, 0.409] & 0.103\newline [0.083, 0.128] & 0.027\newline [0.012, 0.044] & 0.587\newline [0.482, 0.664] & 0.030\newline [0.009, 0.045] \\
 & $\lambda=0.1$ & 0.115 & 0.166$^{\ast}$ & 0.131$^{\ast}$ & 0.113 & 0.060$^{\ast}$ & 0.233$^{\ast}$ & 0.059$^{\ast}$ \\
 & $\lambda=0.3$ & 0.115 & 0.371$^{\ast}$ & 0.207$^{\ast}$ & 0.103 & 0.032 & 0.394$^{\ast}$ & 0.030 \\
 & Neutral type labels & 0.115 & 0.631 & 0.295 & 0.128$^{\ast}$ & 0.016 & 0.684$^{\ast}$ & 0.045 \\
 & $D$-coefficients $0$ & 0.116 & 0.446$^{\ast}$ & 0.237$^{\ast}$ & 0.100 & 0.043 & 0.585 & 0.042 \\
 & Bottle-up $R$ gain $0$ & 0.115 & 0.688$^{\ast}$ & 0.342 & 0.079$^{\ast}$ & 0.021 & 0.633 & 0.045 \\
 & No noon peer effects & 0.116 & 0.794$^{\ast}$ & 0.491$^{\ast}$ & 0.079$^{\ast}$ & 0.029 & 0.614 & 0.020 \\
\midrule
Non-att. & Default rules (7 realizations: mean, [min, max]) & 2 & 6.1\newline [5, 8] & 13.1\newline [12, 15] & 2.4\newline [1, 4] & 5.7\newline [3, 10] & 12.3\newline [10, 15] & 11.4\newline [10, 13] \\
 & $\lambda=0.1$ & 2 & 3$^{\ast}$ & 3$^{\ast}$ & 2 & 2$^{\ast}$ & 4$^{\ast}$ & 3$^{\ast}$ \\
 & $\lambda=0.3$ & 2 & 3$^{\ast}$ & 8$^{\ast}$ & 3 & 2$^{\ast}$ & 11 & 5$^{\ast}$ \\
 & Neutral type labels & 2 & 0$^{\ast}$ & 11$^{\ast}$ & 1 & 6 & 13 & 10 \\
 & $D$-coefficients $0$ & 0 & 3$^{\ast}$ & 5$^{\ast}$ & 1 & 3 & 11 & 9$^{\ast}$ \\
 & Bottle-up $R$ gain $0$ & 2 & 7 & 12 & 3 & 4 & 13 & 10 \\
 & No noon peer effects & 7 & 7 & 14 & 3 & 9 & 17$^{\ast}$ & 16$^{\ast}$ \\
\bottomrule
\end{tabular}
\end{table}

Table~\ref{tab:robust-decision} applies the three pre-specified rules of Section~\ref{sec:robust-rules} to the 30 condition--outcome differences from No AI: (i) the 95\% bootstrap confidence interval over the ten classrooms excludes $0$ and at least nine classrooms share the sign of the mean; (ii) over the seven realizations of the seed-42 block the absolute mean difference is at least twice its standard deviation and every realization has the same sign; (iii) at $\lambda=0.3$ the sign of the difference is preserved and the Spearman correlation of the seven-condition ranking for that outcome with the default-rule ranking is at least $0.8$. A difference is robust when all three hold.

\begin{table}[!htb]
  \centering
  \caption{Pre-specified decision rules applied to the 30 differences from No AI (Section~\ref{sec:robust-rules}). Columns: 95\% CI over classrooms and number of classrooms with the sign of the mean; rule (i); $|\bar d|/\mathrm{SD}$ over the seven realizations and whether all realizations share the sign; rule (ii); sign of the difference at $\lambda=0.3$ relative to the default-rule sign and Spearman $\rho$ of the outcome's seven-condition ranking at $\lambda=0.3$; rule (iii); verdict. Source: \texttt{robustness/decision\_rules.csv}.}
  \label{tab:robust-decision}
  \scriptsize\setlength{\tabcolsep}{3pt}
  \begin{tabular}{@{}llccccccl@{}}
\toprule
Outcome & Condition & 95\% CI; same sign & (i) & $|\bar d|/\mathrm{SD}$; all same sign & (ii) & sign at $\lambda=0.3$; $\rho$ & (iii) & Verdict \\
\midrule
Stress & Affirming & [-0.163, -0.081]; 9/10 & yes & 1.6; yes & no & reversed; 0.54 & no & descriptive \\
 & Listening & [+0.181, +0.240]; 10/10 & yes & 2.8; yes & yes & same; 0.54 & no & descriptive \\
 & Solution-oriented & [-0.052, +0.015]; 7/10 & no & 0.1; no & no & reversed; 0.54 & no & descriptive \\
 & Reality-redirecting & [+0.106, +0.189]; 10/10 & yes & 2.3; yes & yes & same; 0.54 & no & descriptive \\
 & Inciting & [+0.197, +0.262]; 10/10 & yes & 4.4; yes & yes & same; 0.54 & no & descriptive \\
 & Blaming & [+0.259, +0.351]; 10/10 & yes & 4.6; yes & yes & same; 0.54 & no & descriptive \\
\midrule
Happiness & Affirming & [-0.250, -0.192]; 10/10 & yes & 5.7; yes & yes & same; 0.75 & no & descriptive \\
 & Listening & [-0.387, -0.318]; 10/10 & yes & 6.7; yes & yes & same; 0.75 & no & descriptive \\
 & Solution-oriented & [+0.020, +0.105]; 8/10 & no & 0.9; no & no & reversed; 0.75 & no & descriptive \\
 & Reality-redirecting & [-0.235, -0.133]; 10/10 & yes & 2.6; yes & yes & same; 0.75 & no & descriptive \\
 & Inciting & [-0.473, -0.390]; 10/10 & yes & 7.2; yes & yes & same; 0.75 & no & descriptive \\
 & Blaming & [-0.411, -0.322]; 10/10 & yes & 4.0; yes & yes & same; 0.75 & no & descriptive \\
\midrule
Self-reliance & Affirming & [-0.510, -0.445]; 10/10 & yes & 7.5; yes & yes & same; 0.89 & yes & \textbf{robust} \\
 & Listening & [-0.283, -0.220]; 10/10 & yes & 7.0; yes & yes & same; 0.89 & yes & \textbf{robust} \\
 & Solution-oriented & [-0.043, -0.016]; 8/10 & no & 1.2; yes & no & reversed; 0.89 & no & descriptive \\
 & Reality-redirecting & [+0.012, +0.056]; 8/10 & no & 2.6; yes & yes & same; 0.89 & yes & descriptive \\
 & Inciting & [-0.582, -0.486]; 10/10 & yes & 5.8; yes & yes & same; 0.89 & yes & \textbf{robust} \\
 & Blaming & [-0.361, -0.279]; 10/10 & yes & 15.0; yes & yes & same; 0.89 & yes & \textbf{robust} \\
\midrule
AI dep. & Affirming & [+0.472, +0.553]; 10/10 & yes & 7.6; yes & yes & same; 0.93 & yes & \textbf{robust} \\
 & Listening & [+0.147, +0.193]; 10/10 & yes & 4.9; yes & yes & same; 0.93 & yes & \textbf{robust} \\
 & Solution-oriented & [-0.032, -0.007]; 8/10 & no & 0.7; no & no & same; 0.93 & yes & descriptive \\
 & Reality-redirecting & [-0.101, -0.087]; 10/10 & yes & 9.3; yes & yes & same; 0.93 & yes & \textbf{robust} \\
 & Inciting & [+0.494, +0.597]; 10/10 & yes & 6.7; yes & yes & same; 0.93 & yes & \textbf{robust} \\
 & Blaming & [-0.097, -0.077]; 10/10 & yes & 6.4; yes & yes & same; 0.93 & yes & \textbf{robust} \\
\midrule
Non-att. & Affirming & [+2.8, +4.8]; 10/10 & yes & 3.4; yes & yes & same; 0.65 & no & descriptive \\
 & Listening & [+9.5, +11.4]; 10/10 & yes & 9.2; yes & yes & same; 0.65 & no & descriptive \\
 & Solution-oriented & [-0.3, +1.2]; 5/10 & no & 0.4; no & no & same; 0.65 & no & descriptive \\
 & Reality-redirecting & [+3.4, +5.7]; 10/10 & yes & 1.6; yes & no & none; 0.65 & no & descriptive \\
 & Inciting & [+11.7, +13.7]; 10/10 & yes & 5.4; yes & yes & same; 0.65 & no & descriptive \\
 & Blaming & [+8.9, +11.3]; 10/10 & yes & 7.4; yes & yes & same; 0.65 & no & descriptive \\
\bottomrule
\end{tabular}
\end{table}

\end{document}